\documentclass[twocolumn]{aastex631}
\usepackage{amsmath}
\usepackage{booktabs}
\usepackage{amssymb} 
\usepackage{hyperref}
\usepackage{soul}
\setstcolor{red}

\usepackage{tikz}
\usetikzlibrary{shapes,arrows,chains}  

\definecolor{mygreen}{rgb}{0,0.5,0}  

\newcommand{\tsz}{\textnormal{tSZ}}

\usepackage[utf8x]{inputenc}
\shorttitle{The Hydrostatic Mass Bias and the $\sigma_8$ Tension}
\shortauthors{Ibitoye et al.}
\graphicspath{{./}{figures/}}
\hypersetup{
    colorlinks=true,
    linkcolor=blue,
    citecolor=blue,
    filecolor=magenta,      
    urlcolor=blue,
    pdftitle={The Hydrostatic Mass Bias and the $\sigma_8$ Tension: A Multi-Probe Forecast for Stage-IV/V Surveys},
    pdfauthor={Ayodeji Ibitoye, Prabhakar Tiwari},
    pdfsubject={Cosmology, Hydrostatic Mass Bias, CMB-S4, LSST, CSST, SO},
    pdfkeywords={Hydrostatic Mass Bias, tSZ, Galaxy Clustering, Weak Lensing, Fisher Matrix, CMB-S4, LSST, CSST, SO}
}

\begin{document}

\title{The Hydrostatic Mass Bias and the $\sigma_8$ Tension: A Multi-Probe Forecast for Stage-IV/V Surveys}

\author{Ayodeji Ibitoye}
\email{ayodeji.ibitoye@gtiit.edu.cn}
\affiliation{Department of Physics, Guangdong Technion - Israel Institute of Technology, Shantou, Guangdong 515063, P.R. China}
\affiliation{Centre for Space Research, North-West University, Potchefstroom 2520, South Africa}
\affiliation{Department of Physics and Electronics, Adekunle Ajasin University, P. M. B. 001, Akungba-Akoko, Ondo State, Nigeria}

\author{Prabhakar Tiwari}
\email{prabhakar.tiwari@gtiit.edu.cn}
\affiliation{Department of Physics, Guangdong Technion - Israel Institute of Technology, Shantou, Guangdong 515063, P.R. China}

\author{Qi Xiong}
\affiliation{National Astronomical Observatories, Chinese Academy of Sciences, 
\\20A Datun Road, Chaoyang District, Beijing 100101, P. R. China}

\author{Yan Gong}
\affiliation{National Astronomical Observatories, Chinese Academy of Sciences, 
\\20A Datun Road, Chaoyang District, Beijing 100101, P. R. China}

\begin{abstract}

The hydrostatic mass bias ($b_{\mathrm{HSE}}$) is a leading systematic uncertainty in cluster cosmology and a principal source of degeneracy with $\sigma_8$ and $\Omega_m$. We investigate the capability of Stage-IV CMB and optical surveys to calibrate $b_{\mathrm{HSE}}$ using tomographic cross-correlations between the thermal Sunyaev--Zel'dovich (tSZ) effect, galaxy clustering, and weak lensing. We perform a Fisher forecast incorporating realistic survey noise, foreground modeling for clustered CIB and radio sources, and full marginalization over cosmological and astrophysical nuisance parameters, including per-bin galaxy bias perturbations, photometric redshift shifts, intrinsic alignments, and baryonic feedback modeled with HMCode2020. With optimized tomographic binning (10 lens and 5 source bins for LSST; 6 lens and 5 source bins for CSST), we forecast marginalized constraints of $0.98\%$ for SO+LSST, $1.60\%$ for CMB-S4+LSST, and $2.40\%$ for CMB-S4+CSST. Tomography improves $b_{\mathrm{HSE}}$ precision by factors of approximately three relative to non-tomographic analyses, reflecting the role of redshift information in breaking the $b_{\mathrm{HSE}}$--$\sigma_8$ degeneracy. Optical-only probes provide no direct constraint on $b_{\mathrm{HSE}}$, whereas inclusion of tSZ-containing spectra enables percent-level calibration under realistic systematic assumptions. The results demonstrate that multi-probe tomographic analyses with Stage-IV surveys can achieve robust control of hydrostatic mass bias, strengthening cluster-based constraints on structure growth.
\end{abstract}

\keywords{hydrostatic mass bias, tSZ–galaxy cross-correlations, tomography, CMB-S4, LSST, CSST}

\section{Introduction}
\label{sec:introduction}

Galaxy clusters, the most massive collapsed structures in the universe, serve as powerful probes of cosmic growth. Their abundance as a function of mass and redshift is exquisitely sensitive to the parameters governing the background cosmology and the growth of structure, most notably the total matter density $\Omega_m$ and the amplitude of matter fluctuations, $\sigma_8$ \citep{Allen11, Vikhlinin09}. As cosmology enters the Stage-IV and Stage-V era of unprecedented survey precision, the limiting factor in extracting cosmological information from clusters is no longer statistical uncertainty but the control of systematic effects \citep{Nikolina25, Euclid_sys_25, Shiming25, Shiming26}. {In this precision-dominated regime, accurate and internally consistent calibration of cluster mass estimates becomes essential to avoid biased cosmological inference.}

The most significant of these systematics is the so-called hydrostatic mass bias. This parameter quantifies the fractional offset between the true cluster mass and the mass inferred under the assumption of hydrostatic equilibrium (HSE) applied to the hot intracluster medium (ICM), and is denoted by $b_{\mathrm{HSE}}$. The HSE assumption systematically underestimates the true mass ($b_{\mathrm{HSE}} > 0$) because it neglects non-thermal pressure support from sources such as turbulent gas motions, bulk flows, and energy injection from active galactic nuclei \citep{Nagai07, Nelson14}. While often modeled as a constant in cosmological analyses, its value and potential dependencies on redshift and mass remain a primary source of uncertainty. 

This uncertainty in $b_{\mathrm{HSE}}$ is not merely a limitation for cluster counts; it is intrinsically linked to one of the most persistent challenges to the standard $\Lambda$CDM model: the $\sigma_8$ tension. Measurements of the clustering amplitude from the primary Cosmic Microwave Background (CMB), most notably from the \textit{Planck} satellite, favor a higher value of $\sigma_8$ \citep{Planck20}. In contrast, {earlier} low-redshift probes reported lower values \citep{Asgari21, Abbott22}. However, recent re-analyses, including KiDS-Legacy \citep{Wright25}, HSC-Y3 3$\times$2pt \citep{Zhang25}, and HSC-Y3 cosmic shear \citep{Choppin25} yield higher $S_8$ values ($\sim 0.79$--0.81), reducing the historical discrepancy with \textit{Planck} ($S_8 = 0.832$). {Crucially, the tension persists even in independent probes of structure growth: redshift-space distortion (RSD) measurements of $f\sigma_8$, when combined with BAO and SNe Ia, favor $S_8 = 0.762^{+0.030}_{-0.025}$, in $\sim 2.2\sigma$ tension with \textit{Planck} \citep{Nunes21}.} This suggests the discrepancy is not solely attributable to weak lensing systematics. As demonstrated by \citet{Aymerich25} and others \citep{Bolliet20, Raghunathan22a}, even this evolving tension can be reconciled within $\Lambda$CDM if the $b_{\mathrm{HSE}}$ mass bias is significantly higher than canonical simulation calibrations, highlighting a profound degeneracy between $b_{\mathrm{HSE}}$ and $\sigma_8$.Therefore, an independent, precise calibration of $b_{\mathrm{HSE}}$ remains imperative to determine whether the residual discrepancy signifies new physics or an unmitigated astrophysical systematics.

Past observational efforts have only achieved modest precision on $b_{\mathrm{HSE}}$. Weak lensing mass calibration of \textit{Planck} clusters yielded $\sigma(b_{\mathrm{HSE}}) \sim 0.05$--$0.10$ \citep{Planck16, Miyatake19}, while cross-correlation studies between \textit{Planck} tSZ and DES weak lensing reached $\sigma(b_{\mathrm{HSE}}) \approx 0.04$ \citep{Pandey22}. These uncertainties are insufficient for Stage-IV/V cosmology, where sub-percent control over systematics is required to fully exploit the statistical power of next-generation data.

The path to breaking the $b_{\mathrm{HSE}}$–$\sigma_8$ degeneracy lies in multi-wavelength, multi-probe observations that can directly constrain the underlying cluster gravitational potential. The power of cross-correlations to break degeneracies is now well-established in the literature. Recent work by \citet{Zhang22} demonstrated that combining galaxy clustering and weak lensing with CMB lensing can significantly improve cosmological constraints by breaking degeneracies between parameters and systematics, particularly when moving from Stage-III to Stage-IV surveys. Their analysis focused on galaxy-CMB lensing cross-correlations ($\delta_g \kappa_{\rm CMB}$ and $\gamma \kappa_{\rm CMB}$). { Recent joint analyses of tSZ, X-ray, and weak lensing have demonstrated that multi-probe cross-correlations can simultaneously constrain cosmological parameters and intracluster medium (ICM) physics, including the non-thermal pressure support that underlies the hydrostatic mass bias \citep{Shirasaki20}.} Our analysis extends this framework to the thermal Sunyaev–Zeldovich effect, where we show that tSZ–galaxy cross-correlations can obtain $b_{\rm HSE}$ to sub-percent precision, a critical step for resolving the $\sigma_8$ tension. Recent cosmological hydrodynamical simulations, such as FLAMINGO and The Three Hundred Project, have {quantified} $b_{\mathrm{HSE}}$ and its scatter finding a typical value of $1 - b_{\mathrm{HSE}} \sim 0.7$--$0.8$ and a complex dependence on cluster dynamical state \citep{Braspenning25, Corasaniti25}. 

Building on these advances, forecast studies have established the potential of combining Stage-IV surveys using auto- and cross-correlations of CMB lensing, galaxy clustering, and weak lensing to suppress cosmic variance and break parameter degeneracies \citep{Euclid20, Zhang22, Fert24, Xiong25}. These cross-correlations also leverage complementary redshift sensitivity: CMB lensing probes the matter distribution with a kernel peaking around $z \sim 2$ (see Figure 49 in \citet{Qu24}), while optical weak lensing typically peaks at $z \sim 1$ \citep{CSST}, extending our ability to trace structure growth across cosmic time and further breaking parameter degeneracies \citep{Zhang22}. However, many existing forecasts assume coarse or no redshift binning, leaving the critical $b_{\mathrm{HSE}}$–$\sigma_8$ degeneracy largely unbroken. Our analysis demonstrates that non-tomographic analyses degrade constraints on $b_{\mathrm{HSE}}$ by factors of 2.5--3 compared to tomographic approaches, underscoring the essential role of fine redshift binning in achieving the sub-percent precision required for Stage-IV/V cosmology.

The next generation of surveys is poised to provide the data required for this measurement. In the optical/near-infrared, the Vera C.\ Rubin Observatory's Legacy Survey of Space and Time (LSST) will deliver deep, high-resolution weak lensing shear and galaxy density maps over half the sky \citep{LSSTScience09}. The Chinese Space Station Survey Telescope (CSST), with its unique combination of a wide field and high spatial resolution from space, will provide exceptional depth and resolution for weak lensing, complementing ground-based surveys \citep{CSST, CSST_ii, CSST_iii}. In the millimeter-wave, CMB-S4 will produce a deep, high-resolution map of the CMB, enabling a nearly mass-limited catalog of clusters via the thermal Sunyaev--Zeldovich (tSZ) effect and a high-fidelity CMB lensing convergence map \citep{CMBS4_Book, CMBS4_paper}. The Simons Observatory (SO), a precursor to CMB-S4, will already achieve transformative sensitivity for these measurements over a significant sky area \citep{SO_22,SO_25}.

The unique power of our analysis lies in combining these complementary datasets. CMB-S4 and SO provide the tSZ cluster sample and CMB lensing map that is immune to the shape measurement systematics and photo-$z$ uncertainties that affect optical weak lensing. LSST and CSST provide high-signal-to-noise galaxy density and high-resolution weak lensing shear ($\gamma$) maps, with CSST's space-based data offering superior control of systematics on small scales. The cross-correlation between CMB lensing and optical weak lensing shear is especially powerful, as it mitigates the intrinsic alignment systematic that plagues auto-correlation analyses.

In this paper, we present a comprehensive forecast demonstrating that a joint, tomographic analysis of the cross-correlations between the tSZ effect ($y$), galaxy density ($g$), and weak lensing shear ($\gamma$), using all six associated two-point power spectra ($\mathcal{C}_\ell^{yy}, \mathcal{C}_\ell^{gg}, \mathcal{C}_\ell^{\gamma\gamma}, \mathcal{C}_\ell^{y\gamma}, \mathcal{C}_\ell^{yg}, \mathcal{C}_\ell^{\gamma g}$) can constrain $b_{\mathrm{HSE}}$ to sub-percent precision. We build a unified halo model pipeline and employ a Fisher matrix formalism to forecast constraints for combinations of CMB-S4, SO, LSST (Year 10), and CSST. We show that our tomographic method, using five source redshift bins, reduces the uncertainty on $b_{\mathrm{HSE}}$ by a factor of 2.5--3 compared to a non-tomographic analysis, directly addressing the degeneracy-breaking strategy emphasized in the CMB-S4 Science Book. This precision not only mitigates a critical systematic but also enables Stage-IV/V-level constraints on $\sigma_8$ and $\Omega_m$, advancing a core science objective of the CMB-S4 mission.
The full analysis pipeline, from survey specifications to final constraints, is summarized in Figure~\ref{fig:pipeline_vertical_corrected}.

\noindent This paper is organized as follows:  
Section~\ref{sec:theory} outlines the theoretical framework, including the halo model and Limber approximation.  
Section~\ref{sec:survey_specs} details the specifications of CMB-S4, Simons Observatory, LSST, and CSST.  
Section~\ref{sec:forecast} presents the Fisher matrix formalism, power spectrum computation, and parameter space.  
Section~\ref{sec:Result_summary} presents our forecasts for $b_{\mathrm{HSE}}$ and cosmological parameters, discusses survey-specific advantages, and summarizes implications for resolving the $\sigma_8$ tension.

\noindent Throughout, we adopt the \textit{Planck} 2018 cosmology as our fiducial model \citep{Planck20}: $\Omega_m = 0.315$, $\Omega_b = 0.049$, $h = 0.674$, $\sigma_8 = 0.811$, and $n_s = 0.965$.

\begin{figure}
    \centering
    \includegraphics[width=0.45\textwidth]{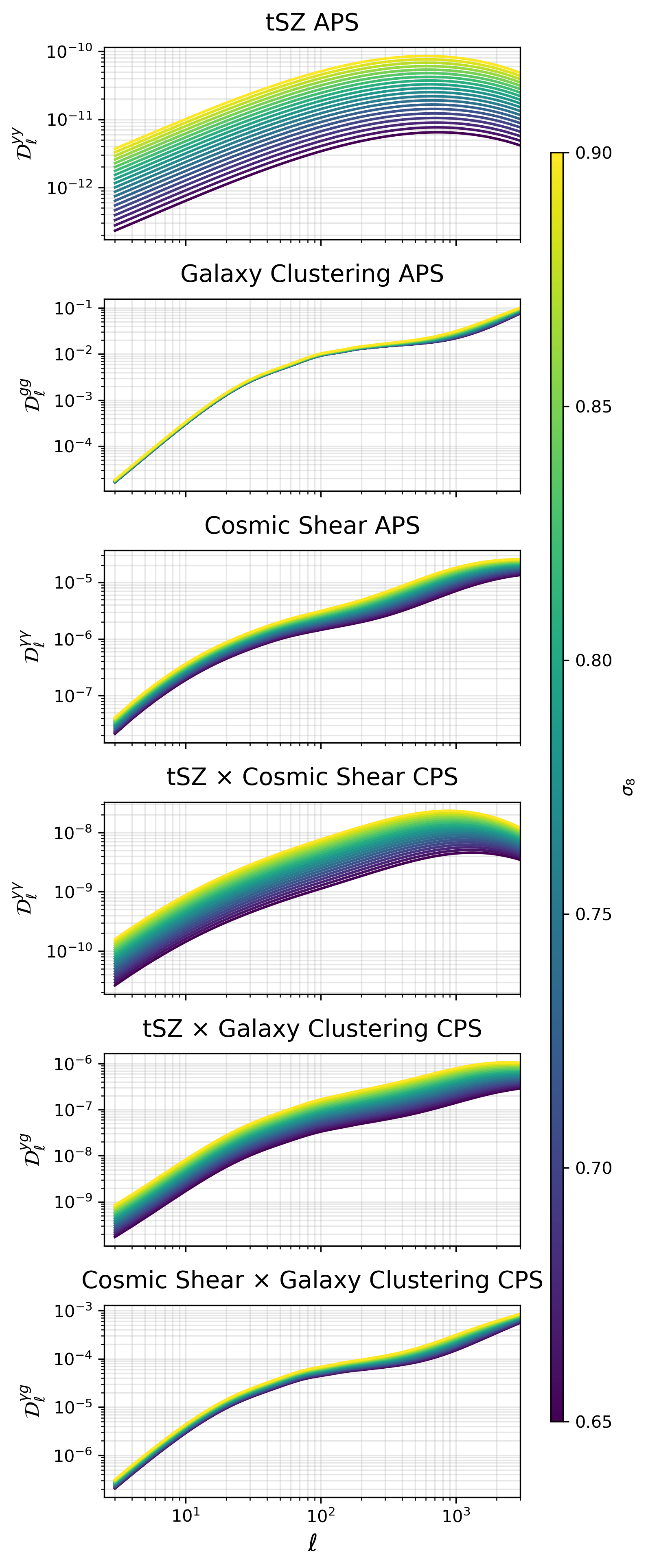}
\caption{Stacked panels showing the sensitivity of each probe and cross-probe combination of the angular power spectra ($\mathcal{D}_\ell= \ell(\ell+1)\mathcal{C}_\ell/(2\pi)$) for auto- and cross-correlations of the thermal Sunyaev-Zel’dovich (tSZ, $y$), galaxy clustering ($g$), and cosmic shear ($\gamma$) to the amplitude of matter density fluctuations ($\sigma_8$), with all other cosmological parameters (e.g., $\Omega_m, h, n_s$) held fixed. APS and CPS represents angular power spectrum and cross-power spectrum, respectively.}
    \label{fig:power_spectra_sigma8}
\end{figure}

\begin{figure*}
\centering
\includegraphics[width=\textwidth]{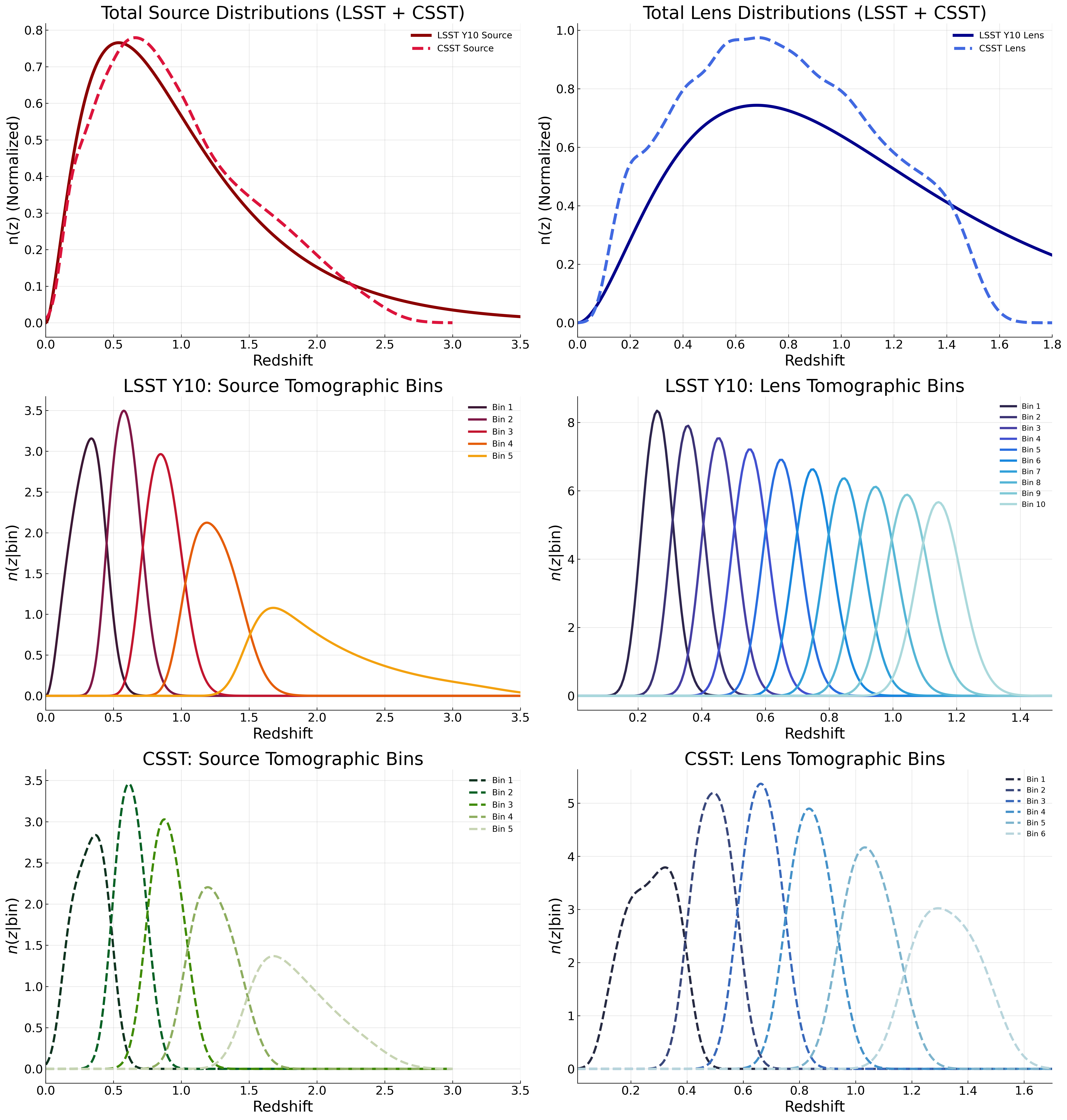}
\caption{
Normalized redshift distributions for LSST Year~10 and CSST. All distributions satisfy $\int n(z)\,dz = 1$, where $n(z)$ denotes the normalized redshift distribution (total: $n(z)$, tomographic bins: $n(z|\mathrm{bin})$). Distributions include photometric redshift uncertainties convolved via the error-function formalism of \citet{Ma06}: $\sigma_z = 0.05(1{+}z)$ (LSST sources), $\sigma_z = 0.03(1{+}z)$ (LSST lenses), and CSST values from mission forecasts ($\sigma_z = 0.05(1{+}z)$ sources, $\sigma_z = 0.03(1{+}z)$ lenses; \citealt{CSST}).
\textit{Top row (Overplotted Totals)}: Direct comparison of total distributions. Left: CSST sources extend to $z \approx 2.8$ with stronger high-$z$ counts than LSST (spanning $z \approx 0$--3.5). Right: Both lens distributions peak at $z \approx 0.5$, with CSST extending to $z \approx 1.8$ (vs. LSST’s focus on $z \approx 0$--1.5).  
\textit{Middle row (LSST Binned)}: Left: 5 equal-count source bins (edges from 1\% peak threshold, ensuring uniform statistical weight). Right: 10 fixed-width lens bins ($z = 0.2$--1.2, $\Delta z = 0.1$) consistent with LSST DESC conventions \citep{DESCSRD}.  
\textit{Bottom row (CSST Binned)}: Left: 5 source bins optimized for tSZ--shear and cosmic shear cross-correlations. Right: 6 lens bins optimized for galaxy--tSZ and clustering cross-correlations, leveraging CSST’s superior photo-$z$ precision.  
LSST distributions use parameters from the Core Cosmology Library (CCL) \citep{Chisari19} while CSST distributions are obtained from simulations following \citet{Xiong25}).
}
\label{fig:dndz_tomographic}
\end{figure*}

\section{Theoretical Framework}
\label{sec:theory}

This section details the theoretical underpinnings and forecasting methodology used to quantify the constraining power of next-generation surveys on the hydrostatic mass bias and cosmological parameters. Our approach is built upon a unified halo model framework that predicts the two-point statistics of three key cosmological observables: the thermal Sunyaev--Zeldovich (tSZ) effect, galaxy number density, and weak gravitational lensing, and their tomographic cross-correlations.

We begin with the theoretical foundation: the {Halo Model} (Section~\ref{subsec:halo_model}), which computes the three-dimensional tracer power spectra $P_{XY}(k,z)$. These 3D spectra are then projected onto the sky using the {Limber Approximation} (Section~\ref{subsec:angular_power_spectra}) to yield observable angular power spectra $C_\ell^{XY}$. To reflect real-world observations, we integrate a detailed {Noise Modeling} component, which is combined with the signal to build a complete {Covariance Matrix}. The final step employs a {Fisher Matrix} analysis to translate this covariance into precise forecasts for parameter constraints.

\subsection{Nonlinear 3D Power Spectra via the Halo Model}
\label{subsec:halo_model}

The halo model provides the foundational framework for describing the non-linear distribution of dark matter and its tracers across all survey combinations. It decomposes the three-dimensional power spectrum for any pair of tracers $X,Y \in \{y, g, \gamma\}$ into two physically distinct components: the 1-halo term, which governs correlations within individual halos, and the 2-halo term, which describes correlations between different halos:
\begin{equation}
    P_{XY}(k,z) = P_{XY}^{\text{1h}}(k,z) + P_{XY}^{\text{2h}}(k,z),
    \label{eq:halo_ps}
\end{equation}
where $k$ is the comoving wavenumber (in units of $h/\text{Mpc}$) and $z$ is the redshift.

\subsubsection{1-Halo and 2-Halo Contributions}
\label{subsubsec:halo_terms}

The 1-halo term ($P_{XY}^{\text{1h}}$) dominates on small angular scales ($\ell \gtrsim 100$) and captures correlations arising from matter within individual dark matter halos. This intra-halo regime is where the hydrostatic mass bias ($b_{\mathrm{HSE}}$) leaves its most pronounced imprint, as it directly modulates the inferred halo mass and its internal structure. {In contrast, the 2-halo term ($P_{XY}^{\text{2h}}$) dominates at large scales ($\ell \lesssim 100$) and encodes correlations between distinct halos driven by linear structure growth.}

{Mathematically, the 1-halo and 2-halo contributions are given by \citep{Cooray02}:}
{\begin{align}
P_{XY}^{\text{1h}}(k,z) &= \int dM \, \frac{dn}{dM} \, \tilde{u}_X(k|M,z) \, \tilde{u}_Y(k|M,z), \\
P_{XY}^{\text{2h}}(k,z) &= \left[ \int dM \, \frac{dn}{dM} \, b_h(M,z) \, \tilde{u}_X(k|M,z) \right] \nonumber \\
&\quad \times \left[ \int dM \, \frac{dn}{dM} \, b_h(M,z) \, \tilde{u}_Y(k|M,z) \right] P_{\text{lin}}(k,z),
\end{align}}
{where $dn/dM$ is the halo mass function \citep{Tinker2008}, $b_h(M,z)$ is the halo bias \citep{Tinker2010}, $\tilde{u}_X(k|M,z)$ denotes the appropriately weighted Fourier-space kernel for tracer $X$ (including the relevant HOD or mass/pressure weighting), and $P_{\text{lin}}(k,z)$ is the linear matter power spectrum.}

The modeling of these terms is tailored to each tracer. For LSST/CSST galaxy clustering, the galaxy--halo connection is described by a Halo Occupation Distribution (HOD) calibrated to the observed lens galaxy densities ($n_g$) and following \citet{Zheng05}. For weak lensing, the matter distribution is modeled with a Navarro--Frenk--White (NFW) density profile \citep{Navarro96}, using the concentration--mass relation from \citet{Duffy2008} and normalized to the effective source galaxy densities ($n_{\rm eff}$). For the tSZ effect, the intracluster medium pressure follows a Generalized NFW (GNFW) profile \citep{Nagai07, Arnaud2010}, the frequency dependence of the tSZ signal is accounted for in the map and noise modeling. 

The 2-halo term is proportional to the linear matter power spectrum $P_{\delta\delta}(k,z)$, computed using \texttt{CAMB} \citep{Lewis2000}, and scales with the large-scale bias of the tracers. The optical galaxy bias is calibrated to the survey-specific redshift distributions ($dN/dz$), which peak at $z \approx 0.8$ (LSST) and $z \approx 1.1$ (CSST). The tSZ bias follows $\Lambda$CDM halo bias predictions, consistent with hydrodynamical simulations \citep{Vikhlinin09}.

{The 1-halo and 2-halo terms are computed using the halo model framework implemented in the Core Cosmology Library (CCL; \citealt{Chisari19}), which follows the standard formalism of \citet{Cooray02}. All power spectra are generated via the \texttt{halomod\_Pk2D} routine,} with the linear matter power spectrum $P_{\text{lin}}(k,z)$ computed using \texttt{CAMB} \citep{Lewis2000}. Halo properties are defined with respect to the mass $M_{\mathrm{200m}}$, enclosed within a radius where the mean enclosed density is 200 times the mean matter density of the universe.

\subsection{Angular Power Spectrum Projection}
\label{subsec:angular_power_spectra}

The 3D power spectra $P_{XY}(k,z)$ are projected onto the celestial sphere to yield observable angular power spectra $C_\ell^{XY}$ using the Limber approximation \citep{Limber53}, which is accurate for $\ell \gtrsim 15$:
\begin{equation}
C_{\ell}^{XY} = \int_{0}^{\chi_{\max}} \frac{d\chi}{\chi^2} W_X(\chi) W_Y(\chi) P_{XY}\left(\frac{\ell + 1/2}{\chi}, z(\chi)\right).
\label{eq:limber}
\end{equation}
Below, we specify the physical interpretation and window function $W_X(\chi)$ for each tracer.

\noindent \textbf{Thermal Sunyaev--Zeldovich effect ($y$)}: 
The Compton-$y$ parameter quantifies the line-of-sight integral of electron pressure in the intracluster medium (ICM), making it a nearly mass-limited tracer of collapsed halos. Unlike tracers with a fixed radial kernel, the redshift dependence of the tSZ signal is not described by a simple line-of-sight projection; instead, it is encoded through the halo-model mass and redshift integrals entering the power spectrum $P_{yy}(k,z)$ and related cross-spectra. The overall tSZ amplitude is fixed to a fiducial value; our analysis varies only $b_{\rm HSE}$, which rescales the inferred halo mass and thus the tSZ signal \citep{Komatsu02, CMBS4_paper}.

\noindent \textbf{Galaxy density ($g$)}: 
This tracer represents the projected overdensity of galaxies from flux-limited optical surveys such as LSST and CSST. Galaxies trace the underlying matter distribution via a linear, scale-independent bias. The corresponding window function is
\begin{equation}
W_g(\chi) = \frac{H(z)}{c} \frac{dN}{dz} b_g(z(\chi)),
\end{equation}
where $dN/dz$ is the normalized redshift distribution and $b_g(z) = 0.95(1+z)$ is our fiducial bias model, consistent with empirical measurements and Stage-IV forecasting conventions. The redshift distributions $dN/dz$ are survey-specific and detailed in Section~\ref{sec:survey_specs}.

\noindent \textbf{Weak Lensing Cosmic Shear ($\gamma$)}: 
Cosmic shear measures the coherent distortions in galaxy shapes induced by gravitational lensing from the total intervening matter distribution (dark and baryonic), making it primarily sensitive to the total matter distribution. In the Limber approximation, the shear power spectrum is equivalent to the convergence spectrum, with window function given by the lensing efficiency kernel,
\begin{equation}
W_\kappa(z) = \frac{3}{2} \Omega_m H_0^2 \frac{\chi(z)}{a(z)} \int_z^{\infty} dz' \frac{dN}{dz'}(z') \frac{\chi(z') - \chi(z)}{\chi(z')},
\label{eq:lensing_efficiency}
\end{equation}
where $a(z) = 1/(1+z)$ is the scale factor and $dN/dz'$ is the redshift distribution of the source galaxies.

The combination of galaxy clustering and weak lensing in a $3 \times 2$pt analysis has been established as a powerful approach for Stage-IV surveys \citep{Euclid20, CMBS4_Book}, and we extend this methodology specifically to the $b_{\mathrm{HSE}}$ problem.

\subsubsection{Hydrostatic Mass Bias Parameterization}
\label{subsubsec:bhse}

A central goal of this work is to constrain the hydrostatic mass bias parameter $b_{\mathrm{HSE}}$, which quantifies the systematic offset between the true halo mass $M$ and the hydrostatic mass $M_{\mathrm{HSE}}$. The latter is inferred from the thermal Sunyaev--Zel'dovich (tSZ) signal under the assumption of hydrostatic equilibrium in the intracluster medium (ICM) and systematically underestimates $M$ due to unaccounted non-thermal pressure support. This relationship is defined as

\begin{equation}
M_{\mathrm{HSE},\Delta} = (1 - b_{\mathrm{HSE}}) M_{\Delta},
\label{eq:bhse_def}
\end{equation}
where $M_{\Delta}$ denotes the true halo mass enclosed within a radius $R_{\Delta}$ inside which the mean density is $\Delta$ times the critical density of the universe, and $M_{\mathrm{HSE},\Delta}$ is the mass inferred under hydrostatic equilibrium. In this work, we adopt $\Delta = 500$ (i.e., $M_{500c}$) for the tSZ mass--observable relation, consistent with tSZ and X-ray cluster analyses. We model $b_{\mathrm{HSE}}$ as a single, redshift- and mass-independent parameter with a fiducial value of $b_{\mathrm{HSE}}^{\rm fid} = 0.20$. This choice is consistent with recent hydrodynamical simulations (e.g., FLAMINGO and The Three Hundred Project), which find $1 - b_{\mathrm{HSE}} \sim 0.7$--$0.8$ \citep{Braspenning25, Corasaniti25}, and aligns with the range of tSZ mass bias constraints summarized in Table~1 of \citet{Ibitoye22}.

Under the self-similar model for cluster formation, the integrated Compton-$y$ signal scales with the true halo mass as $Y \propto M^{5/3}$. Because we assume $b_{\mathrm{HSE}}$ is constant, the observed tSZ signal for a given true mass is suppressed by a fixed factor of $(1 - b_{\mathrm{HSE}})^{5/3}$. This constant scaling propagates directly into the angular power spectra: the tSZ auto-spectrum scales as $C_{\ell}^{yy} \propto (1 - b_{\mathrm{HSE}})^{10/3}$, while the cross-spectra between tSZ and optical tracers scale as $C_{\ell}^{yg},\, C_{\ell}^{y\gamma} \propto (1 - b_{\mathrm{HSE}})^{5/3}$. In contrast, the auto-spectra of galaxy clustering and cosmic shear ($C_{\ell}^{gg},\, C_{\ell}^{\gamma\gamma}$) remain unaffected by $b_{\mathrm{HSE}}$, as they probe the total matter distribution rather than the ICM's thermodynamic state. This differential scaling enables cross-correlations to break the degeneracy between $b_{\mathrm{HSE}}$ and the amplitude of matter fluctuations, $\sigma_8$.

\section{Survey Specifications and Tomographic Binning}
\label{sec:survey_specs}

In this section, we detail the experimental configurations, survey specifications, and tomographic binning strategies for three complementary survey combinations selected for their synergistic potential in constraining the hydrostatic mass bias ($b_{\mathrm{HSE}}$) and key cosmological parameters.

\subsection{Analysis Configurations}
\label{subsec:experimental_config}

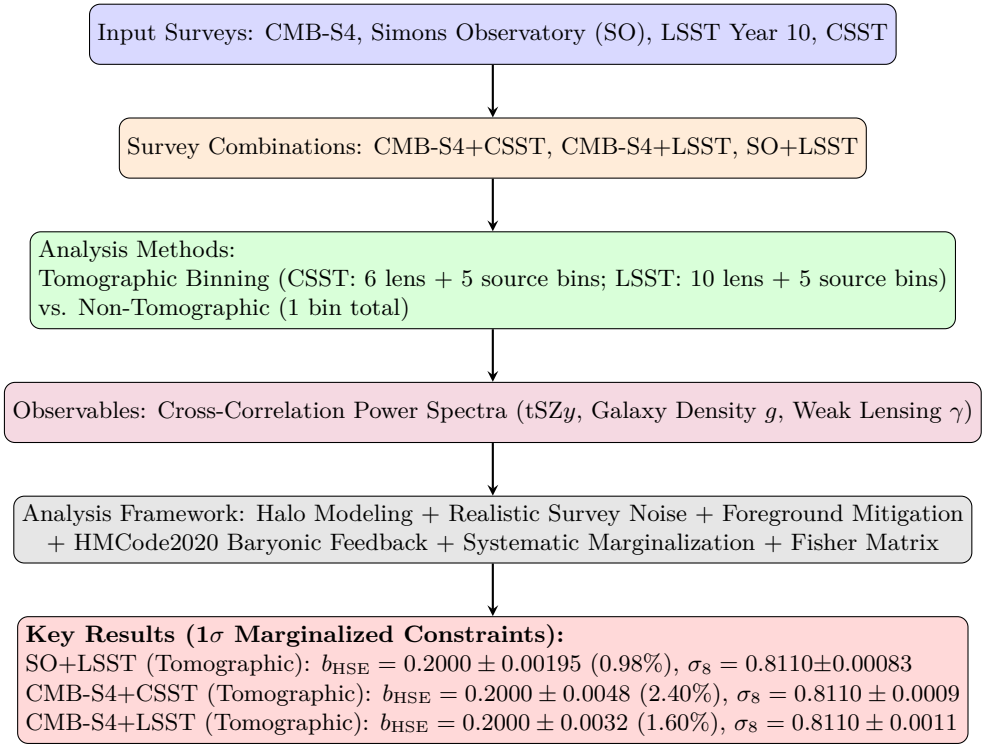
\begin{figure*}[t]
\centering
\begin{tikzpicture}[
    node distance=0.7cm,
    box/.style={draw, rounded corners, minimum width=8cm, minimum height=0.8cm, text centered, align=center},
    arrow/.style={->, >=stealth, thick}
]
\node[box, fill=blue!15] (surveys) {Input Surveys: CMB-S4, Simons Observatory (SO), LSST Year 10, CSST};
\node[box, fill=orange!15, below=of surveys] (combinations) {Survey Combinations: CMB-S4+CSST, CMB-S4+LSST, SO+LSST};
\node[box, fill=green!15, below=of combinations, align=left] (methods) {Analysis Methods: \\
Tomographic Binning (CSST: 6 lens + 5 source bins; LSST: 10 lens + 5 source bins) \\
vs. Non-Tomographic (1 bin total)};
\node[box, fill=purple!15, below=of methods] (observables) {Observables: Cross-Correlation Power Spectra (\tsz $y$, Galaxy Density $g$, Weak Lensing $\gamma$)};
\node[box, fill=gray!20, below=of observables] (framework) {Analysis Framework: Halo Modeling + Realistic Survey Noise + Foreground Mitigation \\+ HMCode2020 Baryonic Feedback + Systematic Marginalization + Fisher Matrix};
\node[box, fill=red!15, below=of framework, align=left] (results) {
    \textbf{Key Results (1$\sigma$ Marginalized Constraints):} \\
    SO+LSST (Tomographic): $b_{\mathrm{HSE}} = 0.2000 \pm {0.00195}$ (0.98\%), $\sigma_8 = 0.8110 {\pm 0.00083}$\\
    CMB-S4+CSST (Tomographic): $b_{\mathrm{HSE}} = 0.2000 \pm 0.0048$ (2.40\%), $\sigma_8 = 0.8110 \pm {0.0009}$ \\
    CMB-S4+LSST (Tomographic): $b_{\mathrm{HSE}} = 0.2000 \pm 0.0032$ (1.60\%), $\sigma_8 = 0.8110 \pm {0.0011}$
};
\draw[arrow] (surveys) -- (combinations);
\draw[arrow] (combinations) -- (methods);
\draw[arrow] (methods) -- (observables);
\draw[arrow] (observables) -- (framework);
\draw[arrow] (framework) -- (results);
\end{tikzpicture}
\caption{Analysis pipeline for forecasting hydrostatic mass bias ($b_{\mathrm{HSE}}$) constraints. The workflow progresses from survey specification through tomographic/non-tomographic analysis of multi-probe power spectra to final marginalized cosmological constraints. Tomographic analyses achieve high-precision $b_{\mathrm{HSE}}$ constraints for SO+LSST {(0.98\% precision)}, with competitive results for CMB-S4+LSST (1.60\%) and CMB-S4+CSST (2.40\%).}
\label{fig:pipeline_vertical_corrected}
\end{figure*}

We analyze three survey combinations designed to exploit complementary strengths of next-generation CMB and optical surveys: {Case A (SO × LSST) leverages SO's enhanced cluster mass calibration with LSST's depth; Case B (CMB-S4 × CSST) pairs CMB-S4's high-fidelity tSZ mapping with CSST's space-based optical precision; and Case C (CMB-S4 × LSST) combines CMB-S4 with LSST's ground-based statistical power.}

These configurations exploit cross-correlations between CMB probes (tSZ) and optical probes (galaxy clustering, cosmic shear) to maximize sensitivity to $b_{\mathrm{HSE}}$ and break degeneracies with $\sigma_8$. For all cases, we assume a common sky fraction $f_{\mathrm{sky}} = 0.4$ and analyze multipoles in the range $\ell \in [30, 3000]$, capturing both linear and non-linear regimes while limiting sensitivity to the most foreground-dominated angular scales.

\begin{table*}[t]
\centering
\caption{Survey Specifications for Hydrostatic Mass Bias Forecasts: CMB-S4 + CSST vs.\ CMB-S4 + LSST vs.\ SO + LSST. Key results include $b_{\rm HSE}$ precision (tomographic/no tomographic) and tomography improvement, calculated as $[1 - (\sigma_{\rm tomo}/\sigma_{\rm no\text{-}tomo})] \times 100\%$. Foreground modeling uses power-law templates calibrated to post-component-separation ILC-cleaned levels. {All $b_{\rm HSE}$ precision values reflect full marginalization over systematic uncertainties (galaxy bias perturbations, photometric redshift shifts, baryonic feedback via HMCode2020).}}
\label{tab:surveys_updated}
\begin{tabular}{l@{\hspace{0.5cm}}ccc} 
\hline\hline
\textbf{Parameter} & \textbf{CMB-S4+CSST} & \textbf{CMB-S4+LSST} & \textbf{SO+LSST} \\
\hline
\multicolumn{4}{c}{\textit{CMB Specifications}} \\
\hline
Experiment & CMB-S4 Wide & CMB-S4 Wide & SO LAT \\
$f_{\rm sky}$ & 0.4 & 0.4 & 0.4 \\
Beam FWHM (145 GHz) & $1.4'$ & $1.4'$ & $1.4'$ \\
Temp. Noise [$\mu$K-arcmin] & 1.0 & 1.0 & 1.2 \\
$\ell$ Range (\tsz) & $[30, 3000]$ & $[30, 3000]$ & $[30, 3000]$ \\
Foregrounds & CIB+radio (ILC-cleaned) & CIB+radio (ILC-cleaned) & CIB+radio (ILC-cleaned) \\
\hline
\multicolumn{4}{c}{\textit{Optical Specifications}} \\
\hline
$n_{\rm source}$ [arcmin$^{-2}$] & 25.8 & 27 & 27 \\
$n_{\rm lens}$ [arcmin$^{-2}$] & 20.8 & 30 & 30 \\
$\sigma_\epsilon$ & 0.26 & 0.3 & 0.3 \\
Source $z$ Range & $0.1$--$2.5$ & $0.2$--$3.5$ & $0.2$--$3.5$ \\
Lens $z$ Range & $0.1$--$1.5$ & $0.2$--$1.2$ & $0.2$--$1.2$ \\
$\sigma_z/(1+z)$ {lenses/sources } & 0.03 & 0.05 & 0.05 \\
\hline
\multicolumn{4}{c}{\textit{Tomography}} \\
\hline
Source Bins & 5 (equal count) & 5 (equal count) & 5 (equal count) \\
Source Bin Edges & [0.10, 0.49, 0.75, 1.04, 1.48, 2.50] & [0.2, 0.5, 0.8, 1.2, 1.8, 3.5] & [0.2, 0.5, 0.8, 1.2, 1.8, 3.5] \\
Lens Bins & 6 (optimized) & 10 (fixed) & 10 (fixed) \\
Lens Bin Edges & [0.10, 0.40, 0.58, 0.75, 0.94, 1.17, 1.50] & [0.2--1.2, 0.1 steps] & [0.2--1.2, 0.1 steps] \\
\hline
\multicolumn{4}{c}{\textit{Forecast Results}} \\
\hline
{$b_{\rm HSE}$ Precision (Tomo)} & {2.40\%} & {1.60\%} & {0.98\%} \\
{$b_{\rm HSE}$ Precision (No Tomo)} & {3.36\%} & {5.30\%} & {4.80\%} \\
{Tomography Improvement} & {28.7\%} & {69.8\%} & {79.7\%} \\
Total Spectra & 78 & 121 & 121 \\
\hline\hline
\end{tabular}
\end{table*}

\subsection{Survey Specifications}
\label{subsec:survey_specs}

\subsubsection{CMB Experiments}  

CMB-S4 is a next-generation CMB observatory, engineered to address foundational questions in cosmology via a hybrid architecture of ground-based and potential satellite assets \citep{CMBS4_24}. For tSZ science central to constraining cluster mass--observable relations, it deploys two 6m Cross-Dragone telescopes in Chile (mapping $\sim$70\% of the sky daily) and a 5m Three Mirror Anastigmat telescope at the South Pole (ultra-deep 3\% sky coverage) \citep{CMBS4_25}. Operating at 95 GHz and 150 GHz (bands optimized to mitigate thermal dust foregrounds), it achieves a beam FWHM of $1.4^{\prime}$ and polarization noise of 1~$\mu$K-arcmin, with detectors cooled to 0.1~Kelvin for maximal sensitivity \citep{CMBS4_24, CMBS4_25}. The experiment will field $\sim$500,000 polarization-sensitive detectors, enabling unprecedented mapping speed for tSZ cluster detection. A key science objective includes improving constraints on cosmological parameters (e.g., $\sigma_8$) and astrophysical effects such as $b_{\rm HSE}$ through cross-correlations between its tSZ maps and optical probes \citep{CMBS4_24}. Its tSZ surveys are expected to reach lower mass threshold than SO, extending cluster detection to higher-redshift systems relevant for tracing structure growth \citep{CMBS4_Book}.

The Simons Observatory (SO) serves as a critical precursor to CMB-S4, deploying a 6m Large Aperture Telescope (LAT) with 93/145 GHz bands to refine cluster cosmology methodologies \citep{SO_25}. It delivers finer angular resolution ($2.2^{\prime}$ at 93 GHz, $1.4^{\prime}$ at 145 GHz) than CMB-S4 but higher noise (8.0/10.0~$\mu$K-arcmin, respectively), trading sensitivity for early calibration power \citep{SO_telescope_specs}. SO's greatest impact for $b_{\rm HSE}$ studies lies in enabling early multi-probe mass calibration: when combined with optical surveys like LSST or DESI, SO's tSZ and CMB lensing maps can constrain the mean cluster mass with $\sigma(M)/M \sim 0.03$ at $z \sim 1$ \citep{SO_25}, providing a benchmark for validating the mass--observable relations that CMB-S4 will rely on. Furthermore, SO's early tSZ cluster catalogs, spanning $z \lesssim 1.5$, will offer empirical constraints on the redshift evolution of $b_{\rm HSE}$, reducing systematic uncertainties in CMB-S4's forecasting frameworks \citep{SO_25, CMBS4_Book}.

\subsubsection{Optical Surveys}
\label{ssec:optical_survey}
LSST Year 10 employs 6-band (\textit{ugrizy}) photometry with effective source galaxy density $n_{\text{source}} = 27~\mathrm{arcmin}^{-2}$ (for weak lensing shear) and shape noise $\sigma_\epsilon = 0.3$ representative of ground-based weak lensing analyses \citep{LSSTScience09}. For galaxy clustering (lens galaxies), we adopt $n_{\text{lens}}=30~\mathrm{arcmin}^{-2}$, a subset of brighter, massive galaxies selected as large-scale structure (LSS) tracers, consistent with LSST DESC forecasting assumptions \citep{DESCSRD, Chang13}. The redshift distribution of the source and lens samples are modeled using Smail-type parameterizations consistent with LSST forecasting conventions; the explicit functional forms and parameters used in our analysis are specified in Section~\ref{subsec:tomography_and_assumptions}. 

CSST conducts a 5-band wide-field survey with effective source galaxy density $n_{\text{source}} = 26~\mathrm{arcmin}^{-2}$ (for weak lensing shear) and superior shape measurement ($\sigma_\epsilon = 0.26$) due to space-based PSF stability \citep{CSST}. For galaxy clustering, we use $n_{\text{lens}}=25~\mathrm{arcmin}^{-2}$ bright, massive lens galaxies aligned with space-based survey clustering forecasts \citep{CSST_NeutrinoCosmo}. Its redshift distribution is modeled with a Smail-type function extending to $z \lesssim 2.5$, consistent with mission forecasts.

For both surveys, $n_{\text{source}}$ accounts for observational losses (blending, masking): LSST Year 10 uses $\sim75\%$ of the raw galaxy density (reflecting 6-band photometry and optimized processing; \citep{DESCSRD, Chang13}), while CSST aligns with cosmic shear simulation results for space-based surveys \citep{CSST}. These choices ensure consistency with our noise model (shape noise $\propto \sigma_\epsilon^2/n_{\text{source}}$ for shear, shot noise $\propto 1/n_{\text{lens}}$ for clustering).

\subsection{Tomographic Binning}
\label{subsec:tomography_and_assumptions}

To break degeneracies between the hydrostatic mass bias $b_{\mathrm{HSE}}$ and cosmological parameters particularly $\sigma_8$, we employ optimized tomographic binning schemes tailored to the redshift precision of each optical survey.

For LSST Year~10, we adopt the DESC Science Requirements Document (SRD) baseline: 10 lens bins spanning $z = 0.263$--$1.149$ (for galaxy clustering) and 5 source bins spanning $z = 0.309$--$2.053$ (for weak lensing), as shown in Figure~\ref{fig:dndz_tomographic}. The lens bins use fixed-width redshift intervals, while the source bins are constructed to contain equal numbers of galaxies, ensuring uniform statistical weight across tomographic slices.

The underlying (true) redshift distribution follows the Smail-type form \citep{DESCSRD}:
\begin{equation}
    n(z) \propto \left( \frac{z}{z_0} \right)^\beta \exp\left[ -\left( \frac{z}{z_0} \right)^\alpha \right].
\end{equation}
We adopt fiducial parameters $z_0 = 0.11$, $\alpha = 0.68$, $\beta = 2.0$ for the source sample (median redshift $z_{\rm med} \approx 1.05$), and $z_0 = 0.28$, $\alpha = 0.90$, $\beta = 2.0$ for the lens sample. Photometric redshift uncertainties are modeled with a Gaussian scatter,
\begin{equation}
    p(z_{\rm ph} \mid z) = \frac{1}{\sqrt{2\pi}\,\sigma_z} \exp\left[ -\frac{(z_{\rm ph} - z)^2}{2\sigma_z^2} \right],
\end{equation}
with $\sigma_z = 0.05(1+z)$ for sources and $\sigma_z = 0.03(1+z)$ for lenses, and with fiducial bias $z_{\rm bias} = 0$. The resulting binned redshift distribution for the $i$-th tomographic bin $[z_i^{\rm min}, z_i^{\rm max}]$ is
\begin{equation}
    n_i(z) = \frac{n(z)}{2} \left[ \mathrm{erf}\left( \frac{z_i^{\rm max} - z}{\sqrt{2}\,\sigma_z} \right) - \mathrm{erf}\left( \frac{z_i^{\rm min} - z}{\sqrt{2}\,\sigma_z} \right) \right],
\end{equation}
which is normalized such that $\int n_i(z)\,dz = 1$, following the photo-$z$ binning formalism of \citet{Ma06}. 

All distributions and binning schemes are implemented using the Core Cosmology Library (CCL) \citep{Chisari19}, ensuring consistency with DESC forecasting standards.

For CSST, we adopt a science-driven tomographic binning scheme with 5 source bins and 6 lens bins, optimized to maximize sensitivity to the redshift-dependent cross-correlations central to our analysis: galaxy--tSZ ($\delta_g y$), tSZ--shear ($y\gamma$), and galaxy--shear ($\delta_g\gamma$). The source bins span $z \approx 0.1$--2.5, while the lens bins cover $z \approx 0.1$--1.5, with edges chosen to align with the characteristic redshift peaks of each observable (e.g., $\delta_g y$ peaks at $z \sim 0.25$--0.6). 

{The CSST redshift distributions are constructed from the Juitian mock galaxy light-cone simulation \citep{Xiong25}, which models CSST’s instrumental characteristics, photometric depth, and survey strategy. We model photometric redshift uncertainties for each individual galaxy} with a Gaussian scatter, $z_{\rm obs} \sim \mathcal{N}(z_{\rm true}, \sigma_z)$, where $\sigma_z = \sigma_{z_0}(1+z)$. We adopt $\sigma_{z_0} = 0.05$ for the source sample and $\sigma_{z_0} = 0.03$ for the lens sample. The redshift distribution for each tomographic bin is constructed by stacking galaxies according to their redshift PDFs and normalized such that $\int n_i(z)\,dz = 1$. These normalized distributions are then used to compute the tracer-specific window functions for our $6\times2$pt cross-correlation forecast.

This tomographic binning strategy maximizes the redshift resolution for galaxy clustering measurements while maintaining sufficient source density for weak lensing, enabling improved constraints on the hydrostatic mass bias through enhanced redshift resolution.

\begin{figure*}
\centering
\includegraphics[width=11cm]{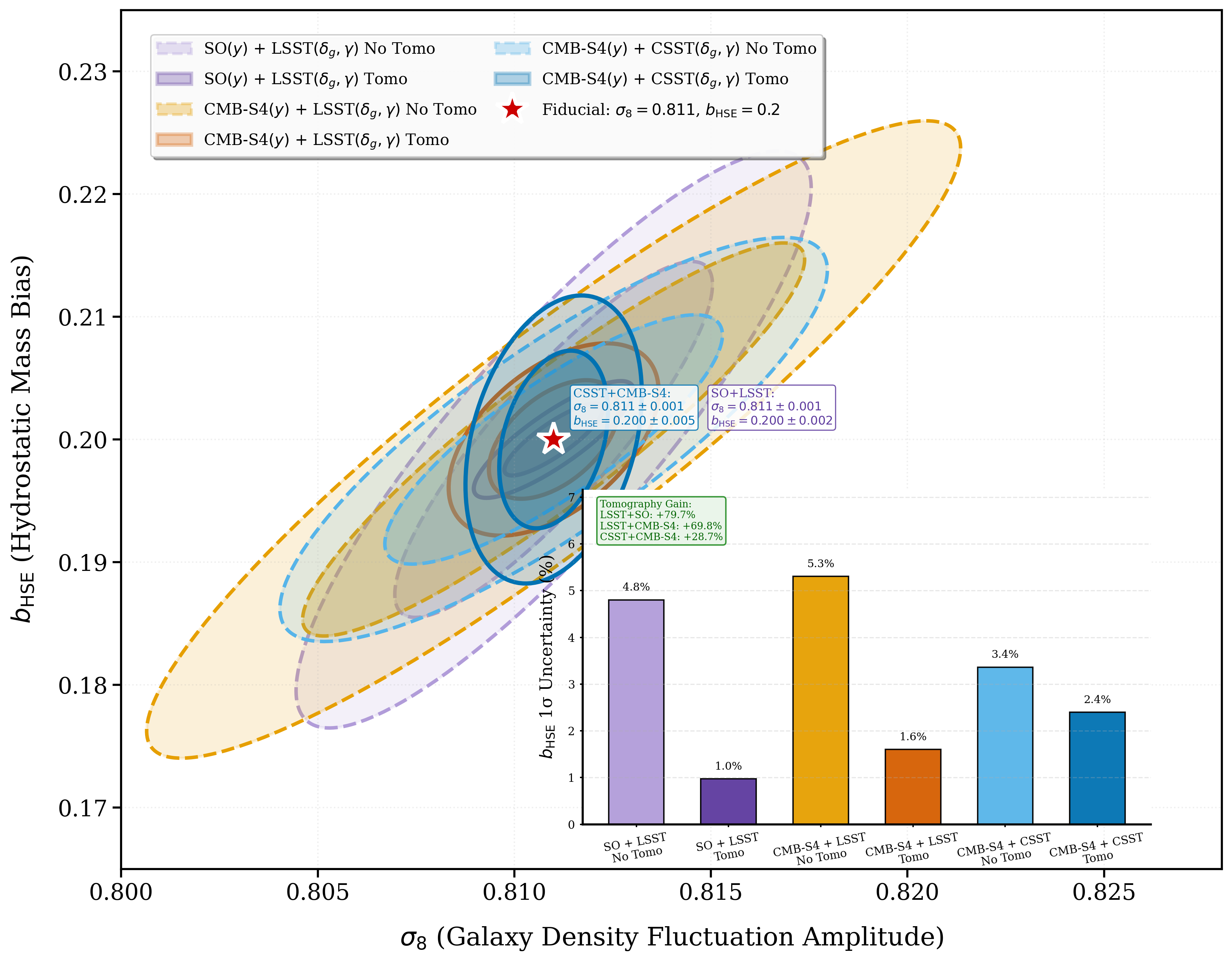}
\caption{
Constraints on hydrostatic mass bias ($b_{\mathrm{HSE}}$) from Fisher matrix forecasts of CMB-optical cross-correlations, comparing CMB-S4 + CSST, CMB-S4 + LSST, and SO + LSST.
\textbf{Main panel}: $68\%$ and $95\%$ confidence contours in the $\sigma_8$--$b_{\mathrm{HSE}}$ plane. Solid/dashed curves denote tomography (CSST: 6+5 bins, LSST: 10+5 bins)/no-tomography configurations. Color coding: CMB-S4 + CSST (blue), CMB-S4 + LSST (orange), SO + LSST (purple). The black star marks the fiducial cosmology ($\sigma_8 = 0.811$, $b_{\mathrm{HSE}} = 0.2$), consistent with the CMB-S4 Science Book's adopted cluster mass model.
\textbf{Inset}: Relative $b_{\mathrm{HSE}}$ precision (1$\sigma$ uncertainty as percentage of fiducial value). SO + LSST tomography achieves 0.98\% precision on $b_{\mathrm{HSE}}$, corresponding to sub-percent precision, and representing a $79.7\%$ reduction in the marginalized 1$\sigma$ uncertainty relative to its no-tomography case (4.80\% $\rightarrow$ 0.98\%).
}
\label{fig:sigma8_bhse_contours}
\end{figure*}

The refined binning strategy provides dense sampling of the redshift evolution of both the halo mass function and structure growth, which is particularly crucial for breaking the strong $b_{\mathrm{HSE}}$--$\sigma_8$ degeneracy. The increased number of tomographic bins significantly enhances the cross-correlation signal between CMB and optical surveys, yielding substantial improvement in $b_{\mathrm{HSE}}$ constraints compared to single-bin analyses, as demonstrated in our results (see Figure~\ref{fig:sigma8_bhse_contours}).

We stress that, in this work, we adopt survey-optimized tomographic binning for each optical experiment, reflecting their distinct photometric precision and science goals. CSST is paired exclusively with CMB-S4, while LSST is analyzed in two configurations: (i) with CMB-S4 and (ii) with SO. The binning schemes are tailored to each combination, ensuring internal consistency within each forecast scenario.

For CMB-S4, we adopt the total temperature noise power spectrum prescribed in the CMB-S4 Science Book (2019 edition; \citet{CMBS4_Book}), consistent with established modeling frameworks for ground-based CMB surveys \citep{Abazajian_GW, Raghunathan22b}. The instrumental noise is parameterized as
\begin{equation}
    N_\ell^{\mathrm{CMB\text{-}S4}} = N_{\mathrm{map}}^2\, B_\ell^{-2} \left[ 1 + \left( \frac{\ell_{\mathrm{knee}}}{\ell} \right)^\gamma \right],
\end{equation}
where $N_{\mathrm{map}} = 1.0\,\mu\mathrm{K}\cdot\mathrm{arcmin}$ is the white-noise level per temperature map at 145 GHz. The beam transfer function is modeled as a Gaussian,
\begin{equation}
    B_\ell = \exp\left(-\frac{\ell(\ell+1)\theta_{\mathrm{FWHM}}^2}{8\ln 2}\right),
\end{equation}
with full-width-at-half-maximum $\theta_{\mathrm{FWHM}} = 1.4^\prime$, corresponding to CMB-S4’s large-aperture telescope. The $1/f$ noise component is characterized by a knee multipole $\ell_{\mathrm{knee}} = 3917$ and slope $\gamma = 3.5$, following the fiducial CMB-S4 calibration \citep{Raghunathan22b}. This yields a white-noise floor of $N_{\ell,\text{white}} \approx 1.2 \times 10^{-7}\,\mu\mathrm{K}^2$ at $\ell \gg \ell_{\mathrm{knee}}$.

Although the tSZ auto-spectrum peaks near $\ell \sim 3000$—where white noise dominates—the inclusion of the full noise model (including low-$\ell$ atmospheric $1/f$ noise) is essential for accurately capturing the covariance of cross-correlations involving CMB lensing and large-scale structure, which are sensitive to modes at $\ell \lesssim 100$.

In addition to instrumental noise, we include residual extragalactic foregrounds in the tSZ auto- and cross-spectra, calibrated to post-component-separation levels expected after multi-frequency internal linear combination (ILC) cleaning. Following \citet{SPT3G}, we model the clustered cosmic infrared background (CIB) and radio point sources as power laws:
\begin{equation}
    D_\ell^{\rm CIB} = A_{\rm CIB} \left( \frac{\ell}{3000} \right)^{\alpha_{\rm CIB}}, \quad
    D_\ell^{\rm radio} = A_{\rm radio} \left( \frac{\ell}{3000} \right)^{\alpha_{\rm radio}},
\end{equation}
where $D_\ell \equiv \ell(\ell+1)C_\ell/(2\pi)$ with amplitudes $A_{\rm CIB} = 5.0 \times 10^{-7}\,\mu\mathrm{K}^2$, $A_{\rm radio} = 3.0 \times 10^{-7}\,\mu\mathrm{K}^2$ at $\ell = 3000$, and indices $\alpha_{\rm CIB} = -0.7$, $\alpha_{\rm radio} = -1.2$. These correspond to residual CMB power of $C_{3000}^{\rm CIB} \approx 3.5 \times 10^{-6}\,\mu\mathrm{K}^2$ and $C_{3000}^{\rm radio} \approx 1.4 \times 10^{-6}\,\mu\mathrm{K}^2$, consistent with CMB-S4 foreground forecasts \citep{CMBS4_Book}.

For the Simons Observatory (SO), we use the same noise parameterization but with SO-specific values: $\ell_{\mathrm{knee}} = 1000$ and $\gamma = 3.5$ \citep{Raghunathan22a, SO_22}, reflecting its distinct atmospheric noise characteristics and baseline LAT configuration. The white-noise level is set by the SO LAT map sensitivity specified in Table~\ref{tab:surveys_updated}.

For galaxy clustering, shot noise in each tomographic bin is modeled as $N_\ell^{gg} = 1/n_g$, where $n_g$ is the angular number density of galaxies (in steradians$^{-1}$) in that bin. Cosmic shear noise follows standard shape-noise prescriptions (see Section~\ref{sec:survey_specs}). The joint analysis assumes a sky overlap of $f_{\mathrm{sky}} = 0.4$ between CMB and optical surveys, which is accounted for in the covariance matrix of the Fisher forecast.

All signal and noise spectra are combined to construct the full covariance matrix for the Fisher analysis (Section~\ref{sec:forecast}). Our modeling employs the halo framework from Section~\ref{sec:theory}, with HOD for galaxies, NFW profiles for lensing, and GNFW pressure models for tSZ. We apply conservative scale cuts: $\ell < 30$ (finite sky effects), $\ell > 3000$ (baryonic uncertainties), and $\ell < 500$ for the tSZ auto-spectrum (foreground mitigation).

\section{Fisher Forecast Methodology}
\label{sec:forecast}

We employ a comprehensive Fisher matrix formalism to forecast constraints on the hydrostatic mass bias $b_{\mathrm{HSE}}$ and cosmological parameters from cross-correlations between CMB and optical surveys. This approach enables rigorous quantification of parameter uncertainties while accounting for all auto- and cross-correlations between tomographic bins. {Our forecasting framework extends the joint tSZ–galaxy–lensing approach of \citet{Shirasaki20} by replacing X-ray with optical weak lensing (more relevant for LSST/CSST), while retaining the core insight: tomographic cross-correlations break the degeneracy between $b_{\mathrm{HSE}}$ and $\sigma_8$ far more efficiently than any single probe.}

\subsection{Mathematical Formalism}
\label{subsec:fisher_formalism}

The Fisher matrix formalism provides a powerful framework for forecasting parameter constraints from future experiments. For Gaussian-distributed observables, the data-driven Fisher matrix element (from survey power spectra and noise) for parameters $\theta_\alpha$ and $\theta_\beta$ is given by:

\begin{equation}
F_{\alpha\beta}^{\rm data} = \sum_{\ell=\ell_{\min}}^{\ell_{\max}} \frac{2\ell+1}{2} f_{\mathrm{sky}} \, \mathrm{Tr}\left( \mathbf{\Gamma}_\ell^{-1} \frac{\partial \mathbf{C}_\ell}{\partial \theta_\alpha} \mathbf{\Gamma}_\ell^{-1} \frac{\partial \mathbf{C}_\ell}{\partial \theta_\beta} \right)
\label{eq:fisher_matrix}
\end{equation}

where $\mathbf{C}_\ell$ is the signal covariance matrix containing all auto- and cross-power spectra between different probes and tomographic bins, $\mathbf{\Gamma}_\ell = \mathbf{C}_\ell + \mathbf{N}_\ell$ is the total covariance including noise $\mathbf{N}_\ell$, $f_{\mathrm{sky}}$ is the sky fraction, and the factor $(2\ell+1)/2$ accounts for the number of independent modes per multipole $\ell$.

{To incorporate uncorrelated Gaussian priors on individual parameters (summarized in Table \ref{tab:priors_expanded}), we add the prior contribution to the diagonal of the data-driven Fisher matrix to obtain the total Fisher matrix:}
{\begin{equation}
F_{\alpha\alpha}^{\rm total} = F_{\alpha\alpha}^{\rm data} + \frac{1}{\sigma_{\rm prior,\alpha}^2}
\label{eq:fisher_prior}
\end{equation}}
The Off-diagonal elements of the total Fisher matrix remain identical to the data-driven Fisher matrix ($F_{\alpha\beta}^{\rm total} = F_{\alpha\beta}^{\rm data}$ for $\alpha \neq \beta$), {as our priors are Gaussian and uncorrelated. No prior contribution is added for the target parameter $b_{\rm HSE}$, as we impose no prior on this parameter. All other parameters include survey-specific or community-calibrated Gaussian priors (summarized in Table \ref{tab:priors_expanded}), with weak relative priors for CMB-S4 and tighter absolute priors for SO to match each survey’s capabilities.}

The parameter covariance matrix is obtained by inverting the total Fisher matrix:
\begin{equation}
\mathrm{Cov}(\theta_\alpha, \theta_\beta) \geq [\mathbf{F}^{\rm total,-1}]_{\alpha\beta}
\label{eq:parameter_covariance}
\end{equation}
with the equality holding in the limit of Gaussian likelihoods and linear parameter dependence. The 1$\sigma$ uncertainty on parameter $\theta_\alpha$ is then $\sigma(\theta_\alpha) = \sqrt{[\mathbf{F}^{\rm total,-1}]_{\alpha\alpha}}$.

\subsection{Power Spectrum Covariance and Survey Implementation}
\label{subsec:covariance_survey}

The signal covariance matrix $\mathbf{C}_\ell$ contains all auto- and cross-correlations between three primary probes: thermal Sunyaev-Zel'dovich effect ($y$), galaxy clustering ($\delta_g$), and cosmic shear ($\gamma$), across multiple tomographic bins. For a single tSZ map ($N_y =1$), $N_g$ lens bins, and $N_s$ source bins, the total number of spectra is:

\begin{equation}
N_{\mathrm{spec}} = \frac{(1 + N_g + N_s)(2 + N_g + N_s)}{2}
\label{eq:total_spectra}
\end{equation}

The covariance matrix block structure is:

\begin{equation}
\mathbf{C}_\ell = \begin{pmatrix}
C_\ell^{yy} & C_\ell^{y\delta_g} & C_\ell^{y\gamma} \\
C_\ell^{\delta_g y} & C_\ell^{\delta_g\delta_g} & C_\ell^{\delta_g\gamma} \\
C_\ell^{\gamma y} & C_\ell^{\gamma\delta_g} & C_\ell^{\gamma\gamma}
\end{pmatrix}
\label{eq:covariance_structure}
\end{equation}

where each entry represents a sub-matrix spanning the relevant tomographic bin combinations.
The noise covariance $\mathbf{N}_\ell$ includes experimental noise contributions:

\begin{align}
N_\ell^{yy} &= N_\ell^{\rm{tSZ}} + N_\ell^{\rm{fg}} \quad \text{(tSZ noise \& foregrounds)} \\
N_\ell^{\delta_g\delta_g} &= \frac{1}{n_g^i} \delta_{ij} \quad \text{(shot noise per lens bin)} \\
N_\ell^{\gamma\gamma} &= \frac{\sigma_\epsilon^2}{n_s^i} \delta_{ij} \quad \text{(shape noise per source bin)}
\end{align}
with $n_g^i$ and $n_s^i$ representing the galaxy density in the $i$-th lens bin and source bin, respectively, and cross-correlation noise terms are neglected, assuming no correlated instrumental noise between probes. Foregrounds are modeled as a power-law combination of cosmic infrared background (CIB) and radio emission, calibrated to the residual levels expected after multi-frequency ILC cleaning ensuring their contribution to tSZ–galaxy and tSZ–shear cross-correlations is consistent with realistic survey performance.

We consider three survey combinations with specifications detailed in Table~\ref{tab:surveys_updated}. The tomographic binning strategy is optimized for each survey: CMB-S4 + CSST uses 6 lens bins and 5 source bins (optimized for sensitivity to redshift-dependent cross-correlations), leveraging CSST's superior photometric redshift precision as specified in Section~\ref{subsec:tomography_and_assumptions}; CMB-S4 + LSST employs 10 lens bins (fixed redshift intervals: $z \in [0.2, 1.2]$) and 5 source bins (equal counts), utilizing LSST's high source density and wide redshift coverage; and SO + LSST uses identical tomographic binning to CMB-S4 + LSST, enabling direct comparison of CMB experiment capabilities. The multipole range $\ell \in [30, 3000]$ captures the relevant scales for cluster and large-scale structure science, while limiting sensitivity to very small scales where baryonic effects and non-linear modeling uncertainties become significant.

\subsection{Parameter Space and Halo Model Framework}
\label{subsec:parameters_halo}

Our Fisher analysis employs a comprehensive parameter space that fully marginalizes over key systematic uncertainties:

{\begin{equation}
\begin{aligned}
\theta = {} & \underbrace{\{\Omega_m, \Omega_b,\sigma_8, h, n_s, \}}_{\Lambda\text{CDM}} 
\cup \underbrace{\{b_{\rm HSE}, \log_{10}(T_{\rm AGN}/{\rm K}), A_{\rm IA}, \eta_{\rm IA}\}}_{\text{Astrophysics}} \\
& \cup \underbrace{\{b_{g,i}, \Delta z_{{\rm lens},i}, \Delta z_{{\rm src},j}\}}_{\text{Nuisance}}
\label{eq:parameter_vector}
\end{aligned}
\end{equation}}

The hydrostatic mass bias $b_{\mathrm{HSE}}$ enters the tSZ power spectra through the mass-pressure relation:
\begin{equation}
C_\ell^{yy} \propto (1 - b_{\mathrm{HSE}})^{10/3}, \quad C_\ell^{y\delta_g} \propto (1 - b_{\mathrm{HSE}})^{5/3},
\end{equation}
reflecting its impact on both tSZ auto-correlation and cross-correlation with galaxies.

{Baryonic feedback is modeled via HMCode2020 with $\log_{10}(T_{\rm AGN}/{\rm K})$ as a free parameter (fiducial 7.8, prior $\sigma=0.3$), calibrated to BAHAMAS hydrodynamical simulations. Intrinsic alignments use the NLA model with redshift evolution parameter $\eta_{\rm IA}$ (fiducial 0, prior $\sigma=0.5$), per DESC SRD (2024). Nuisance parameters include:}

{\noindent Galaxy bias perturbations $b_{g,i} = b_g^{\rm fid}(z) \times (1 + \delta b_{g,i})$ for each lens bin ($i=1,\dots,10$ for LSST; $i=1,\dots,6$ for CSST), with Gaussian priors $\sigma(\delta b_{g,i}) = 0.05$,}

{\noindent photo-$z$ mean shift parameters $\Delta z_{{\rm lens},i}$ and $\Delta z_{{\rm src},j}$ for all tomographic bins, with DESC SRD priors $\sigma(\Delta z_{\rm lens}) = 0.002(1+z)$ and $\sigma(\Delta z_{\rm src}) = 0.005(1+z)$.}

All power spectra are computed using the halo model framework in CCL. We employ the \citet{Tinker2008} halo mass function (200m definition) and \citet{Tinker2010} halo bias. Galaxy clustering uses an HOD model with fiducial bias $b_g^{\rm fid}(z) = 0.95(1+z)$, weak lensing uses NFW profiles with \citet{Duffy2008} concentration, and tSZ uses generalized NFW pressure profiles with \citet{Arnaud2010} parameters.

We adopt a Planck 2018 $\Lambda$CDM cosmology as fiducial: $\{\Omega_m = 0.315, \sigma_8 = 0.811, h = 0.674, n_s = 0.9649, \Omega_b = 0.0486\}$, with $b_{\rm HSE} = 0.2$. The Fisher matrix is computed via central finite differences ($\Delta\theta_\alpha = 0.01\theta_\alpha$), with covariance including survey noise, foregrounds, and tomographic binning via block matrix assembly. No prior is imposed on $b_{\rm HSE}$.

\begin{figure*}[htbp]
    \centering
    \includegraphics[width=\textwidth]{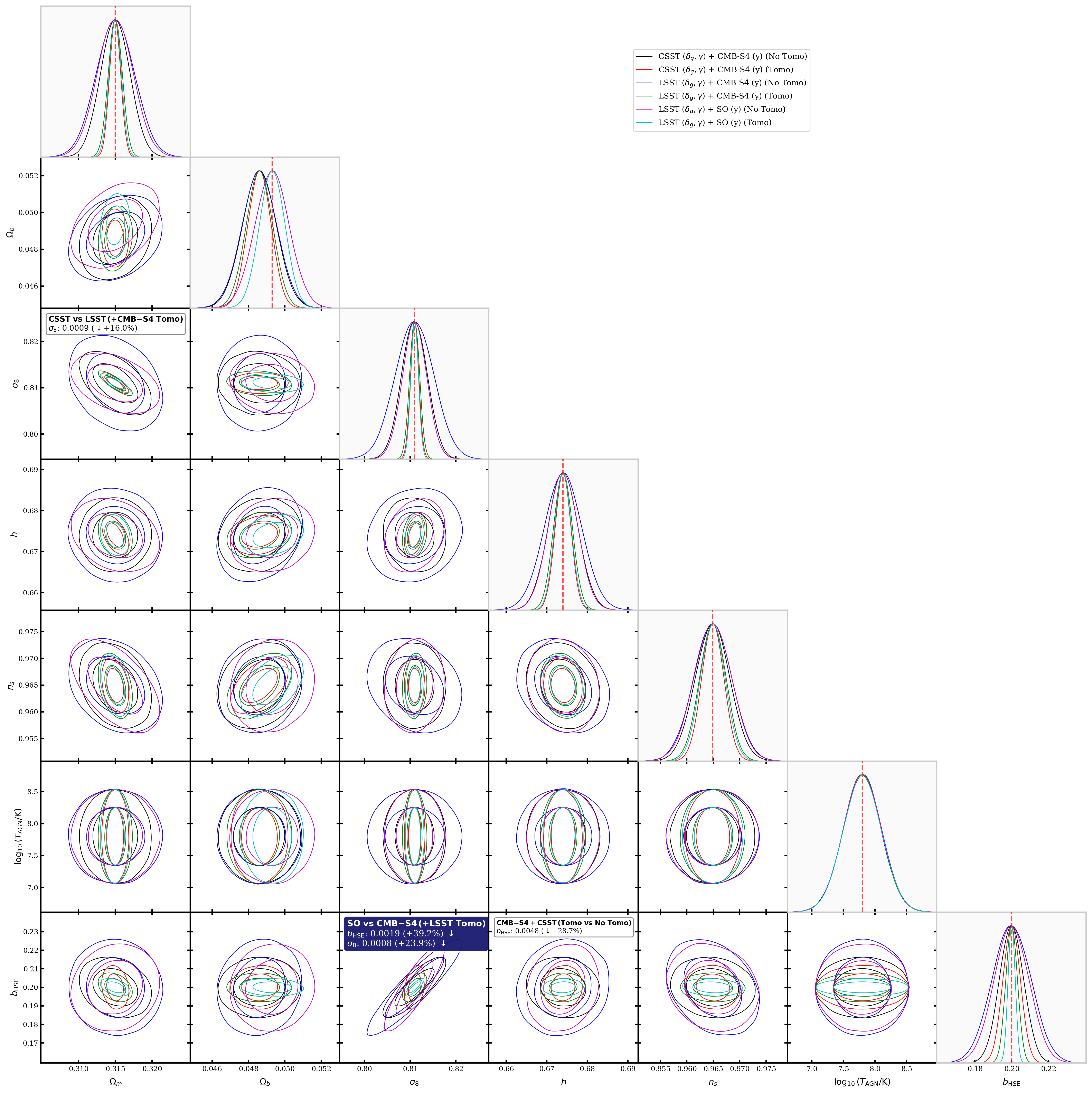}
    \caption{Multi-parameter constraints from cross-correlation analyses combining CMB and optical surveys. The corner plot shows 1$\sigma$ and 2$\sigma$ confidence contours for five selected cosmological and astrophysical parameters: the matter fluctuation amplitude $\sigma_8$, hydrostatic mass bias $b_{\mathrm{HSE}}$, matter density $\Omega_m$, Hubble constant $h$, and intrinsic alignment amplitude $A_{\mathrm{IA}}$. Four survey configurations are compared: CMB-S4 + CSST without tomography (blue dashed), CMB-S4 + CSST with tomography (blue solid), CMB-S4 + LSST with tomography (orange solid), and SO + LSST with tomography (purple solid). The significant tightening of constraints with tomography demonstrates the power of redshift binning in breaking parameter degeneracies, particularly for $b_{\mathrm{HSE}}$ which shows precision improvements from 29\% (CMB-S4+CSST) to 79.7\% (SO+LSST) with tomographic analyses.}
    \label{fig:corner_all_surveys}
\end{figure*}

\subsection{Probe Decomposition Analysis}
\label{subsec:probe_decomp}

{To quantify the synergy between multi-probe observables, we perform a probe decomposition analysis for all survey combinations, isolating contributions from tSZ auto-correlations ($C_\ell^{yy}$), tSZ--galaxy cross-correlations ($C_\ell^{gy}$), and galaxy clustering ($C_\ell^{gg}$). For all tomographic configurations, the joint $yy+gy+gg$ combination delivers the tightest constraints on $b_{\rm HSE}$, with $gy$ cross-correlations playing a key role in mitigating degeneracies between $b_{\rm HSE}$ and $\sigma_8$.} 

{A tSZ auto-spectrum-only analysis (with Planck priors) yields $\sigma(b_{\rm HSE}) = 0.0255$ (12.8\%), demonstrating that cross-correlations with LSST large-scale structure are essential to achieve percent-level calibration, even when external CMB information is available. Optical-only probes, including galaxy clustering ($gg$), cosmic shear ($\gamma\gamma$), and their combination in the standard 3$\times$2pt analysis (i.e., $gg + \gamma\gamma + g\gamma$), provide no significant constraint on $b_{\rm HSE}$, as they lack direct sensitivity to the thermal pressure field. By contrast, these same optical probes deliver strong constraints on $\sigma_8$: 3$\times$2pt alone achieves $\sigma(\sigma_8) = 0.00106$ (0.13\%), comparable to CMB-only forecasts. The inclusion of tSZ cross-correlations further refines this to $\sigma(\sigma_8) = 0.00083$ (0.10\%) in the full combination, while simultaneously enabling the first sub-percent constraint on $b_{\rm HSE}$.}

{Although the tSZ auto-spectrum exhibits the largest absolute response to changes in $\sigma_8$ (see top panel of Figure~\ref{fig:power_spectra_sigma8}),  its constraining power on $\sigma_8$ is limited by cosmic variance and residual foregrounds even with SO’s ultra-low instrumental noise (1.2 $\mu$K-arcmin). In contrast, galaxy clustering (gg) and 3$\times$2pt achieve tighter $\sigma_8$ constraints due to superior signal-to-noise at low multipoles, where $\sigma_8$ is most sensitively probed. This illustrates that final parameter precision depends on the interplay of sensitivity, noise, and cosmic variance rather than sensitivity alone. The $gy$ cross-correlations resolve the $b_{\rm HSE}$ limitation by linking the tSZ signal (a direct mass tracer) to galaxy overdensities (a large-scale structure tracer). For SO + LSST tomography, $gy$ contributes more than 60\% of the Fisher information for $b_{\rm HSE}$, yielding $\sigma(b_{\rm HSE}) = 0.00195$ (0.98\%) in the full combination. Tomography amplifies this synergy by reducing redshift-dependent degeneracies across multiple bins.}

\subsection{Systematics and Prior Choices}
\label{subsec:systematics_priors}

{We marginalize over all dominant astrophysical and observational systematics relevant to Stage-IV CMB and optical surveys, with priors calibrated to community standards \citep{DESCSRD} and validated by DES Y1 results \citep{DESY1}. Baryonic feedback is modeled via HMCode2020 \citep{Mead21}, parameterized by $\log_{10}(T_{\rm AGN}/{\rm K})$ (fiducial = 7.8, Gaussian prior $\sigma=0.3$ from BAHAMAS simulations). The GNFW pressure profile parameters ($P_0, c_{500}$) are fixed to X-ray-calibrated universal values \citep{Arnaud2010}, as our forecasts show no significant improvement on the $\log_{10}(T_{\rm AGN}/{\rm K})$ prior across all survey combinations. TSZ cross-correlations probe ICM thermal pressure, whose uncertainty is absorbed by $b_{\rm HSE}$ and GNFW parameters rather than matter clustering amplitude, which aligns with \citep{DESCSRD} that found weak baryonic constraints from cosmic shear alone. While hydrodynamical simulations indicate correlated suppression of small-scale clustering and Compton-$y$ signal under stronger AGN feedback \citep{Battaglia12,LeBrun15,Harnois15,McCarthy17}, a self-consistent joint model linking matter and pressure profiles remains an open challenge, so we adopt independent modeling following current multi-probe practice \citet{Mead21,DESCSRD}. For intrinsic alignments, we use the NLA model with two parameters: amplitude $A_{\rm IA}$ and redshift evolution $\eta_{\rm IA}$, applying Gaussian priors $\sigma(A_{\rm IA})=0.50$ and $\sigma(\eta_{\rm IA})=0.50$ (both surveys) consistent with DESC SRD recommendations but simplified by omitting luminosity and high-redshift scaling terms, a standard approach for baseline forecasts. For observational systematics, we include per-tomographic-bin galaxy bias perturbations ($\delta b_{g,i}$) and photometric redshift shifts ($\Delta z_{\rm lens}, \Delta z_{\rm src}$) with conservative Gaussian priors: $\sigma(\delta b_{g,i})=0.05$, $\sigma(\Delta z_{\rm lens})=0.002(1+z)$, and $\sigma(\Delta z_{\rm src})=0.005(1+z)$, which matches the SRD’s Y10 Cluster requirement CL2. Fiducial values (e.g., $b_{\rm HSE}=0.2$, $\sigma_8=0.811$) follow Planck 2018 $\Lambda$CDM \citep{Planck:2018vyg} cluster mass calibration studies. Marginalized constraints on nuisance parameters are shown in Appendix~\ref{app:photo_z_bias_plots}.
Secondary systematics (e.g., cluster misalignments, pressure profile deviations, CMB beam and optical shear calibration errors) are subdominant for Stage-IV performance and expected to contribute $<0.5\%$ additional uncertainty to $b_{\rm HSE}$. All priors are intentionally wide to ensure constraints are driven by data, not assumptions, consistent with \citet{DESCSRD} and \citet{DESY1}.}

\subsection{Robustness Tests}
\label{subsec:robustness}

{To ensure our constraints are not driven by optimistic assumptions, we perform five robustness checks on our fiducial SO + LSST tomographic forecast (Section~\ref{subsec:bHSE_constraints}).  
First, varying the maximum multipole from the baseline $\ell_{\rm max} = 3000$ to $\ell_{\rm max} = 2000$ or $1500$ degrades $\sigma(b_{\rm HSE})$ from $0.98\%$ to $1.48\%$ (+51\%) and $2.00\%$ (+104\%), respectively. This degradation occurs because the tSZ power spectrum peaks at $\ell \sim 2000\text{--}3000$, where most of the Fisher information on $b_{\rm HSE}$ resides. The high-precision calibration of hydrostatic bias thus relies critically on nonlinear, small-scale modes precisely those accessible to SO’s low-noise, high-resolution measurements.}

{Second, widening or tightening the $\log_{10}(T_{\rm AGN}/{\rm K})$ prior by a factor of two changes $\sigma(b_{\rm HSE})$ by only $\sim 0\%$ (since $\log_{10}(T_{\rm AGN})$ is unconstrained), indicating minimal sensitivity to baryonic feedback modeling.}  
{Third, doubling or halving the photometric redshift shift priors ($\Delta z_{\rm lens}, \Delta z_{\rm src}$) alters $\sigma(b_{\rm HSE})$ by $+2\%$ to $-4\%$, identifying photo-$z$ calibration as a negligible systematic for $b_{\rm HSE}$.  
Fourth, removing all external Planck CMB priors on $\Lambda$CDM parameters degrades $\sigma(b_{\rm HSE})$ from $0.98\%$ to $1.00\%$ (+2\%). Crucially, the constraint remains at the sub-percent level, demonstrating that tSZ--optical cross-correlations not external CMB information provide the foundational calibration of hydrostatic mass bias. Planck priors offer negligible refinement for $b_{\rm HSE}$.  
Finally, alternative nuisance parameter specifications (e.g., $\sigma(\Delta z_{\rm src}) = 0.003$ vs.\ $0.005$) yield consistent results, confirming prior robustness.  
Together, these tests confirm that our sub-percent $b_{\rm HSE}$ constraint is data-driven, stable, and anchored in the multi-probe synergy of SO and LSST.}

\section{Results \& Summary}
\label{sec:Result_summary}

Our Fisher forecast analysis demonstrates that tomographic cross-correlations between next-generation CMB (CMB-S4, Simons Observatory/SO) and optical (LSST, CSST) surveys will deliver improved constraints on the hydrostatic mass bias parameter $b_{\rm HSE}$ and core cosmological parameters. These results are derived from a halo model framework incorporating realistic noise, foregrounds, and photometric redshift uncertainties \citep{CMBS4_Book, DESCSRD, Xiong25, SO_25}.

\subsection{Hydrostatic Mass Bias Constraints}
\label{subsec:bHSE_constraints}

The most striking finding is the substantial improvement in $b_{\rm HSE}$ precision enabled by tomography, though the magnitude of gain varies across survey combinations (Table~\ref{tab:surveys_updated}). Our forecasts yield 1$\sigma$ constraints on $b_{\rm HSE}$ (fiducial = 0.20) {with the tightest constraint achieved by SO + LSST tomography at 0.98\%, down to 5.30\% for CMB-S4 + LSST without tomography.}

Tomography drives precision gains of {28.6\%} for CMB-S4 + CSST {(3.36\% $\to$ 2.40\%)}, {69.8\%} for CMB-S4 + LSST {(5.30\% $\to$ 1.60\%)}, and {79.7\%} for SO + LSST {(4.80\% $\to$ 0.98\%)}. The higher galaxy density of LSST (27 arcmin$^{-2}$ vs. CSST’s 22 arcmin$^{-2}$) and its 10 lens/5 source bins amplify tomography gains, while SO’s ultra-low tSZ noise (1.2~$\mu$K-arcmin at 145~GHz) enables sub-percent mass calibration when combined with LSST tomography.

Notably, the SO + LSST tomographic constraint ({$\sigma(b_{\rm HSE}) = 0.00195$}) achieves {0.98\%} precision—sufficient to meet the CMB-S4 Science Book's requirement for sub-percent cluster mass calibration and help resolve the $\sigma_8$ tension between CMB and large-scale structure probes \citep{Planck20}. This surpasses even optimistic projections for current cluster surveys, marking a leap from $b_{\rm HSE}$ as a limiting systematic to a precision-constrained parameter.
{Given SO’s ongoing operations and competitive performance, our results establish SO + LSST as the leading near-term pathway to sub-percent hydrostatic mass bias calibration via tSZ cross-correlations.}

\subsection{Cosmological Parameter Constraints}

The sensitivity of individual probes and their cross-correlations to $\sigma_8$ is illustrated in Figure~\ref{fig:power_spectra_sigma8}, which shows that tSZ--shear and galaxy--tSZ cross-spectra provide complementary leverage across multipoles. All tomographic configurations achieve Stage-IV/V-level constraints on core cosmological parameters (Table~\ref{tab:cosmo_constraints}). When marginalizing over $b_{\rm HSE}$, {baryonic feedback $\log_{10}(T_{\rm AGN}/{\rm K})$}, and intrinsic alignments ($A_{\rm IA}, \eta_{\rm IA}$), we find:
{SO + LSST delivers the tightest constraints with $\sigma(\sigma_8) = 0.00083$ (0.10\%), $\sigma(\Omega_m) = 0.00083$ (0.26\%), and $\sigma(h) = 0.00204$ (0.30\%)};
CMB-S4 + CSST achieves {$\sigma(\sigma_8) = 0.00092$, $\sigma(\Omega_m) = 0.00075$, $\sigma(h) = 0.00193$};
and CMB-S4 + LSST yields {$\sigma(\sigma_8) = 0.00109$, $\sigma(\Omega_m) = 0.00095$, $\sigma(h) = 0.00231$}.
These results are competitive with standalone CMB or optical surveys, but with the added robustness of multi-probe synergy {and full systematic marginalization.}

Intrinsic alignment ({$A_{\rm IA}, \eta_{\rm IA}$}) is tightly controlled in tomographic setups, {with marginalized uncertainties reduced by factors of 3–5} compared to non-tomographic analyses. This highlights tomography’s unique ability to isolate lensing signals from redshift-dependent systematics, a critical advantage for weak lensing cosmology that also strengthens constraints on cluster-based cosmological probes like the tSZ effect.

\begin{table}[h]
\centering
\caption{{Forecasted 1$\sigma$ constraints on key cosmological parameters from \textit{tomographic} multi-probe analyses, fully marginalized over $b_{\rm HSE}$ (fiducial = 0.20), $\log_{10}(T_{\rm AGN}/{\rm K})$, intrinsic alignment parameters ($A_{\rm IA}, \eta_{\rm IA}$), galaxy bias perturbations, and photometric redshift shifts. All uncertainty values are reported in units of $10^{-3}$}.}
\label{tab:cosmo_constraints}
\renewcommand{\arraystretch}{1.1} 
\begin{tabular}{lccc}
\toprule
\multicolumn{4}{c}{\textit{Tomographic Survey Combinations}} \\ 
\midrule
Survey Combination & ${\sigma(\sigma_8) \times 10^3}$ & ${\sigma(\Omega_m) \times 10^3}$ & ${\sigma(h) \times 10^3}$ \\
\midrule
SO + LSST     & {0.83} & {0.80} & {2.04} \\
CMB-S4 + CSST & {0.92} & {0.75} & {1.93} \\
CMB-S4 + LSST & {1.09} & {0.95} & {2.31} \\
\bottomrule
\end{tabular}
\end{table}

\subsection{Degeneracy Breaking, Parameter Correlations}

The multi-parameter corner plot (Figure~\ref{fig:corner_all_surveys}) reveals critical degeneracy patterns, with tomography playing a pivotal role in breaking the fundamental $b_{\rm HSE}$–$\sigma_8$ correlation. For SO + LSST, tomographic cross-correlations reduce the 2D contour area for this pair by $\sim 80\%$ compared to non-tomographic analyses, enabling simultaneous precision constraints on both parameters.

A defining feature of our constraints is the evolution of key parameter degeneracies, driven by tomographic information and synergy between probes. In the $\Omega_m$–$\sigma_8$ plane, no-tomography analyses produce tight, negatively sloped anticorrelations a signature of averaging signals across redshifts. Tomography tightens this degeneracy via redshift-resolved distance information, preserving its anticorrelated direction for CMB-S4+CSST/LSST. Notably, the SO+LSST no-tomography case weakens this anticorrelation: SO’s tSZ signal (sensitive to total cluster abundance) introduces a competing positive correlation between $\Omega_m$ and $\sigma_8$, partially canceling the negative clustering anticorrelation. This competition vanishes in SO+LSST tomography, where LSST’s redshift-resolved data dominates, restoring the standard $\Omega_m$–$\sigma_8$ anticorrelation while significantly improving precision.

The $b_{\rm HSE}$–$\sigma_8$ plane exhibits an even more striking degeneracy breakdown. In no-tomography cases, $b_{\rm HSE}$ and $\sigma_8$ are nearly indistinguishable their effects on the observed signal are conflated when averaging across redshifts. Tomography resolves this ambiguity by leveraging redshift-dependent $b_{\rm HSE}$ behavior: hydrostatic mass bias varies with cluster mass (and thus redshift), while $\sigma_8$ is redshift-independent. By comparing signals across tomographic bins, we separate the redshift-varying apparent amplitude (from $b_{\rm HSE}$) from the redshift-constant intrinsic amplitude (from $\sigma_8$), thereby substantially reducing the degeneracy.

The $b_{\rm HSE}$–$\Omega_m$ degeneracy exhibits survey-specific behavior: most combinations show moderate anti-correlation, while SO + LSST significantly weakens this anti-correlation. This unique performance stems from SO’s ultra-low tSZ noise and LSST’s high-density tomography, which provide independent constraints on $\Omega_m$ (via galaxy clustering) and $b_{\rm HSE}$ (via redshift-dependent tSZ–shear cross-correlations).

Notably, non-tomographic configurations show contrasting $A_{\rm IA}$–$\Omega_m$ correlations: CMB-S4 + LSST (No Tomo) exhibits anti-correlation, while CMB-S4 + CSST (No Tomo) shows weak correlation. These differences arise from survey-specific properties. LSST’s wider redshift coverage and CSST’s tighter photometric redshift precision, which cause non-tomographic analyses to partially conflate intrinsic alignments (redshift-dependent) with cosmological structure growth. Tomography eliminates this ambiguity, with all tomographic setups showing minimal $A_{\rm IA}$–cosmology correlations, indicating self-calibration of intrinsic alignments without degrading constraints.

These results underscore two critical insights: (1) probe synergy (e.g., SO+LSST) can modify degeneracies even without tomography, and (2) tomography’s unique power lies in prioritizing redshift-dependent information disentangling systematic effects (like $b_{\rm HSE}$) from cosmological parameters (like $\Omega_m$ and $\sigma_8$), reasserting physical parameter relations, and maximizing precision for precision cosmology.

The total information content quantified by the number of independent power spectra, ranges from 78 (CSST tomography: 6 lens × 5 source + tSZ-auto + cross terms) to 121 (LSST tomography: 10 lens × 5 source + tSZ terms), reflecting the leverage of tomographic binning in breaking degeneracies.

\subsection{Survey-Specific Advantages and Cosmological Implications}
Each combination demonstrates distinct strengths with profound implications for cluster cosmology. For instance, the CMB-S4 + CSST benefits from CSST’s space-based advantages (superior image quality, tighter photo-$z$ uncertainties: $\sigma_z/(1+z) = {0.03}$) to deliver competitive $b_{\rm HSE}$ precision ({2.40\% tomographic}) with modest tomography gains ({28.7\%}), positioning it as a cost-effective complement to ground-based surveys. Although CMB-S4 + LSST uses LSST's high galaxy density and wide redshift coverage to achieve {1.60\% $b_{\rm HSE}$ precision}, with {69.8\% tomography improvement}, it excels in balancing cluster bias and cosmological constraints. However, SO + LSST: SO’s ultra-low tSZ noise (1.2 $\mu$K-arcmin at 145 GHz) and optimized frequency coverage combine with LSST’s deep tomography to set the gold standard, with {0.98\% $b_{\rm HSE}$ precision}. Even without tomography, SO + LSST achieves a competitive {4.80\% constraint on $b_{\rm HSE}$ comparable to the non-tomographic performance of CMB-S4 + LSST (5.30\%) and already sufficient for Stage-IV science. When combined with tomography, SO + LSST reaches 0.98\%, surpassing even CMB-S4-based configurations.}

These results mark a paradigm shift: {high-precision $b_{\rm HSE}$ constraints (0.98–2.40\% for tomographic configurations)} will reduce cluster mass calibration uncertainties from a dominant systematic to a sub-dominant one, enabling $\sigma_8$ measurements from cluster abundances {with sub-percent statistical uncertainty after marginalizing over key systematics.} Tomographic cross-correlations provide a self-calibration pathway for $b_{\rm HSE}$ independent of X-ray or SZ scaling relations, mitigating circularity in mass calibration.

For the $\sigma_8$ tension, SO + LSST’s {0.98\% $b_{\rm HSE}$ precision and 0.00083 $\sigma(\sigma_8)$} will enable discrimination between competing explanations (e.g., new physics vs. systematic mismodeling). Meanwhile, CSST and LSST’s complementary coverage offers cross-validation, ensuring robustness of results.

The varying but substantial tomography gains ({28.7–79.7\%}) underscore redshift binning as a foundational tool for breaking degeneracies, paving the way for a new era of precision in structure formation and inference of cosmological parameters. { Although we adopt a constant hydrostatic mass bias $b_{\rm HSE}$ for simplicity, physically motivated models based on hydrodynamical simulations indicate that the non-thermal pressure fraction and thus $b_{\rm HSE}$ depends on halo mass accretion history and radius. Crucially, when the radial scale is normalized to $r_{200m}$ (the radius enclosing 200 times the mean matter density), this dependence exhibits minimal redshift evolution, as demonstrated in Figure~1 of \citet{Nelson14} and further developed in \citet{Shi15,Shi16}. Moreover, recent high-resolution simulations by \citet{Barranco26} reveal that $b_{\rm HSE}$ undergoes a characteristic time-dependent evolution during major mergers exhibiting a sharp negative dip at virial crossing, followed by a transient positive overshoot (indicating temporary mass overestimation), before relaxing back to a negative baseline. This behavior is primarily driven by merger-induced morphological reconfigurations of the gas density profile, rather than thermodynamic changes. Incorporating such a physically grounded, mass-, redshift-, and dynamical-state-dependent model in future forecasts would enable more realistic marginalization over astrophysical uncertainties and test the robustness of our constraints.}

In summary, tomographic cross-correlations between CMB and optical surveys will transform cluster cosmology. SO + LSST tomography leads with {0.98\% $b_{\rm HSE}$ precision}, CMB-S4 + LSST balances sensitivity and accessibility, and CMB-S4 + CSST provides a viable alternative with space-based advantages.

\section*{acknowledgments}
We thank the CMB-S4, Simons Observatory, LSST/DESC, and CSST collaborations for making public their science requirements, redshift distributions, and instrumental specifications. This work was supported by Guangdong Technion startup funds. YG acknowledges the support from National Key R\&D Program of China grant Nos. 2022YFF0503404 and 2020SKA0110402, ,the CAS Project for Young Scientists in Basic Re- search (No. YSBR-092), and science research grants from the China Manned Space Project with grant Nos. CMS-CSST- 2025-A02. All results presented here are theoretical forecasts based on published survey specifications; no observational data were used.

\appendix

\section{Supplementary Methodological Details}
\label{app:supp_methods}

\subsection{Prior Specifications for Fisher Forecast Parameters}
\label{app:priors_table}

Full prior specifications for all parameters in our Fisher forecast analysis are provided in Table~\ref{tab:priors_expanded}. The methodology and physical motivation for these choices are detailed in Section~\ref{subsec:systematics_priors}.


\begin{table*}
\centering
\caption{Prior specifications for Fisher forecast parameters (CMB-S4 + LSST/CSST surveys). 
All priors are absolute unless noted as relative (scaled to fiducial value or redshift). 
References: HMCode2020 \citep{Mead21}, DESC SRD 2018 \citep{DESCSRD}, DES Y1 \citep{DESY1}, Planck 2018 \citep{Planck:2018vyg}.}
\label{tab:priors_expanded}

\setlength{\tabcolsep}{6pt} 
\renewcommand{\arraystretch}{1.1} 

\begin{tabular}{llll}
\toprule
\textbf{Parameter} & \textbf{Fiducial} & \textbf{Prior $\sigma$ (Baseline / Tested)} & \textbf{Source/Justification} \\
\midrule
\multicolumn{4}{l}{\textit{$\Lambda$CDM Cosmology} \citep{Planck:2018vyg}} \\
$\Omega_m$                & 0.315  & 0.007 (abs; SO) / 1\% rel (CMB-S4) & Planck 2018 / CMB-S4 Sci. Book \\
$\Omega_b$                & 0.0493 & 0.001 (abs) / ---                  & Planck 2018 / BBN \\
$\sigma_8$                & 0.811  & 0.006 (abs; SO) / 1\% rel (CMB-S4) & Planck 2018 / CMB-S4 Sci. Book \\
$h$                       & 0.674  & 1\% rel (CMB-S4) / 0.01 (abs; SO)  & Planck 2018 / CMB-S4 Sci. Book \\
$n_s$                     & 0.9649 & 0.004 (abs) / ---                  & Planck 2018 \\
\midrule
\multicolumn{4}{l}{\textit{Astrophysical Systematics}} \\
$b_{\rm HSE}$             & 0.2    & None (target) / ---                & CMB cluster cal. / HMCode2020 \\
$\log_{10}(T_{\rm AGN}/{\rm K})$ & 7.8 & 0.3 (abs) / 0.15, 0.60         & HMCode2020 / BAHAMAS sims \\
$A_{\rm IA}$              & 1.0    & 10\% rel (0.10) / 0.50 (abs)       & NLA model (DES Y1 / DESC SRD) \\
$\eta_{\rm IA}$           & 0.0    & 0.50 (abs) / ---                   & NLA model (DES Y1 / DESC SRD) \\
\midrule
\multicolumn{4}{l}{\textit{Nuisance Parameters (per tomographic bin)}} \\
$\delta b_g^{(i)}$        & 0.0    & 0.05 (abs) / ---                   & DESC SRD 2018 \\
$\Delta z_{\rm lens}^{(i)}$ & 0.0  & $0.002(1+z)$ / 0.001, 0.004    & DESC SRD 2018 (LSS1 Y10) \\
$\Delta z_{\rm src}^{(j)}$  & 0.0  & $0.003(1+z)$ / 0.005, 0.010    & DESC SRD 2018 (WL1 Y10) \\
\bottomrule
\end{tabular}
\end{table*}

\subsection{Galaxy Bias and Photometric Redshift Uncertainty Models}
\label{app:photo_z_bias_plots}

\begin{figure*}
    \centering
    \includegraphics[width=\textwidth]{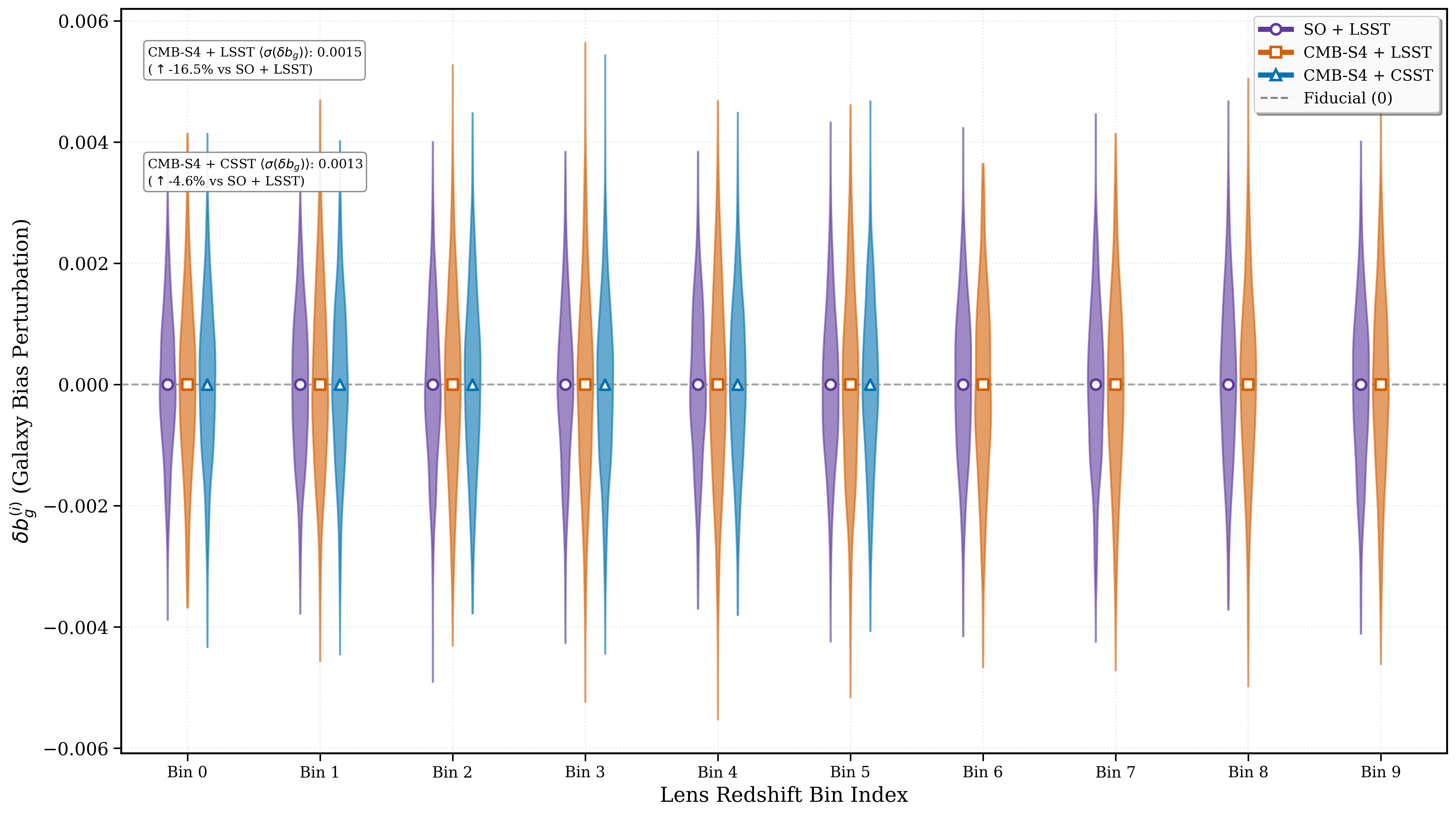}
    \caption{
        Constraints on galaxy bias perturbation nuisance parameters $\delta b_g^{(i)}$ for tomographic survey combinations: SO+LSST (purple), CMB-S4+LSST (orange), and CMB-S4+CSST (blue).
        Results are shown for 10 lens redshift bins, with violin plots representing 1$\sigma$ uncertainty distributions.
        Annotations in the top-left report the average uncertainty $\langle\sigma(\delta b_g)\rangle$ for each survey and its percentage change relative to the SO+LSST baseline (purple).
        The horizontal gray dashed line marks the fiducial zero value for all bias perturbations.
    }
    
    \label{fig:galaxy_bias}
\end{figure*}

\begin{figure*}
    \centering
    \includegraphics[width=\textwidth]{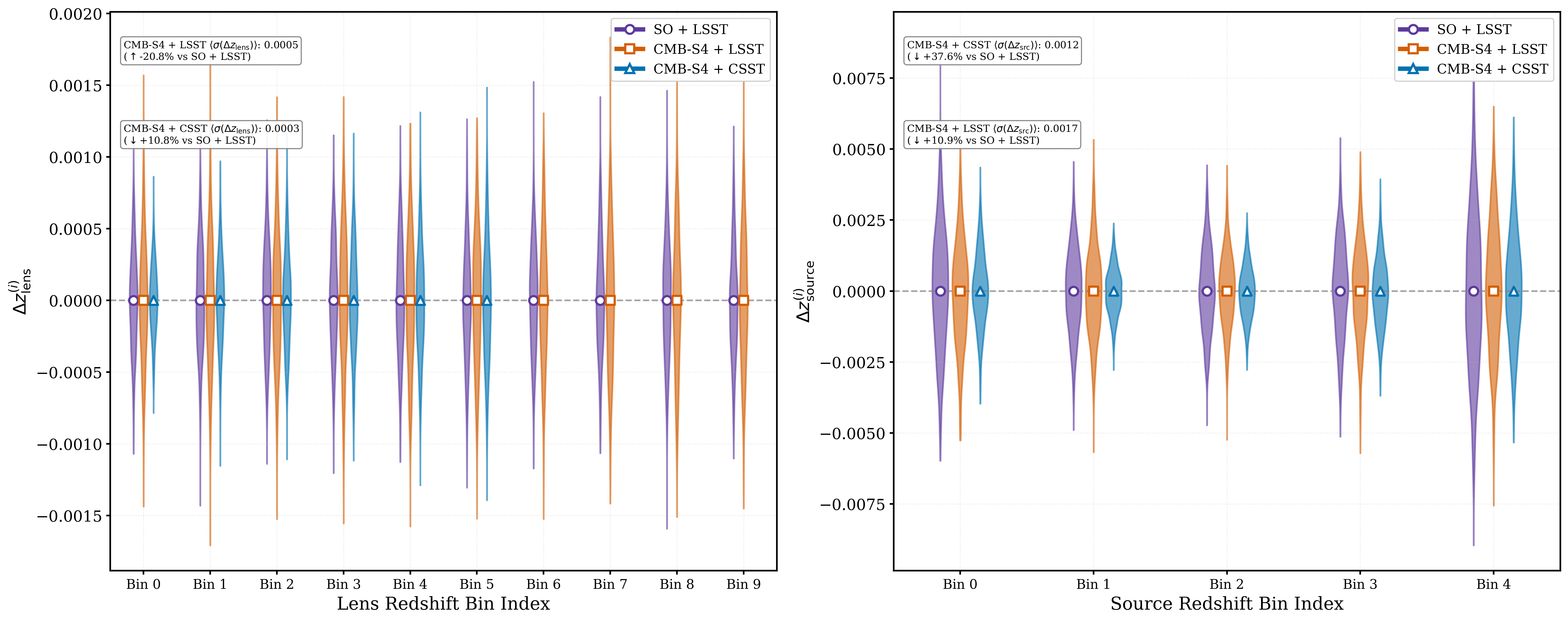}
    \caption{
        Constraints on photometric redshift shift nuisance parameters for tomographic survey combinations: SO+LSST (purple), CMB-S4+LSST (orange), and CMB-S4+CSST (blue).
        Violin plots show 1$\sigma$ uncertainty distributions for each redshift bin.
        \textit{Left}: Lens photo-z shifts $\Delta z_{\rm lens}^{(i)}$ across 10 lens redshift bins.
        \textit{Right}: Source photo-z shifts $\Delta z_{\rm source}^{(i)}$ across 5 source redshift bins.
        Annotations in the top-left of each panel report the average uncertainty $\langle\sigma(\Delta z)\rangle$ and percentage changes relative to the SO+LSST baseline.
        The horizontal gray dashed lines mark the fiducial zero shift value.
    }
    \label{fig:photoz_shifts}
\end{figure*}

{Figures~\ref{fig:galaxy_bias} and~\ref{fig:photoz_shifts} show marginalized 1$\sigma$ constraints on tomographic nuisance parameters for the three survey combinations: SO+LSST (baseline), CMB-S4+LSST, and CMB-S4+CSST. Violin plots represent the full uncertainty distribution per redshift bin, with white markers indicating the mean constraint.} 

{For galaxy bias perturbations ($\delta b_g^{(i)}$), SO+LSST achieves the tightest average constraint ($\langle\sigma(\delta b_g)\rangle = 0.00128$), outperforming CMB-S4+LSST ($0.00150$, $-16.5\%$) and CMB-S4+CSST ($0.00134$, $-4.6\%$). For photometric redshift shifts, SO+LSST also sets the benchmark: lens photo-$z$ shifts are constrained to $\langle\sigma(\Delta z_{\rm lens})\rangle = 0.00038$, while source photo-$z$ shifts reach $\langle\sigma(\Delta z_{\rm src})\rangle = 0.00191$. All results reflect full marginalization over cosmological and astrophysical parameters, demonstrating the robustness of our systematics modeling.}

\bibliography{cited_references}{}

@ARTICLE{Vikhlinin09,
       author = {{Vikhlinin}, A. and {Kravtsov}, A.~V. and {Burenin}, R.~A. and {Ebeling}, H. and {Forman}, W.~R. and {Hornstrup}, A. and {Jones}, C. and {Murray}, S.~S. and {Nagai}, D. and {Quintana}, H. and {Voevodkin}, A.},
        title = "{Chandra Cluster Cosmology Project III: Cosmological Parameter Constraints}",
      journal = {\apj},
         year = 2009,
        month = feb,
       volume = {692},
       number = {2},
        pages = {1060-1074},
          doi = {10.1088/0004-637X/692/2/1060},
archivePrefix = {arXiv},
       eprint = {0812.2720},
 primaryClass = {astro-ph},
       adsurl = {https://ui.adsabs.harvard.edu/abs/2009ApJ...692.1060V}
}

@ARTICLE{CSST_ii,
       author = {{Gong}, Yan and {Miao}, Haitao and {Zhou}, Xingchen and {Xiong}, Qi and {Song}, Yingxiao and {Jiang}, Yuer and {Wang}, Minglin and {Yan}, Junhui and {Wu}, Beichen and {Deng}, Furen and {Chen}, Xuelei and {Fan}, Zuhui and {Jing}, Yipeng and {Yang}, Xiaohu and {Zhan}, Hu},
        title = "{Future cosmology: New physics and opportunity from the China Space Station Telescope (CSST)}",
      journal = {Science China Physics, Mechanics, and Astronomy},
         year = 2025,
        month = aug,
       volume = {68},
       number = {8},
          eid = {280402},
        pages = {280402},
          doi = {10.1007/s11433-025-2646-2},
archivePrefix = {arXiv},
       eprint = {2501.15023},
 primaryClass = {astro-ph.CO},
       adsurl = {https://ui.adsabs.harvard.edu/abs/2025SCPMA..6880402G}
}

@ARTICLE{CSST_iii,
       author = {{CSST Collaboration} and {Gong}, Yan and {Miao}, Haitao and {Zhan}, Hu and {Li}, Zhao-Yu and {Shangguan}, Jinyi and {Li}, Haining and {Liu}, Chao and {Chen}, Xuefei and {Yuan}, Haibo and {Zhou}, Jilin and {Liu}, Hui-Gen and {Yu}, Cong and {Ji}, Jianghui and {Qi}, Zhaoxiang and {Liu}, Jiacheng and {Dai}, Zigao and {Wang}, Xiaofeng and {Zheng}, Zhenya and {Hao}, Lei and {Dou}, Jiangpei and {Ao}, Yiping and {Lin}, Zhenhui and {Zhang}, Kun and {Wang}, Wei and {Sun}, Guotong and {Li}, Ran and {Li}, Guoliang and {Xu}, Youhua and {Li}, Xinfeng and {Li}, Shengyang and {Wu}, Peng and {Zhang}, Jiuxing and {Wang}, Bo and {Bai}, Jinming and {Cai}, Yi-Fu and {Cai}, Zheng and {Cao}, Jie and {Chan}, Kwan Chuen and {Chang}, Jin and {Chen}, Xiaodian and {Chen}, Xuelei and {Chen}, Yuqin and {Chen}, Yun and {Cui}, Wei and {Dong}, Subo and {Du}, Pu and {Duan}, Wenying and {Fan}, Junhui and {Fan}, LuLu and {Fan}, Zhou and {Fan}, Zuhui and {Fang}, Taotao and {Fu}, Jianning and {Fu}, Liping and {Fu}, Zhensen and {Gao}, Jian and {Gu}, Shenghong and {Gu}, Yidong and {Guo}, Qi and {Han}, Zhanwen and {Hu}, Bin and {Huang}, Zhiqi and {Ho}, Luis C. and {Jiang}, Linhua and {Jiang}, Ning and {Jing}, Yipeng and {Kang}, Xi and {Kong}, Xu and {Li}, Cheng and {Li}, Chengyuan and {Li}, Di and {Li}, Jing and {Li}, Nan and {Li}, Yang A. and {Liao}, Shilong and {Lin}, Weipeng and {Liu}, Fengshan and {Liu}, Jifeng and {Liu}, Xiangkun and {Liu}, Zhuokai and {Mao}, Ruiqing and {Mao}, Shude and {Meng}, Xianmin and {Pang}, Xiaoying and {Peng}, Xiyan and {Peng}, Yingjie and {Shan}, Huanyuan and {Shen}, Juntai and {Shen}, Shiyin and {Shen}, Zhiqiang and {Shi}, Sheng-Cai and {Shi}, Yong and {Tan}, Siyuan and {Tian}, Hao and {Wang}, Jianmin and {Wang}, Jun-Xian and {Wang}, Xin and {Wang}, Yuting and {Wu}, Hong and {Wu}, Jingwen and {Wu}, Xuebing and {Xu}, Chun and {Xue}, Xiang-Xiang and {Xue}, Yongquan and {Yang}, Ji and {Yang}, Xiaohu and {Yao}, Qijun and {Yuan}, Fangting and {Yuan}, Zhen and {Zhang}, Jun and {Zhang}, Pengjie and {Zhang}, Tianmeng and {Zhang}, Wei and {Zhang}, Xin and {Zhao}, Gang and {Zhao}, Gongbo and {Zhong}, Hongen and {Zhong}, Jing and {Zhou}, Liyong and {Zhu}, Wei and {Zu}, Ying},
        title = "{Introduction to the Chinese Space Station Survey Telescope (CSST)}",
      journal = {arXiv e-prints},
         year = 2025,
        month = jul,
          eid = {arXiv:2507.04618},
        pages = {arXiv:2507.04618},
          doi = {10.48550/arXiv.2507.04618},
archivePrefix = {arXiv},
       eprint = {2507.04618},
 primaryClass = {astro-ph.IM},
       adsurl = {https://ui.adsabs.harvard.edu/abs/2025arXiv250704618C}
}

@ARTICLE{Zhang22,
       author = {{Zhang}, Zhuoqi (Jackie) and {Chang}, Chihway and {Larsen}, Patricia and {Secco}, Lucas F. and {Zuntz}, Joe and {LSST Dark Energy Science Collaboration}},
        title = "{Transitioning from Stage-III to Stage-IV: cosmology from galaxy{\texttimes}CMB lensing and shear{\texttimes}CMB lensing}",
      journal = {\mnras},
         year = 2022,
        month = aug,
       volume = {514},
       number = {2},
        pages = {2181-2197},
          doi = {10.1093/mnras/stac1407},
archivePrefix = {arXiv},
       eprint = {2111.04917},
 primaryClass = {astro-ph.CO},
       adsurl = {https://ui.adsabs.harvard.edu/abs/2022MNRAS.514.2181Z}
}

@ARTICLE{Nagai07,
       author = {{Nagai}, Daisuke and {Vikhlinin}, Alexey and {Kravtsov}, Andrey V.},
        title = "{Testing X-Ray Measurements of Galaxy Clusters with Cosmological Simulations}",
      journal = {\apj},
         year = 2007,
        month = jan,
       volume = {655},
       number = {1},
        pages = {98-108},
          doi = {10.1086/509868},
archivePrefix = {arXiv},
       eprint = {astro-ph/0609247},
 primaryClass = {astro-ph},
       adsurl = {https://ui.adsabs.harvard.edu/abs/2007ApJ...655...98N}
}

@ARTICLE{Nelson14,
       author = {{Nelson}, Kaylea and {Lau}, Erwin T. and {Nagai}, Daisuke},
        title = "{Hydrodynamic Simulation of Non-thermal Pressure Profiles of Galaxy Clusters}",
      journal = {\apj},
         year = 2014,
        month = sep,
       volume = {792},
       number = {1},
          eid = {25},
        pages = {25},
          doi = {10.1088/0004-637X/792/1/25},
archivePrefix = {arXiv},
       eprint = {1404.4636},
 primaryClass = {astro-ph.CO},
       adsurl = {https://ui.adsabs.harvard.edu/abs/2014ApJ...792...25N}
}

@ARTICLE{Braspenning25,
       author = {{Braspenning}, Joey and {Schaye}, Joop and {Schaller}, Matthieu and {Kugel}, Roi and {Kay}, Scott T.},
        title = "{Hydrostatic mass bias for galaxy groups and clusters in the FLAMINGO simulations}",
      journal = {\mnras},
         year = 2025,
        month = feb,
       volume = {536},
       number = {4},
        pages = {3784-3802},
          doi = {10.1093/mnras/stae2798},
archivePrefix = {arXiv},
       eprint = {2409.07849},
 primaryClass = {astro-ph.CO},
       adsurl = {https://ui.adsabs.harvard.edu/abs/2025MNRAS.536.3784B}
}

@ARTICLE{Planck20,
       author = {{Planck Collaboration} and {Aghanim}, N. and {Akrami}, Y. and {Ashdown}, M. and {Aumont}, J. and {Baccigalupi}, C. and {Ballardini}, M. and {Banday}, A.~J. and {Barreiro}, R.~B. and {Bartolo}, N. and {Basak}, S. and {Battye}, R. and {Benabed}, K. and {Bernard}, J. -P. and {Bersanelli}, M. and {Bielewicz}, P. and {Bock}, J.~J. and {Bond}, J.~R. and {Borrill}, J. and {Bouchet}, F.~R. and {Boulanger}, F. and {Bucher}, M. and {Burigana}, C. and {Butler}, R.~C. and {Calabrese}, E. and {Cardoso}, J. -F. and {Carron}, J. and {Challinor}, A. and {Chiang}, H.~C. and {Chluba}, J. and {Colombo}, L.~P.~L. and {Combet}, C. and {Contreras}, D. and {Crill}, B.~P. and {Cuttaia}, F. and {de Bernardis}, P. and {de Zotti}, G. and {Delabrouille}, J. and {Delouis}, J. -M. and {Di Valentino}, E. and {Diego}, J.~M. and {Dor{\'e}}, O. and {Douspis}, M. and {Ducout}, A. and {Dupac}, X. and {Dusini}, S. and {Efstathiou}, G. and {Elsner}, F. and {En{\ss}lin}, T.~A. and {Eriksen}, H.~K. and {Fantaye}, Y. and {Farhang}, M. and {Fergusson}, J. and {Fernandez-Cobos}, R. and {Finelli}, F. and {Forastieri}, F. and {Frailis}, M. and {Fraisse}, A.~A. and {Franceschi}, E. and {Frolov}, A. and {Galeotta}, S. and {Galli}, S. and {Ganga}, K. and {G{\'e}nova-Santos}, R.~T. and {Gerbino}, M. and {Ghosh}, T. and {Gonz{\'a}lez-Nuevo}, J. and {G{\'o}rski}, K.~M. and {Gratton}, S. and {Gruppuso}, A. and {Gudmundsson}, J.~E. and {Hamann}, J. and {Handley}, W. and {Hansen}, F.~K. and {Herranz}, D. and {Hildebrandt}, S.~R. and {Hivon}, E. and {Huang}, Z. and {Jaffe}, A.~H. and {Jones}, W.~C. and {Karakci}, A. and {Keih{\"a}nen}, E. and {Keskitalo}, R. and {Kiiveri}, K. and {Kim}, J. and {Kisner}, T.~S. and {Knox}, L. and {Krachmalnicoff}, N. and {Kunz}, M. and {Kurki-Suonio}, H. and {Lagache}, G. and {Lamarre}, J. -M. and {Lasenby}, A. and {Lattanzi}, M. and {Lawrence}, C.~R. and {Le Jeune}, M. and {Lemos}, P. and {Lesgourgues}, J. and {Levrier}, F. and {Lewis}, A. and {Liguori}, M. and {Lilje}, P.~B. and {Lilley}, M. and {Lindholm}, V. and {L{\'o}pez-Caniego}, M. and {Lubin}, P.~M. and {Ma}, Y. -Z. and {Mac{\'\i}as-P{\'e}rez}, J.~F. and {Maggio}, G. and {Maino}, D. and {Mandolesi}, N. and {Mangilli}, A. and {Marcos-Caballero}, A. and {Maris}, M. and {Martin}, P.~G. and {Martinelli}, M. and {Mart{\'\i}nez-Gonz{\'a}lez}, E. and {Matarrese}, S. and {Mauri}, N. and {McEwen}, J.~D. and {Meinhold}, P.~R. and {Melchiorri}, A. and {Mennella}, A. and {Migliaccio}, M. and {Millea}, M. and {Mitra}, S. and {Miville-Desch{\^e}nes}, M. -A. and {Molinari}, D. and {Montier}, L. and {Morgante}, G. and {Moss}, A. and {Natoli}, P. and {N{\o}rgaard-Nielsen}, H.~U. and {Pagano}, L. and {Paoletti}, D. and {Partridge}, B. and {Patanchon}, G. and {Peiris}, H.~V. and {Perrotta}, F. and {Pettorino}, V. and {Piacentini}, F. and {Polastri}, L. and {Polenta}, G. and {Puget}, J. -L. and {Rachen}, J.~P. and {Reinecke}, M. and {Remazeilles}, M. and {Renzi}, A. and {Rocha}, G. and {Rosset}, C. and {Roudier}, G. and {Rubi{\~n}o-Mart{\'\i}n}, J.~A. and {Ruiz-Granados}, B. and {Salvati}, L. and {Sandri}, M. and {Savelainen}, M. and {Scott}, D. and {Shellard}, E.~P.~S. and {Sirignano}, C. and {Sirri}, G. and {Spencer}, L.~D. and {Sunyaev}, R. and {Suur-Uski}, A. -S. and {Tauber}, J.~A. and {Tavagnacco}, D. and {Tenti}, M. and {Toffolatti}, L. and {Tomasi}, M. and {Trombetti}, T. and {Valenziano}, L. and {Valiviita}, J. and {Van Tent}, B. and {Vibert}, L. and {Vielva}, P. and {Villa}, F. and {Vittorio}, N. and {Wandelt}, B.~D. and {Wehus}, I.~K. and {White}, M. and {White}, S.~D.~M. and {Zacchei}, A. and {Zonca}, A.},
        title = "{Planck 2018 results. VI. Cosmological parameters}",
      journal = {\aap},
         year = 2020,
        month = sep,
       volume = {641},
          eid = {A6},
        pages = {A6},
          doi = {10.1051/0004-6361/201833910},
archivePrefix = {arXiv},
       eprint = {1807.06209},
 primaryClass = {astro-ph.CO},
       adsurl = {https://ui.adsabs.harvard.edu/abs/2020A&A...641A...6P}
}

@ARTICLE{Miyatake19,
       author = {{Miyatake}, Hironao and {Battaglia}, Nicholas and {Hilton}, Matt and {Medezinski}, Elinor and {Nishizawa}, Atsushi J. and {More}, Surhud and {Aiola}, Simone and {Bahcall}, Neta and {Bond}, J. Richard and {Calabrese}, Erminia and {Choi}, Steve K. and {Devlin}, Mark J. and {Dunkley}, Joanna and {Dunner}, Rolando and {Fuzia}, Brittany and {Gallardo}, Patricio and {Gralla}, Megan and {Hasselfield}, Matthew and {Halpern}, Mark and {Hikage}, Chiaki and {Hill}, J. Colin and {Hincks}, Adam D. and {Hlo{\v{z}}ek}, Ren{\'e}e and {Huffenberger}, Kevin and {Hughes}, John P. and {Koopman}, Brian and {Kosowsky}, Arthur and {Louis}, Thibaut and {Madhavacheril}, Mathew S. and {McMahon}, Jeff and {Mandelbaum}, Rachel and {Marriage}, Tobias A. and {Maurin}, Lo{\"\i}c and {Miyazaki}, Satoshi and {Moodley}, Kavilan and {Murata}, Ryoma and {Naess}, Sigurd and {Newburgh}, Laura and {Niemack}, Michael D. and {Nishimichi}, Takahiro and {Okabe}, Nobuhiro and {Oguri}, Masamune and {Osato}, Ken and {Page}, Lyman and {Partridge}, Bruce and {Robertson}, Naomi and {Sehgal}, Neelima and {Sherwin}, Blake and {Shirasaki}, Masato and {Sievers}, Jonathan and {Sif{\'o}n}, Crist{\'o}bal and {Simon}, Sara and {Spergel}, David N. and {Staggs}, Suzanne T. and {Stein}, George and {Takada}, Masahiro and {Trac}, Hy and {Umetsu}, Keiichi and {van Engelen}, Alex and {Wollack}, Edward J.},
        title = "{Weak-lensing Mass Calibration of ACTPol Sunyaev-Zel{\textquoteright}dovich Clusters with the Hyper Suprime-Cam Survey}",
      journal = {\apj},
         year = 2019,
        month = apr,
       volume = {875},
       number = {1},
          eid = {63},
        pages = {63},
          doi = {10.3847/1538-4357/ab0af0},
archivePrefix = {arXiv},
       eprint = {1804.05873},
 primaryClass = {astro-ph.CO},
       adsurl = {https://ui.adsabs.harvard.edu/abs/2019ApJ...875...63M}
}

@ARTICLE{Abbott22,
       author = {{Abbott}, T.~M.~C. and {Aguena}, M. and {Alarcon}, A. and {Allam}, S. and {Alves}, O. and {Amon}, A. and {Andrade-Oliveira}, F. and {Annis}, J. and {Avila}, S. and {Bacon}, D. and {Baxter}, E. and {Bechtol}, K. and {Becker}, M.~R. and {Bernstein}, G.~M. and {Bhargava}, S. and {Birrer}, S. and {Blazek}, J. and {Brandao-Souza}, A. and {Bridle}, S.~L. and {Brooks}, D. and {Buckley-Geer}, E. and {Burke}, D.~L. and {Camacho}, H. and {Campos}, A. and {Carnero Rosell}, A. and {Carrasco Kind}, M. and {Carretero}, J. and {Castander}, F.~J. and {Cawthon}, R. and {Chang}, C. and {Chen}, A. and {Chen}, R. and {Choi}, A. and {Conselice}, C. and {Cordero}, J. and {Costanzi}, M. and {Crocce}, M. and {da Costa}, L.~N. and {da Silva Pereira}, M.~E. and {Davis}, C. and {Davis}, T.~M. and {De Vicente}, J. and {DeRose}, J. and {Desai}, S. and {Di Valentino}, E. and {Diehl}, H.~T. and {Dietrich}, J.~P. and {Dodelson}, S. and {Doel}, P. and {Doux}, C. and {Drlica-Wagner}, A. and {Eckert}, K. and {Eifler}, T.~F. and {Elsner}, F. and {Elvin-Poole}, J. and {Everett}, S. and {Evrard}, A.~E. and {Fang}, X. and {Farahi}, A. and {Fernandez}, E. and {Ferrero}, I. and {Fert{\'e}}, A. and {Fosalba}, P. and {Friedrich}, O. and {Frieman}, J. and {Garc{\'\i}a-Bellido}, J. and {Gatti}, M. and {Gaztanaga}, E. and {Gerdes}, D.~W. and {Giannantonio}, T. and {Giannini}, G. and {Gruen}, D. and {Gruendl}, R.~A. and {Gschwend}, J. and {Gutierrez}, G. and {Harrison}, I. and {Hartley}, W.~G. and {Herner}, K. and {Hinton}, S.~R. and {Hollowood}, D.~L. and {Honscheid}, K. and {Hoyle}, B. and {Huff}, E.~M. and {Huterer}, D. and {Jain}, B. and {James}, D.~J. and {Jarvis}, M. and {Jeffrey}, N. and {Jeltema}, T. and {Kovacs}, A. and {Krause}, E. and {Kron}, R. and {Kuehn}, K. and {Kuropatkin}, N. and {Lahav}, O. and {Leget}, P.-F. and {Lemos}, P. and {Liddle}, A.~R. and {Lidman}, C. and {Lima}, M. and {Lin}, H. and {MacCrann}, N. and {Maia}, M.~A.~G. and {Marshall}, J.~L. and {Martini}, P. and {McCullough}, J. and {Melchior}, P. and {Mena-Fern{\'a}ndez}, J. and {Menanteau}, F. and {Miquel}, R. and {Mohr}, J.~J. and {Morgan}, R. and {Muir}, J. and {Myles}, J. and {Nadathur}, S. and {Navarro-Alsina}, A. and {Nichol}, R.~C. and {Ogando}, R.~L.~C. and {Omori}, Y. and {Palmese}, A. and {Pandey}, S. and {Park}, Y. and {Paz-Chinch{\'o}n}, F. and {Petravick}, D. and {Pieres}, A. and {Plazas Malag{\'o}n}, A.~A. and {Porredon}, A. and {Prat}, J. and {Raveri}, M. and {Rodriguez-Monroy}, M. and {Rollins}, R.~P. and {Romer}, A.~K. and {Roodman}, A. and {Rosenfeld}, R. and {Ross}, A.~J. and {Rykoff}, E.~S. and {Samuroff}, S. and {S{\'a}nchez}, C. and {Sanchez}, E. and {Sanchez}, J. and {Sanchez Cid}, D. and {Scarpine}, V. and {Schubnell}, M. and {Scolnic}, D. and {Secco}, L.~F. and {Serrano}, S. and {Sevilla-Noarbe}, I. and {Sheldon}, E. and {Shin}, T. and {Smith}, M. and {Soares-Santos}, M. and {Suchyta}, E. and {Swanson}, M.~E.~C. and {Tabbutt}, M. and {Tarle}, G. and {Thomas}, D. and {To}, C. and {Troja}, A. and {Troxel}, M.~A. and {Tucker}, D.~L. and {Tutusaus}, I. and {Varga}, T.~N. and {Walker}, A.~R. and {Weaverdyck}, N. and {Wechsler}, R. and {Weller}, J. and {Yanny}, B. and {Yin}, B. and {Zhang}, Y. and {Zuntz}, J. and {DES Collaboration}},
        title = "{Dark Energy Survey Year 3 results: Cosmological constraints from galaxy clustering and weak lensing}",
      journal = {\prd},
         year = 2022,
        month = jan,
       volume = {105},
       number = {2},
          eid = {023520},
        pages = {023520},
          doi = {10.1103/PhysRevD.105.023520},
archivePrefix = {arXiv},
       eprint = {2105.13549},
 primaryClass = {astro-ph.CO},
       adsurl = {https://ui.adsabs.harvard.edu/abs/2022PhRvD.105b3520A}
}

@ARTICLE{Pandey22,
       author = {{Pandey}, S. and {Gatti}, M. and {Baxter}, E. and {Hill}, J.~C. and {Fang}, X. and {Doux}, C. and {Giannini}, G. and {Raveri}, M. and {DeRose}, J. and {Huang}, H. and {Moser}, E. and {Battaglia}, N. and {Alarcon}, A. and {Amon}, A. and {Becker}, M. and {Campos}, A. and {Chang}, C. and {Chen}, R. and {Choi}, A. and {Eckert}, K. and {Elvin-Poole}, J. and {Everett}, S. and {Ferte}, A. and {Harrison}, I. and {Maccrann}, N. and {Mccullough}, J. and {Myles}, J. and {Navarro Alsina}, A. and {Prat}, J. and {Rollins}, R.~P. and {Sanchez}, C. and {Shin}, T. and {Troxel}, M. and {Tutusaus}, I. and {Yin}, B. and {Aguena}, M. and {Allam}, S. and {Andrade-Oliveira}, F. and {Bernstein}, G.~M. and {Bertin}, E. and {Bolliet}, B. and {Bond}, J.~R. and {Brooks}, D. and {Calabrese}, E. and {Carnero Rosell}, A. and {Carrasco Kind}, M. and {Carretero}, J. and {Cawthon}, R. and {Costanzi}, M. and {Crocce}, M. and {da Costa}, L.~N. and {Pereira}, M.~E.~S. and {De Vicente}, J. and {Desai}, S. and {Diehl}, H.~T. and {Dietrich}, J.~P. and {Doel}, P. and {Dunkley}, J. and {Everett}, S. and {Evrard}, A.~E. and {Ferraro}, S. and {Ferrero}, I. and {Flaugher}, B. and {Fosalba}, P. and {Garc{\'\i}a-Bellido}, J. and {Gaztanaga}, E. and {Gerdes}, D.~W. and {Giannantonio}, T. and {Gruen}, D. and {Gruendl}, R.~A. and {Gschwend}, J. and {Gutierrez}, G. and {Herner}, K. and {Hincks}, A.~D. and {Hinton}, S.~R. and {Hollowood}, D.~L. and {Honscheid}, K. and {Hughes}, J.~P. and {Huterer}, D. and {Jain}, B. and {James}, D.~J. and {Jeltema}, T. and {Krause}, E. and {Kuehn}, K. and {Lahav}, O. and {Lima}, M. and {Lokken}, M. and {Madhavacheril}, M.~S. and {Maia}, M.~A.~G. and {Mcmahon}, J.~J. and {Melchior}, P. and {Menanteau}, F. and {Miquel}, R. and {Mohr}, J.~J. and {Moodley}, K. and {Morgan}, R. and {Nati}, F. and {Niemack}, M.~D. and {Page}, L. and {Palmese}, A. and {Paz-Chinch{\'o}n}, F. and {Pieres}, A. and {Plazas Malag{\'o}n}, A.~A. and {Rodriguez-Monroy}, M. and {Romer}, A.~K. and {Sanchez}, E. and {Scarpine}, V. and {Schaan}, E. and {Serrano}, S. and {Sevilla-Noarbe}, I. and {Sheldon}, E. and {Sherwin}, B.~D. and {Sif{\'o}n}, C. and {Smith}, M. and {Soares-Santos}, M. and {Spergel}, D. and {Suchyta}, E. and {Swanson}, M.~E.~C. and {Tarle}, G. and {Thomas}, D. and {To}, C. and {Varga}, T.~N. and {Weller}, J. and {Wollack}, E.~J. and {Xu}, Z. and {DES} and {ACT Collaboration}},
        title = "{Cross-correlation of Dark Energy Survey Year 3 lensing data with ACT and P l a n c k thermal Sunyaev-Zel'dovich effect observations. II. Modeling and constraints on halo pressure profiles}",
      journal = {\prd},
         year = 2022,
        month = jun,
       volume = {105},
       number = {12},
          eid = {123526},
        pages = {123526},
          doi = {10.1103/PhysRevD.105.123526},
archivePrefix = {arXiv},
       eprint = {2108.01601},
 primaryClass = {astro-ph.CO},
       adsurl = {https://ui.adsabs.harvard.edu/abs/2022PhRvD.105l3526P}
}

@ARTICLE{Komatsu02,
       author = {{Komatsu}, E. and {Seljak}, U.},
        title = "{The Sunyaev-Zel'dovich angular power spectrum as a probe of cosmological parameters}",
      journal = {\mnras},
         year = 2002,
        month = nov,
       volume = {336},
       number = {4},
        pages = {1256-1270},
          doi = {10.1046/j.1365-8711.2002.05889.x},
archivePrefix = {arXiv},
       eprint = {astro-ph/0205468},
 primaryClass = {astro-ph},
       adsurl = {https://ui.adsabs.harvard.edu/abs/2002MNRAS.336.1256K}
}

@INPROCEEDINGS{CMBS4_25,
       author = {{Gallardo}, Patricio A. and {Harrington}, Kathleen and {Puddu}, Roberto and 
                 {Benson}, Bradford and {Carlstrom}, John and {Emerson}, Nick and 
                 {McMahon}, Jeff and {Natoli}, Tyler and {Nagy}, Johanna M. and 
                 {Niemack}, Michael D. and {Ruhl}, John},
        title = "{Overview of the optical design of the CMB-S4 large aperture telescopes and camera optics}",
    booktitle = {Ground-based and Airborne Telescopes X},
         year = 2024,
       editor = {{Marshall}, Heather K. and {Spyromilio}, Jason and {Usuda}, Tomonori},
       series = {Society of Photo-Optical Instrumentation Engineers (SPIE) Conference Series},
       volume = {13094},
        month = aug,
          eid = {130942F},
        pages = {130942F},
          doi = {10.1117/12.3020608},
archivePrefix = {arXiv},
       eprint = {2406.13854},
 primaryClass = {astro-ph.IM},
       adsurl = {https://ui.adsabs.harvard.edu/abs/2024SPIE13094E..2FG}
}

@ARTICLE{Chisari19,
       author = {{Chisari}, Nora Elisa and {Alonso}, David and {Krause}, Elisabeth and {Leonard}, C. Danielle and {Bull}, Philip and {Neveu}, J{\'e}r{\'e}my and {Villarreal}, Antonia Sierra and {Singh}, Sukhdeep and {McClintock}, Thomas and {Ellison}, John and {Du}, Zilong and {Zuntz}, Joe and {Mead}, Alexander and {Joudaki}, Shahab and {Lorenz}, Christiane S. and {Tr{\"o}ster}, Tilman and {Sanchez}, Javier and {Lanusse}, Francois and {Ishak}, Mustapha and {Hlozek}, Ren{\'e}e and {Blazek}, Jonathan and {Campagne}, Jean-Eric and {Almoubayyed}, Husni and {Eifler}, Tim and {Kirby}, Matthew and {Kirkby}, David and {Plaszczynski}, St{\'e}phane and {Slosar}, An{\v{z}}e and {Vrastil}, Michal and {Wagoner}, Erika L. and {LSST Dark Energy Science Collaboration}},
        title = "{Core Cosmology Library: Precision Cosmological Predictions for LSST}",
      journal = {\apjs},
         year = 2019,
        month = may,
       volume = {242},
       number = {1},
          eid = {2},
        pages = {2},
          doi = {10.3847/1538-4365/ab1658},
archivePrefix = {arXiv},
       eprint = {1812.05995},
 primaryClass = {astro-ph.CO},
       adsurl = {https://ui.adsabs.harvard.edu/abs/2019ApJS..242....2C}
}

@ARTICLE{Ma06,
       author = {{Ma}, Zhaoming and {Hu}, Wayne and {Huterer}, Dragan},
        title = "{Effects of Photometric Redshift Uncertainties on Weak-Lensing Tomography}",
      journal = {\apj},
         year = 2006,
        month = jan,
       volume = {636},
       number = {1},
        pages = {21-29},
          doi = {10.1086/497068},
archivePrefix = {arXiv},
       eprint = {astro-ph/0506614},
 primaryClass = {astro-ph},
       adsurl = {https://ui.adsabs.harvard.edu/abs/2006ApJ...636...21M}
}

@ARTICLE{Xiong25,
       author = {{Xiong}, Qi and {Gong}, Yan and {Zhou}, Xingchen and {Lin}, Hengjie and {Deng}, Furen and {Li}, Ziwei and {Ibitoye}, Ayodeji and {Chen}, Xuelei and {Fan}, Zuhui and {Guo}, Qi and {Li}, Ming and {Liu}, Yun and {Pei}, Wenxiang},
        title = "{Exploring Cosmological Constraints of the Weak Gravitational Lensing and Galaxy Clustering Joint Analysis in the CSST Photometric Survey}",
      journal = {\apj},
         year = 2025,
        month = may,
       volume = {985},
       number = {1},
          eid = {131},
        pages = {131},
          doi = {10.3847/1538-4357/adcb44},
archivePrefix = {arXiv},
       eprint = {2410.19388},
 primaryClass = {astro-ph.CO},
       adsurl = {https://ui.adsabs.harvard.edu/abs/2025ApJ...985..131X}
}

@ARTICLE{CMBS4_24,
       author = {{Gallardo}, Patricio A. and {Puddu}, Roberto and {Harrington}, Kathleen and {Benson}, Bradford and {Carlstrom}, John E. and {Dicker}, Simon R. and {Emerson}, Nick and {Gudmundsson}, Jon E. and {Limon}, Michele and {McMahon}, Jeff and {Nagy}, Johanna M. and {Natoli}, Tyler and {Niemack}, Michael D. and {Padin}, Stephen and {Ruhl}, John and {Simon}, Sara M. and {CMB-S4 Collaboration}},
        title = "{Freeform three-mirror anastigmatic large-aperture telescope and receiver optics for CMB-S4}",
      journal = {\ao},
         year = 2024,
        month = jan,
       volume = {63},
       number = {2},
        pages = {310},
          doi = {10.1364/AO.501744},
archivePrefix = {arXiv},
       eprint = {2307.12931},
 primaryClass = {astro-ph.IM},
       adsurl = {https://ui.adsabs.harvard.edu/abs/2024ApOpt..63..310G}
}

@ARTICLE{Asgari21,
       author = {{Asgari}, Marika and {Lin}, Chieh-An and {Joachimi}, Benjamin and {Giblin}, Benjamin and {Heymans}, Catherine and {Hildebrandt}, Hendrik and {Kannawadi}, Arun and {St{\"o}lzner}, Benjamin and {Tr{\"o}ster}, Tilman and {van den Busch}, Jan Luca and {Wright}, Angus H. and {Bilicki}, Maciej and {Blake}, Chris and {de Jong}, Jelte and {Dvornik}, Andrej and {Erben}, Thomas and {Getman}, Fedor and {Hoekstra}, Henk and {K{\"o}hlinger}, Fabian and {Kuijken}, Konrad and {Miller}, Lance and {Radovich}, Mario and {Schneider}, Peter and {Shan}, HuanYuan and {Valentijn}, Edwin},
        title = "{KiDS-1000 cosmology: Cosmic shear constraints and comparison between two point statistics}",
      journal = {\aap},
         year = 2021,
        month = jan,
       volume = {645},
          eid = {A104},
        pages = {A104},
          doi = {10.1051/0004-6361/202039070},
archivePrefix = {arXiv},
       eprint = {2007.15633},
 primaryClass = {astro-ph.CO},
       adsurl = {https://ui.adsabs.harvard.edu/abs/2021A&A...645A.104A}
}

@ARTICLE{Fert24,
       author = {{Fert{\'e}}, Agn{\`e}s and {Hong}, Kevin},
        title = "{Results and forecasts on cosmic inflation from weak lensing}",
      journal = {\prd},
         year = 2024,
        month = may,
       volume = {109},
       number = {10},
          eid = {103502},
        pages = {103502},
          doi = {10.1103/PhysRevD.109.103502},
archivePrefix = {arXiv},
       eprint = {2310.10731},
 primaryClass = {astro-ph.CO},
       adsurl = {https://ui.adsabs.harvard.edu/abs/2024PhRvD.109j3502F}
}

@ARTICLE{CSST_NeutrinoCosmo,
       author = {{Lin}, Hengjie and {Gong}, Yan and {Chen}, Xuelei and {Chan}, Kwan Chuen and {Fan}, Zuhui and {Zhan}, Hu},
        title = "{Forecast of neutrino cosmology from the CSST photometric galaxy clustering and cosmic shear surveys}",
      journal = {\mnras},
         year = 2022,
        month = oct,
       volume = {515},
       number = {4},
        pages = {5743-5757},
          doi = {10.1093/mnras/stac2126},
archivePrefix = {arXiv},
       eprint = {2203.11429},
 primaryClass = {astro-ph.CO},
       adsurl = {https://ui.adsabs.harvard.edu/abs/2022MNRAS.515.5743L}
}

@ARTICLE{Chang13,
       author = {{Chang}, C. and {Jarvis}, M. and {Jain}, B. and {Kahn}, S.~M. and {Kirkby}, D. and {Connolly}, A. and {Krughoff}, S. and {Peng}, E.-H. and {Peterson}, J.~R.},
        title = "{The effective number density of galaxies for weak lensing measurements in the LSST project}",
      journal = {\mnras},
         year = 2013,
        month = sep,
       volume = {434},
       number = {3},
        pages = {2121-2135},
          doi = {10.1093/mnras/stt1156},
archivePrefix = {arXiv},
       eprint = {1305.0793},
 primaryClass = {astro-ph.CO},
       adsurl = {https://ui.adsabs.harvard.edu/abs/2013MNRAS.434.2121C}
}

@ARTICLE{LSSTScience09,
       author = {{LSST Science Collaboration} and {Abell}, Paul A. and {Allison}, Julius and {Anderson}, Scott F. and {Andrew}, John R. and {Angel}, J. Roger P. and {Armus}, Lee and {Arnett}, David and {Asztalos}, S.~J. and {Axelrod}, Tim S. and {Bailey}, Stephen and {Ballantyne}, D.~R. and {Bankert}, Justin R. and {Barkhouse}, Wayne A. and {Barr}, Jeffrey D. and {Barrientos}, L. Felipe and {Barth}, Aaron J. and {Bartlett}, James G. and {Becker}, Andrew C. and {Becla}, Jacek and {Beers}, Timothy C. and {Bernstein}, Joseph P. and {Biswas}, Rahul and {Blanton}, Michael R. and {Bloom}, Joshua S. and {Bochanski}, John J. and {Boeshaar}, Pat and {Borne}, Kirk D. and {Bradac}, Marusa and {Brandt}, W.~N. and {Bridge}, Carrie R. and {Brown}, Michael E. and {Brunner}, Robert J. and {Bullock}, James S. and {Burgasser}, Adam J. and {Burge}, James H. and {Burke}, David L. and {Cargile}, Phillip A. and {Chandrasekharan}, Srinivasan and {Chartas}, George and {Chesley}, Steven R. and {Chu}, You-Hua and {Cinabro}, David and {Claire}, Mark W. and {Claver}, Charles F. and {Clowe}, Douglas and {Connolly}, A.~J. and {Cook}, Kem H. and {Cooke}, Jeff and {Cooray}, Asantha and {Covey}, Kevin R. and {Culliton}, Christopher S. and {de Jong}, Roelof and {de Vries}, Willem H. and {Debattista}, Victor P. and {Delgado}, Francisco and {Dell'Antonio}, Ian P. and {Dhital}, Saurav and {Di Stefano}, Rosanne and {Dickinson}, Mark and {Dilday}, Benjamin and {Djorgovski}, S.~G. and {Dobler}, Gregory and {Donalek}, Ciro and {Dubois-Felsmann}, Gregory and {Durech}, Josef and {Eliasdottir}, Ardis and {Eracleous}, Michael and {Eyer}, Laurent and {Falco}, Emilio E. and {Fan}, Xiaohui and {Fassnacht}, Christopher D. and {Ferguson}, Harry C. and {Fernandez}, Yanga R. and {Fields}, Brian D. and {Finkbeiner}, Douglas and {Figueroa}, Eduardo E. and {Fox}, Derek B. and {Francke}, Harold and {Frank}, James S. and {Frieman}, Josh and {Fromenteau}, Sebastien and {Furqan}, Muhammad and {Galaz}, Gaspar and {Gal-Yam}, A. and {Garnavich}, Peter and {Gawiser}, Eric and {Geary}, John and {Gee}, Perry and {Gibson}, Robert R. and {Gilmore}, Kirk and {Grace}, Emily A. and {Green}, Richard F. and {Gressler}, William J. and {Grillmair}, Carl J. and {Habib}, Salman and {Haggerty}, J.~S. and {Hamuy}, Mario and {Harris}, Alan W. and {Hawley}, Suzanne L. and {Heavens}, Alan F. and {Hebb}, Leslie and {Henry}, Todd J. and {Hileman}, Edward and {Hilton}, Eric J. and {Hoadley}, Keri and {Holberg}, J.~B. and {Holman}, Matt J. and {Howell}, Steve B. and {Infante}, Leopoldo and {Ivezic}, Zeljko and {Jacoby}, Suzanne H. and {Jain}, Bhuvnesh and {R} and {Jedicke} and {Jee}, M. James and {Garrett Jernigan}, J. and {Jha}, Saurabh W. and {Johnston}, Kathryn V. and {Jones}, R. Lynne and {Juric}, Mario and {Kaasalainen}, Mikko and {Styliani} and {Kafka} and {Kahn}, Steven M. and {Kaib}, Nathan A. and {Kalirai}, Jason and {Kantor}, Jeff and {Kasliwal}, Mansi M. and {Keeton}, Charles R. and {Kessler}, Richard and {Knezevic}, Zoran and {Kowalski}, Adam and {Krabbendam}, Victor L. and {Krughoff}, K. Simon and {Kulkarni}, Shrinivas and {Kuhlman}, Stephen and {Lacy}, Mark and {Lepine}, Sebastien and {Liang}, Ming and {Lien}, Amy and {Lira}, Paulina and {Long}, Knox S. and {Lorenz}, Suzanne and {Lotz}, Jennifer M. and {Lupton}, R.~H. and {Lutz}, Julie and {Macri}, Lucas M. and {Mahabal}, Ashish A. and {Mandelbaum}, Rachel and {Marshall}, Phil and {May}, Morgan and {McGehee}, Peregrine M. and {Meadows}, Brian T. and {Meert}, Alan and {Milani}, Andrea and {Miller}, Christopher J. and {Miller}, Michelle and {Mills}, David and {Minniti}, Dante and {Monet}, David and {Mukadam}, Anjum S. and {Nakar}, Ehud and {Neill}, Douglas R. and {Newman}, Jeffrey A. and {Nikolaev}, Sergei and {Nordby}, Martin and {O'Connor}, Paul and {Oguri}, Masamune and {Oliver}, John and {Olivier}, Scot S. and {Olsen}, Julia K. and {Olsen}, Knut and {Olszewski}, Edward W. and {Oluseyi}, Hakeem and {Padilla}, Nelson D. and {Parker}, Alex and {Pepper}, Joshua and {Peterson}, John R. and {Petry}, Catherine and {Pinto}, Philip A. and {Pizagno}, James L. and {Popescu}, Bogdan and {Prsa}, Andrej and {Radcka}, Veljko and {Raddick}, M. Jordan and {Rasmussen}, Andrew and {Rau}, Arne and {Rho}, Jeonghee and {Rhoads}, James E. and {Richards}, Gordon T. and {Ridgway}, Stephen T. and {Robertson}, Brant E. and {Roskar}, Rok and {Saha}, Abhijit and {Sarajedini}, Ata and {Scannapieco}, Evan and {Schalk}, Terry and {Schindler}, Rafe and {Schmidt}, Samuel},
        title = "{LSST Science Book, Version 2.0}",
      journal = {arXiv e-prints},
         year = 2009,
        month = dec,
          eid = {arXiv:0912.0201},
        pages = {arXiv:0912.0201},
          doi = {10.48550/arXiv.0912.0201},
archivePrefix = {arXiv},
       eprint = {0912.0201},
 primaryClass = {astro-ph.IM},
       adsurl = {https://ui.adsabs.harvard.edu/abs/2009arXiv0912.0201L}
}

@ARTICLE{DESCSRD,
       author = {{The LSST Dark Energy Science Collaboration} and {Mandelbaum}, Rachel and {Eifler}, Tim and {Hlo{\v{z}}ek}, Ren{\'e}e and {Collett}, Thomas and {Gawiser}, Eric and {Scolnic}, Daniel and {Alonso}, David and {Awan}, Humna and {Biswas}, Rahul and {Blazek}, Jonathan and {Burchat}, Patricia and {Chisari}, Nora Elisa and {Dell'Antonio}, Ian and {Digel}, Seth and {Frieman}, Josh and {Goldstein}, Daniel A. and {Hook}, Isobel and {Ivezi{\'c}}, {\v{Z}}eljko and {Kahn}, Steven M. and {Kamath}, Sowmya and {Kirkby}, David and {Kitching}, Thomas and {Krause}, Elisabeth and {Leget}, Pierre-Fran{\c{c}}ois and {Marshall}, Philip J. and {Meyers}, Joshua and {Miyatake}, Hironao and {Newman}, Jeffrey A. and {Nichol}, Robert and {Rykoff}, Eli and {Sanchez}, F. Javier and {Slosar}, An{\v{z}}e and {Sullivan}, Mark and {Troxel}, M.~A.},
        title = "{The LSST Dark Energy Science Collaboration (DESC) Science Requirements Document}",
      journal = {arXiv e-prints},
         year = 2018,
        month = sep,
          eid = {arXiv:1809.01669},
        pages = {arXiv:1809.01669},
          doi = {10.48550/arXiv.1809.01669},
archivePrefix = {arXiv},
       eprint = {1809.01669},
 primaryClass = {astro-ph.CO},
       adsurl = {https://ui.adsabs.harvard.edu/abs/2018arXiv180901669T}
}

@ARTICLE{CSST,
       author = {{Gong}, Yan and {Liu}, Xiangkun and {Cao}, Ye and {Chen}, Xuelei and {Fan}, Zuhui and {Li}, Ran and {Li}, Xiao-Dong and {Li}, Zhigang and {Zhang}, Xin and {Zhan}, Hu},
        title = "{Cosmology from the Chinese Space Station Optical Survey (CSS-OS)}",
      journal = {\apj},
         year = 2019,
        month = oct,
       volume = {883},
       number = {2},
          eid = {203},
        pages = {203},
          doi = {10.3847/1538-4357/ab391e},
archivePrefix = {arXiv},
       eprint = {1901.04634},
 primaryClass = {astro-ph.CO},
       adsurl = {https://ui.adsabs.harvard.edu/abs/2019ApJ...883..203G}
}

@ARTICLE{Euclid_sys_25,
       author = {{Fumagalli}, A. and {Costanzi}, M. and {Castro}, T. and {Saro}, A. and {Borgani}, S. and {Romanello}, M. and {Marulli}, F. and {Tsaprazi}, E. and {Monaco}, P. and {Altieri}, B. and {Amara}, A. and {Amendola}, L. and {Andreon}, S. and {Auricchio}, N. and {Baccigalupi}, C. and {Baldi}, M. and {Balestra}, A. and {Bardelli}, S. and {Biviano}, A. and {Branchini}, E. and {Brescia}, M. and {Camera}, S. and {Ca{\~n}as-Herrera}, G. and {Capobianco}, V. and {Carbone}, C. and {Carretero}, J. and {Casas}, S. and {Castellano}, M. and {Castignani}, G. and {Cavuoti}, S. and {Chambers}, K.~C. and {Cimatti}, A. and {Colodro-Conde}, C. and {Congedo}, G. and {Conversi}, L. and {Copin}, Y. and {Courbin}, F. and {Courtois}, H.~M. and {Da Silva}, A. and {Degaudenzi}, H. and {de la Torre}, S. and {De Lucia}, G. and {Di Giorgio}, A.~M. and {Dole}, H. and {Douspis}, M. and {Dubath}, F. and {Duncan}, C.~A.~J. and {Dupac}, X. and {Dusini}, S. and {Escoffier}, S. and {Farina}, M. and {Farinelli}, R. and {Faustini}, F. and {Ferriol}, S. and {Finelli}, F. and {Fosalba}, P. and {Fourmanoit}, N. and {Frailis}, M. and {Franceschi}, E. and {Fumana}, M. and {Galeotta}, S. and {George}, K. and {Gillis}, B. and {Giocoli}, C. and {Gracia-Carpio}, J. and {Grazian}, A. and {Grupp}, F. and {Guzzo}, L. and {Haugan}, S.~V.~H. and {Holmes}, W. and {Hormuth}, F. and {Hornstrup}, A. and {Jahnke}, K. and {Jhabvala}, M. and {Joachimi}, B. and {Keih{\"a}nen}, E. and {Kermiche}, S. and {Kiessling}, A. and {Kubik}, B. and {K{\"u}mmel}, M. and {Kunz}, M. and {Kurki-Suonio}, H. and {Le Brun}, A.~M.~C. and {Ligori}, S. and {Lilje}, P.~B. and {Lindholm}, V. and {Lloro}, I. and {Mainetti}, G. and {Maino}, D. and {Maiorano}, E. and {Mansutti}, O. and {Marggraf}, O. and {Martinelli}, M. and {Martinet}, N. and {Massey}, R.~J. and {Medinaceli}, E. and {Mei}, S. and {Mellier}, Y. and {Meneghetti}, M. and {Merlin}, E. and {Meylan}, G. and {Mohr}, J.~J. and {Mora}, A. and {Moresco}, M. and {Moscardini}, L. and {Munari}, E. and {Nakajima}, R. and {Neissner}, C. and {Niemi}, S.-M. and {Padilla}, C. and {Paltani}, S. and {Pasian}, F. and {Pedersen}, K. and {Pettorino}, V. and {Pires}, S. and {Polenta}, G. and {Poncet}, M. and {Popa}, L.~A. and {Pozzetti}, L. and {Raison}, F. and {Rebolo}, R. and {Renzi}, A. and {Rhodes}, J. and {Riccio}, G. and {Romelli}, E. and {Roncarelli}, M. and {Rosset}, C. and {Saglia}, R. and {Sakr}, Z. and {S{\'a}nchez}, A.~G. and {Sapone}, D. and {Sartoris}, B. and {Schneider}, P. and {Schrabback}, T. and {Secroun}, A. and {Sefusatti}, E. and {Seidel}, G. and {Seiffert}, M. and {Serrano}, S. and {Simon}, P. and {Sirignano}, C. and {Sirri}, G. and {Spurio Mancini}, A. and {Stanco}, L. and {Steinwagner}, J. and {Tallada-Cresp{\'\i}}, P. and {Tavagnacco}, D. and {Taylor}, A.~N. and {Tereno}, I. and {Tessore}, N. and {Toft}, S. and {Toledo-Moreo}, R. and {Torradeflot}, F. and {Tutusaus}, I. and {Valenziano}, L. and {Valiviita}, J. and {Vassallo}, T. and {Verdoes Kleijn}, G. and {Veropalumbo}, A. and {Wang}, Y. and {Weller}, J. and {Zamorani}, G. and {Zerbi}, F.~M. and {Zucca}, E. and {Burigana}, C. and {Gabarra}, L. and {Maturi}, M. and {Porciani}, C. and {Scottez}, V. and {Sereno}, M. and {Viel}, M.},
        title = "{Euclid: Exploring observational systematics in cluster cosmology -- a comprehensive analysis of cluster counts and clustering}",
      journal = {arXiv e-prints},
         year = 2025,
        month = oct,
          eid = {arXiv:2510.13509},
        pages = {arXiv:2510.13509},
          doi = {10.48550/arXiv.2510.13509},
archivePrefix = {arXiv},
       eprint = {2510.13509},
 primaryClass = {astro-ph.CO},
       adsurl = {https://ui.adsabs.harvard.edu/abs/2025arXiv251013509F}
}

@ARTICLE{Shirasaki20,
       author = {{Shirasaki}, Masato and {Lau}, Erwin T. and {Nagai}, Daisuke},
        title = "{Probing cosmology and cluster astrophysics with multiwavelength surveys - I. Correlation statistics}",
      journal = {\mnras},
         year = 2020,
        month = jan,
       volume = {491},
       number = {1},
        pages = {235-253},
          doi = {10.1093/mnras/stz3021},
archivePrefix = {arXiv},
       eprint = {1909.02179},
 primaryClass = {astro-ph.CO},
       adsurl = {https://ui.adsabs.harvard.edu/abs/2020MNRAS.491..235S}
}

@ARTICLE{Shi16,
       author = {{Shi}, Xun and {Komatsu}, Eiichiro and {Nagai}, Daisuke and {Lau}, Erwin T.},
        title = "{Analytical model for non-thermal pressure in galaxy clusters - III. Removing the hydrostatic mass bias}",
      journal = {\mnras},
         year = 2016,
        month = jan,
       volume = {455},
       number = {3},
        pages = {2936-2944},
          doi = {10.1093/mnras/stv2504},
archivePrefix = {arXiv},
       eprint = {1507.04338},
 primaryClass = {astro-ph.CO},
       adsurl = {https://ui.adsabs.harvard.edu/abs/2016MNRAS.455.2936S}
}

@ARTICLE{Shi15,
       author = {{Shi}, Xun and {Komatsu}, Eiichiro and {Nelson}, Kaylea and {Nagai}, Daisuke},
        title = "{Analytical model for non-thermal pressure in galaxy clusters - II. Comparison with cosmological hydrodynamics simulation}",
      journal = {\mnras},
         year = 2015,
        month = mar,
       volume = {448},
       number = {1},
        pages = {1020-1029},
          doi = {10.1093/mnras/stv036},
archivePrefix = {arXiv},
       eprint = {1408.3832},
 primaryClass = {astro-ph.CO},
       adsurl = {https://ui.adsabs.harvard.edu/abs/2015MNRAS.448.1020S}
}

@ARTICLE{Zheng05,
       author = {{Zheng}, Zheng and {Berlind}, Andreas A. and {Weinberg}, David H. and {Benson}, Andrew J. and {Baugh}, Carlton M. and {Cole}, Shaun and {Dav{\'e}}, Romeel and {Frenk}, Carlos S. and {Katz}, Neal and {Lacey}, Cedric G.},
        title = "{Theoretical Models of the Halo Occupation Distribution: Separating Central and Satellite Galaxies}",
      journal = {\apj},
         year = 2005,
        month = nov,
       volume = {633},
       number = {2},
        pages = {791-809},
          doi = {10.1086/466510},
archivePrefix = {arXiv},
       eprint = {astro-ph/0408564},
 primaryClass = {astro-ph},
       adsurl = {https://ui.adsabs.harvard.edu/abs/2005ApJ...633..791Z}
}

@ARTICLE{Nunes21,
       author = {{Nunes}, Rafael C. and {Vagnozzi}, Sunny},
        title = "{Arbitrating the S$_{8}$ discrepancy with growth rate measurements from redshift-space distortions}",
      journal = {\mnras},
         year = 2021,
        month = aug,
       volume = {505},
       number = {4},
        pages = {5427-5437},
          doi = {10.1093/mnras/stab1613},
archivePrefix = {arXiv},
       eprint = {2106.01208},
 primaryClass = {astro-ph.CO},
       adsurl = {https://ui.adsabs.harvard.edu/abs/2021MNRAS.505.5427N}
}

@ARTICLE{Cooray02,
       author = {{Cooray}, Asantha and {Sheth}, Ravi},
        title = "{Halo models of large scale structure}",
      journal = {\physrep},
         year = 2002,
        month = dec,
       volume = {372},
       number = {1},
        pages = {1-129},
          doi = {10.1016/S0370-1573(02)00276-4},
archivePrefix = {arXiv},
       eprint = {astro-ph/0206508},
 primaryClass = {astro-ph},
       adsurl = {https://ui.adsabs.harvard.edu/abs/2002PhR...372....1C}
}

@ARTICLE{Wright25,
       author = {{Wright}, Angus H. and {St{\"o}lzner}, Benjamin and {Asgari}, Marika and {Bilicki}, Maciej and {Giblin}, Benjamin and {Heymans}, Catherine and {Hildebrandt}, Hendrik and {Hoekstra}, Henk and {Joachimi}, Benjamin and {Kuijken}, Konrad and {Li}, Shun-Sheng and {Reischke}, Robert and {von Wietersheim-Kramsta}, Maximilian and {Yoon}, Mijin and {Burger}, Pierre and {Chisari}, Nora Elisa and {de Jong}, Jelte and {Dvornik}, Andrej and {Georgiou}, Christos and {Harnois-D{\'e}raps}, Joachim and {Jalan}, Priyanka and {William}, Anjitha John and {Joudaki}, Shahab and {Lesci}, Giorgio Francesco and {Linke}, Laila and {Loureiro}, Arthur and {Mahony}, Constance and {Maturi}, Matteo and {Miller}, Lance and {Moscardini}, Lauro and {Napolitano}, Nicola R. and {Porth}, Lucas and {Radovich}, Mario and {Schneider}, Peter and {Tr{\"o}ster}, Tilman and {Valentijn}, Edwin and {Wittje}, Anna and {Yan}, Ziang and {Zhang}, Yun-Hao},
        title = "{KiDS-Legacy: Cosmological constraints from cosmic shear with the complete Kilo-Degree Survey}",
      journal = {\aap},
         year = 2025,
        month = nov,
       volume = {703},
          eid = {A158},
        pages = {A158},
          doi = {10.1051/0004-6361/202554908},
archivePrefix = {arXiv},
       eprint = {2503.19441},
 primaryClass = {astro-ph.CO},
       adsurl = {https://ui.adsabs.harvard.edu/abs/2025A&A...703A.158W}
}

@ARTICLE{Choppin25,
       author = {{Choppin de Janvry}, J. and {Dai}, B. and {Gontcho}, S. Gontcho A and {Seljak}, U. and {Zhang}, T.},
        title = "{Cosmic Shear constraints from HSC Year 3 with clustering calibration of the tomographic redshift distributions from DESI}",
      journal = {arXiv e-prints},
         year = 2025,
        month = nov,
          eid = {arXiv:2511.18134},
        pages = {arXiv:2511.18134},
          doi = {10.48550/arXiv.2511.18134},
archivePrefix = {arXiv},
       eprint = {2511.18134},
 primaryClass = {astro-ph.CO},
       adsurl = {https://ui.adsabs.harvard.edu/abs/2025arXiv251118134C}
}

@ARTICLE{Zhang25,
       author = {{Zhang}, Tianqing and {Li}, Xiangchong and {Sugiyama}, Sunao and {Mandelbaum}, Rachel and {More}, Surhud and {Dalal}, Roohi and {Kannawadi}, Arun and {Miyatake}, Hironao and {Nishizawa}, Atsushi J. and {Nishimichi}, Takahiro and {Oguri}, Masamune and {Osato}, Ken and {Rau}, Markus M. and {Shirasaki}, Masato and {Sunayama}, Tomomi and {Takada}, Masahiro},
        title = "{Cosmology and Source Redshift Constraints from Galaxy Clustering and Tomographic Weak Lensing with HSC Y3 and SDSS using the Point-Mass Correction Model}",
      journal = {arXiv e-prints},
         year = 2025,
        month = jul,
          eid = {arXiv:2507.01386},
        pages = {arXiv:2507.01386},
          doi = {10.48550/arXiv.2507.01386},
archivePrefix = {arXiv},
       eprint = {2507.01386},
 primaryClass = {astro-ph.CO},
       adsurl = {https://ui.adsabs.harvard.edu/abs/2025arXiv250701386Z}
}

@ARTICLE{DESY1,
       author = {{To}, C. and {Krause}, E. and {Rozo}, E. and {Wu}, H. and {Gruen}, D. and {Wechsler}, R.~H. and {Eifler}, T.~F. and {Rykoff}, E.~S. and {Costanzi}, M. and {Becker}, M.~R. and {Bernstein}, G.~M. and {Blazek}, J. and {Bocquet}, S. and {Bridle}, S.~L. and {Cawthon}, R. and {Choi}, A. and {Crocce}, M. and {Davis}, C. and {DeRose}, J. and {Drlica-Wagner}, A. and {Elvin-Poole}, J. and {Fang}, X. and {Farahi}, A. and {Friedrich}, O. and {Gatti}, M. and {Gaztanaga}, E. and {Giannantonio}, T. and {Hartley}, W.~G. and {Hoyle}, B. and {Jarvis}, M. and {MacCrann}, N. and {McClintock}, T. and {Miranda}, V. and {Pereira}, M.~E.~S. and {Park}, Y. and {Porredon}, A. and {Prat}, J. and {Rau}, M.~M. and {Ross}, A.~J. and {Samuroff}, S. and {S{\'a}nchez}, C. and {Sevilla-Noarbe}, I. and {Sheldon}, E. and {Troxel}, M.~A. and {Varga}, T.~N. and {Vielzeuf}, P. and {Zhang}, Y. and {Zuntz}, J. and {Abbott}, T.~M.~C. and {Aguena}, M. and {Amon}, A. and {Annis}, J. and {Avila}, S. and {Bertin}, E. and {Bhargava}, S. and {Brooks}, D. and {Burke}, D.~L. and {Carnero Rosell}, A. and {Carrasco Kind}, M. and {Carretero}, J. and {Chang}, C. and {Conselice}, C. and {da Costa}, L.~N. and {Davis}, T.~M. and {Desai}, S. and {Diehl}, H.~T. and {Dietrich}, J.~P. and {Everett}, S. and {Evrard}, A.~E. and {Ferrero}, I. and {Flaugher}, B. and {Fosalba}, P. and {Frieman}, J. and {Garc{\'\i}a-Bellido}, J. and {Gruendl}, R.~A. and {Gutierrez}, G. and {Hinton}, S.~R. and {Hollowood}, D.~L. and {Honscheid}, K. and {Huterer}, D. and {James}, D.~J. and {Jeltema}, T. and {Kron}, R. and {Kuehn}, K. and {Kuropatkin}, N. and {Lima}, M. and {Maia}, M.~A.~G. and {Marshall}, J.~L. and {Menanteau}, F. and {Miquel}, R. and {Morgan}, R. and {Muir}, J. and {Myles}, J. and {Palmese}, A. and {Paz-Chinch{\'o}n}, F. and {Plazas}, A.~A. and {Romer}, A.~K. and {Roodman}, A. and {Sanchez}, E. and {Santiago}, B. and {Scarpine}, V. and {Serrano}, S. and {Smith}, M. and {Suchyta}, E. and {Swanson}, M.~E.~C. and {Tarle}, G. and {Thomas}, D. and {Tucker}, D.~L. and {Weller}, J. and {Wester}, W. and {Wilkinson}, R.~D. and {DES Collaboration}},
        title = "{Dark Energy Survey Year 1 Results: Cosmological Constraints from Cluster Abundances, Weak Lensing, and Galaxy Correlations}",
      journal = {\prl},
         year = 2021,
        month = apr,
       volume = {126},
       number = {14},
          eid = {141301},
        pages = {141301},
          doi = {10.1103/PhysRevLett.126.141301},
archivePrefix = {arXiv},
       eprint = {2010.01138},
 primaryClass = {astro-ph.CO},
       adsurl = {https://ui.adsabs.harvard.edu/abs/2021PhRvL.126n1301T}
}

@ARTICLE{Shiming26,
       author = {{Gu}, Shiming and {van Waerbeke}, Ludovic and {Bernardeau}, Francis and {Fabbro}, S{\'e}bastien},
        title = "{Mitigating nonlinear systematics in weak lensing surveys. II. Stability and diagnostics with intrinsic alignment}",
      journal = {\prd},
         year = 2026,
        month = jan,
       volume = {113},
       number = {2},
          eid = {023528},
        pages = {023528},
          doi = {10.1103/y7c6-t42s},
archivePrefix = {arXiv},
       eprint = {2511.09544},
 primaryClass = {astro-ph.CO},
       adsurl = {https://ui.adsabs.harvard.edu/abs/2026PhRvD.113b3528G}
}

@ARTICLE{Shiming25,
       author = {{Gu}, Shiming and {van Waerbeke}, Ludovic and {Bernardeau}, Francis and {Dalal}, Roohi},
        title = "{Mitigating nonlinear systematics in weak lensing surveys: The Bernardeau-Nishimichi-Taruya approach}",
      journal = {\prd},
         year = 2025,
        month = apr,
       volume = {111},
       number = {8},
          eid = {083530},
        pages = {083530},
          doi = {10.1103/PhysRevD.111.083530},
archivePrefix = {arXiv},
       eprint = {2412.14704},
 primaryClass = {astro-ph.CO},
       adsurl = {https://ui.adsabs.harvard.edu/abs/2025PhRvD.111h3530G}
}

@ARTICLE{Nikolina25,
       author = {{{\v{S}}ar{\v{c}}evi{\'c}}, Nikolina and {Leonard}, C. Danielle and {Rau}, Markus M. and {the LSST Dark Energy Science Collaboration}},
        title = "{Joint modelling of astrophysical systematics for cosmology with LSST cosmic shear}",
      journal = {\mnras},
         year = 2025,
        month = feb,
       volume = {537},
       number = {2},
        pages = {1924-1948},
          doi = {10.1093/mnras/staf156},
archivePrefix = {arXiv},
       eprint = {2406.03352},
 primaryClass = {astro-ph.CO},
       adsurl = {https://ui.adsabs.harvard.edu/abs/2025MNRAS.537.1924S}
}

@ARTICLE{LeBrun15,
       author = {{Le Brun}, Amandine M.~C. and {McCarthy}, Ian G. and {Melin}, Jean-Baptiste},
        title = "{Testing Sunyaev-Zel'dovich measurements of the hot gas content of dark matter haloes using synthetic skies}",
      journal = {\mnras},
         year = 2015,
        month = aug,
       volume = {451},
       number = {4},
        pages = {3868-3881},
          doi = {10.1093/mnras/stv1172},
archivePrefix = {arXiv},
       eprint = {1501.05666},
 primaryClass = {astro-ph.CO},
       adsurl = {https://ui.adsabs.harvard.edu/abs/2015MNRAS.451.3868L}
}

@ARTICLE{McCarthy17,
       author = {{McCarthy}, Ian G. and {Schaye}, Joop and {Bird}, Simeon and {Le Brun}, Amandine M.~C.},
        title = "{The BAHAMAS project: calibrated hydrodynamical simulations for large-scale structure cosmology}",
      journal = {\mnras},
         year = 2017,
        month = mar,
       volume = {465},
       number = {3},
        pages = {2936-2965},
          doi = {10.1093/mnras/stw2792},
archivePrefix = {arXiv},
       eprint = {1603.02702},
 primaryClass = {astro-ph.CO},
       adsurl = {https://ui.adsabs.harvard.edu/abs/2017MNRAS.465.2936M}
}

@ARTICLE{Battaglia12,
       author = {{Battaglia}, N. and {Bond}, J.~R. and {Pfrommer}, C. and {Sievers}, J.~L.},
        title = "{On the Cluster Physics of Sunyaev-Zel'dovich and X-Ray Surveys. II. Deconstructing the Thermal SZ Power Spectrum}",
      journal = {\apj},
         year = 2012,
        month = oct,
       volume = {758},
       number = {2},
          eid = {75},
        pages = {75},
          doi = {10.1088/0004-637X/758/2/75},
archivePrefix = {arXiv},
       eprint = {1109.3711},
 primaryClass = {astro-ph.CO},
       adsurl = {https://ui.adsabs.harvard.edu/abs/2012ApJ...758...75B}
}

@ARTICLE{Harnois15,
       author = {{Harnois-D{\'e}raps}, Joachim and {van Waerbeke}, Ludovic and {Viola}, Massimo and {Heymans}, Catherine},
        title = "{Baryons, neutrinos, feedback and weak gravitational lensing}",
      journal = {\mnras},
         year = 2015,
        month = jun,
       volume = {450},
       number = {2},
        pages = {1212-1223},
          doi = {10.1093/mnras/stv646},
archivePrefix = {arXiv},
       eprint = {1407.4301},
 primaryClass = {astro-ph.CO},
       adsurl = {https://ui.adsabs.harvard.edu/abs/2015MNRAS.450.1212H}
}

@ARTICLE{CMBS4_Book,
       author = {{Abazajian}, Kevork and {Addison}, Graeme and {Adshead}, Peter and {Ahmed}, Zeeshan and {Allen}, Steven W. and {Alonso}, David and {Alvarez}, Marcelo and {Anderson}, Adam and {Arnold}, Kam S. and {Baccigalupi}, Carlo and {Bailey}, Kathy and {Barkats}, Denis and {Barron}, Darcy and {Barry}, Peter S. and {Bartlett}, James G. and {Basu Thakur}, Ritoban and {Battaglia}, Nicholas and {Baxter}, Eric and {Bean}, Rachel and {Bebek}, Chris and {Bender}, Amy N. and {Benson}, Bradford A. and {Berger}, Edo and {Bhimani}, Sanah and {Bischoff}, Colin A. and {Bleem}, Lindsey and {Bocquet}, Sebastian and {Boddy}, Kimberly and {Bonato}, Matteo and {Bond}, J. Richard and {Borrill}, Julian and {Bouchet}, Fran{\c{c}}ois R. and {Brown}, Michael L. and {Bryan}, Sean and {Burkhart}, Blakesley and {Buza}, Victor and {Byrum}, Karen and {Calabrese}, Erminia and {Calafut}, Victoria and {Caldwell}, Robert and {Carlstrom}, John E. and {Carron}, Julien and {Cecil}, Thomas and {Challinor}, Anthony and {Chang}, Clarence L. and {Chinone}, Yuji and {Cho}, Hsiao-Mei Sherry and {Cooray}, Asantha and {Crawford}, Thomas M. and {Crites}, Abigail and {Cukierman}, Ari and {Cyr-Racine}, Francis-Yan and {de Haan}, Tijmen and {de Zotti}, Gianfranco and {Delabrouille}, Jacques and {Demarteau}, Marcel and {Devlin}, Mark and {Di Valentino}, Eleonora and {Dobbs}, Matt and {Duff}, Shannon and {Duivenvoorden}, Adriaan and {Dvorkin}, Cora and {Edwards}, William and {Eimer}, Joseph and {Errard}, Josquin and {Essinger-Hileman}, Thomas and {Fabbian}, Giulio and {Feng}, Chang and {Ferraro}, Simone and {Filippini}, Jeffrey P. and {Flauger}, Raphael and {Flaugher}, Brenna and {Fraisse}, Aurelien A. and {Frolov}, Andrei and {Galitzki}, Nicholas and {Galli}, Silvia and {Ganga}, Ken and {Gerbino}, Martina and {Gilchriese}, Murdock and {Gluscevic}, Vera and {Green}, Daniel and {Grin}, Daniel and {Grohs}, Evan and {Gualtieri}, Riccardo and {Guarino}, Victor and {Gudmundsson}, Jon E. and {Habib}, Salman and {Haller}, Gunther and {Halpern}, Mark and {Halverson}, Nils W. and {Hanany}, Shaul and {Harrington}, Kathleen and {Hasegawa}, Masaya and {Hasselfield}, Matthew and {Hazumi}, Masashi and {Heitmann}, Katrin and {Henderson}, Shawn and {Henning}, Jason W. and {Hill}, J. Colin and {Hlozek}, Ren{\'e}e and {Holder}, Gil and {Holzapfel}, William and {Hubmayr}, Johannes and {Huffenberger}, Kevin M. and {Huffer}, Michael and {Hui}, Howard and {Irwin}, Kent and {Johnson}, Bradley R. and {Johnstone}, Doug and {Jones}, William C. and {Karkare}, Kirit and {Katayama}, Nobuhiko and {Kerby}, James and {Kernovsky}, Sarah and {Keskitalo}, Reijo and {Kisner}, Theodore and {Knox}, Lloyd and {Kosowsky}, Arthur and {Kovac}, John and {Kovetz}, Ely D. and {Kuhlmann}, Steve and {Kuo}, Chao-lin and {Kurita}, Nadine and {Kusaka}, Akito and {Lahteenmaki}, Anne and {Lawrence}, Charles R. and {Lee}, Adrian T. and {Lewis}, Antony and {Li}, Dale and {Linder}, Eric and {Loverde}, Marilena and {Lowitz}, Amy and {Madhavacheril}, Mathew S. and {Mantz}, Adam and {Matsuda}, Frederick and {Mauskopf}, Philip and {McMahon}, Jeff and {McQuinn}, Matthew and {Meerburg}, P. Daniel and {Melin}, Jean-Baptiste and {Meyers}, Joel and {Millea}, Marius and {Mohr}, Joseph and {Moncelsi}, Lorenzo and {Mroczkowski}, Tony and {Mukherjee}, Suvodip and {M{\"u}nchmeyer}, Moritz and {Nagai}, Daisuke and {Nagy}, Johanna and {Namikawa}, Toshiya and {Nati}, Federico and {Natoli}, Tyler and {Negrello}, Mattia and {Newburgh}, Laura and {Niemack}, Michael D. and {Nishino}, Haruki and {Nordby}, Martin and {Novosad}, Valentine and {O'Connor}, Paul and {Obied}, Georges and {Padin}, Stephen and {Pandey}, Shivam and {Partridge}, Bruce and {Pierpaoli}, Elena and {Pogosian}, Levon and {Pryke}, Clement and {Puglisi}, Giuseppe and {Racine}, Benjamin and {Raghunathan}, Srinivasan and {Rahlin}, Alexandra and {Rajagopalan}, Srini and {Raveri}, Marco and {Reichanadter}, Mark and {Reichardt}, Christian L. and {Remazeilles}, Mathieu and {Rocha}, Graca and {Roe}, Natalie A. and {Roy}, Anirban and {Ruhl}, John and {Salatino}, Maria and {Saliwanchik}, Benjamin and {Schaan}, Emmanuel and {Schillaci}, Alessandro and {Schmittfull}, Marcel M. and {Scott}, Douglas and {Sehgal}, Neelima and {Shandera}, Sarah and {Sheehy}, Christopher and {Sherwin}, Blake D. and {Shirokoff}, Erik and {Simon}, Sara M. and {Slosar}, Anze and {Somerville}, Rachel and {Spergel}, David and {Staggs}, Suzanne T. and {Stark}, Antony and {Stompor}, Radek and {Story}, Kyle T. and {Stoughton}, Chris and {Suzuki}, Aritoki},
        title = "{CMB-S4 Science Case, Reference Design, and Project Plan}",
      journal = {arXiv e-prints},
         year = 2019,
        month = jul,
          eid = {arXiv:1907.04473},
        pages = {arXiv:1907.04473},
          doi = {10.48550/arXiv.1907.04473},
archivePrefix = {arXiv},
       eprint = {1907.04473},
 primaryClass = {astro-ph.IM},
       adsurl = {https://ui.adsabs.harvard.edu/abs/2019arXiv190704473A}
}

@ARTICLE{CMBS4_paper,
       author = {{Abazajian}, Kevork and {Abdulghafour}, Arwa and {Addison}, Graeme E. and {Adshead}, Peter and {Ahmed}, Zeeshan and {Ajello}, Marco and {Akerib}, Daniel and {Allen}, Steven W. and {Alonso}, David and {Alvarez}, Marcelo and {Amin}, Mustafa A. and {Amiri}, Mandana and {Anderson}, Adam and {Ansarinejad}, Behzad and {Archipley}, Melanie and {Arnold}, Kam S. and {Ashby}, Matt and {Aung}, Han and {Baccigalupi}, Carlo and {Baker}, Carina and {Bakshi}, Abhishek and {Bard}, Debbie and {Barkats}, Denis and {Barron}, Darcy and {Barry}, Peter S. and {Bartlett}, James G. and {Barton}, Paul and {Basu Thakur}, Ritoban and {Battaglia}, Nicholas and {Beall}, Jim and {Bean}, Rachel and {Beck}, Dominic and {Belkner}, Sebastian and {Benabed}, Karim and {Bender}, Amy N. and {Benson}, Bradford A. and {Besuner}, Bobby and {Bethermin}, Matthieu and {Bhimani}, Sanah and {Bianchini}, Federico and {Biquard}, Simon and {Birdwell}, Ian and {Bischoff}, Colin A. and {Bleem}, Lindsey and {Bocaz}, Paulina and {Bock}, James J. and {Bocquet}, Sebastian and {Boddy}, Kimberly K. and {Bond}, J. Richard and {Borrill}, Julian and {Bouchet}, Francois R. and {Brinckmann}, Thejs and {Brown}, Michael L. and {Bryan}, Sean and {Buza}, Victor and {Byrum}, Karen and {Calabrese}, Erminia and {Calafut}, Victoria and {Caldwell}, Robert and {Carlstrom}, John E. and {Carron}, Julien and {Cecil}, Thomas and {Challinor}, Anthony and {Chan}, Victor and {Chang}, Clarence L. and {Chapman}, Scott and {Charles}, Eric and {Chauvin}, Eric and {Cheng}, Cheng and {Chesmore}, Grace and {Cheung}, Kolen and {Chinone}, Yuji and {Chluba}, Jens and {Cho}, Hsiao-Mei Sherry and {Choi}, Steve and {Clancy}, Justin and {Clark}, Susan and {Cooray}, Asantha and {Coppi}, Gabriele and {Corlett}, John and {Coulton}, Will and {Crawford}, Thomas M. and {Crites}, Abigail and {Cukierman}, Ari and {Cyr-Racine}, Francis-Yan and {Dai}, Wei-Ming and {Daley}, Cail and {Dart}, Eli and {Daues}, Gregorg and {de Haan}, Tijmen and {Deaconu}, Cosmin and {Delabrouille}, Jacques and {Derylo}, Greg and {Devlin}, Mark and {Di Valentino}, Eleonora and {Dierickx}, Marion and {Dober}, Brad and {Doriese}, Randy and {Duff}, Shannon and {Dutcher}, Daniel and {Dvorkin}, Cora and {D{\"u}nner}, Rolando and {Eftekhari}, Tarraneh and {Eimer}, Joseph and {El Bouhargani}, Hamza and {Elleflot}, Tucker and {Emerson}, Nick and {Errard}, Josquin and {Essinger-Hileman}, Thomas and {Fabbian}, Giulio and {Fanfani}, Valentina and {Fasano}, Alessandro and {Feng}, Chang and {Ferraro}, Simone and {Filippini}, Jeffrey P. and {Flauger}, Raphael and {Flaugher}, Brenna and {Fraisse}, Aurelien A. and {Frisch}, Josef and {Frolov}, Andrei and {Galitzki}, Nicholas and {Gallardo}, Patricio A. and {Galli}, Silvia and {Ganga}, Ken and {Gerbino}, Martina and {Giannakopoulos}, Christos and {Gilchriese}, Murdock and {Gluscevic}, Vera and {Goeckner-Wald}, Neil and {Goldfinger}, David and {Green}, Daniel and {Grimes}, Paul and {Grin}, Daniel and {Grohs}, Evan and {Gualtieri}, Riccardo and {Guarino}, Vic and {Gudmundsson}, Jon E. and {Gullett}, Ian and {Guns}, Sam and {Habib}, Salman and {Haller}, Gunther and {Halpern}, Mark and {Halverson}, Nils W. and {Hanany}, Shaul and {Hand}, Emma and {Harrington}, Kathleen and {Hasegawa}, Masaya and {Hasselfield}, Matthew and {Hazumi}, Masashi and {Heitmann}, Katrin and {Henderson}, Shawn and {Hensley}, Brandon and {Herbst}, Ryan and {Hervias-Caimapo}, Carlos and {Hill}, J. Colin and {Hills}, Richard and {Hivon}, Eric and {Hlozek}, Ren{\'e}e and {Ho}, Anna and {Holder}, Gil and {Hollister}, Matt and {Holzapfel}, William and {Hood}, John and {Hotinli}, Selim and {Hryciuk}, Alec and {Hubmayr}, Johannes and {Huffenberger}, Kevin M. and {Hui}, Howard and {Ib{\'a} nez}, Roberto and {Ibitoye}, Ayodeji and {Ikape}, Margaret and {Irwin}, Kent and {Jacobus}, Cooper and {Jeong}, Oliver and {Johnson}, Bradley R. and {Johnstone}, Doug and {Jones}, William C. and {Joseph}, John and {Jost}, Baptiste and {Kang}, Jae Hwan and {Kaplan}, Ari and {Karkare}, Kirit S. and {Katayama}, Nobuhiko and {Keskitalo}, Reijo and {King}, Cesiley and {Kisner}, Theodore and {Klein}, Matthias and {Knox}, Lloyd and {Koopman}, Brian J. and {Kosowsky}, Arthur and {Kovac}, John and {Kovetz}, Ely D. and {Krolewski}, Alex and {Kubik}, Donna and {Kuhlmann}, Steve and {Kuo}, Chao-Lin and {Kusaka}, Akito and {L{\"a}hteenm{\"a}ki}, Anne and {Lau}, Kenny and {Lawrence}, Charles R.},
        title = "{Snowmass 2021 CMB-S4 White Paper}",
      journal = {arXiv e-prints},
         year = 2022,
        month = mar,
          eid = {arXiv:2203.08024},
        pages = {arXiv:2203.08024},
          doi = {10.48550/arXiv.2203.08024},
archivePrefix = {arXiv},
       eprint = {2203.08024},
 primaryClass = {astro-ph.CO},
       adsurl = {https://ui.adsabs.harvard.edu/abs/2022arXiv220308024A}
}

@ARTICLE{Euclid20,
       author = {{Euclid Collaboration} and {Blanchard}, A. and {Camera}, S. and {Carbone}, C. and {Cardone}, V.~F. and {Casas}, S. and {Clesse}, S. and {Ili{\'c}}, S. and {Kilbinger}, M. and {Kitching}, T. and {Kunz}, M. and {Lacasa}, F. and {Linder}, E. and {Majerotto}, E. and {Markovi{\v{c}}}, K. and {Martinelli}, M. and {Pettorino}, V. and {Pourtsidou}, A. and {Sakr}, Z. and {S{\'a}nchez}, A.~G. and {Sapone}, D. and {Tutusaus}, I. and {Yahia-Cherif}, S. and {Yankelevich}, V. and {Andreon}, S. and {Aussel}, H. and {Balaguera-Antol{\'\i}nez}, A. and {Baldi}, M. and {Bardelli}, S. and {Bender}, R. and {Biviano}, A. and {Bonino}, D. and {Boucaud}, A. and {Bozzo}, E. and {Branchini}, E. and {Brau-Nogue}, S. and {Brescia}, M. and {Brinchmann}, J. and {Burigana}, C. and {Cabanac}, R. and {Capobianco}, V. and {Cappi}, A. and {Carretero}, J. and {Carvalho}, C.~S. and {Casas}, R. and {Castander}, F.~J. and {Castellano}, M. and {Cavuoti}, S. and {Cimatti}, A. and {Cledassou}, R. and {Colodro-Conde}, C. and {Congedo}, G. and {Conselice}, C.~J. and {Conversi}, L. and {Copin}, Y. and {Corcione}, L. and {Coupon}, J. and {Courtois}, H.~M. and {Cropper}, M. and {Da Silva}, A. and {de la Torre}, S. and {Di Ferdinando}, D. and {Dubath}, F. and {Ducret}, F. and {Duncan}, C.~A.~J. and {Dupac}, X. and {Dusini}, S. and {Fabbian}, G. and {Fabricius}, M. and {Farrens}, S. and {Fosalba}, P. and {Fotopoulou}, S. and {Fourmanoit}, N. and {Frailis}, M. and {Franceschi}, E. and {Franzetti}, P. and {Fumana}, M. and {Galeotta}, S. and {Gillard}, W. and {Gillis}, B. and {Giocoli}, C. and {G{\'o}mez-Alvarez}, P. and {Graci{\'a}-Carpio}, J. and {Grupp}, F. and {Guzzo}, L. and {Hoekstra}, H. and {Hormuth}, F. and {Israel}, H. and {Jahnke}, K. and {Keihanen}, E. and {Kermiche}, S. and {Kirkpatrick}, C.~C. and {Kohley}, R. and {Kubik}, B. and {Kurki-Suonio}, H. and {Ligori}, S. and {Lilje}, P.~B. and {Lloro}, I. and {Maino}, D. and {Maiorano}, E. and {Marggraf}, O. and {Martinet}, N. and {Marulli}, F. and {Massey}, R. and {Medinaceli}, E. and {Mei}, S. and {Mellier}, Y. and {Metcalf}, B. and {Metge}, J.~J. and {Meylan}, G. and {Moresco}, M. and {Moscardini}, L. and {Munari}, E. and {Nichol}, R.~C. and {Niemi}, S. and {Nucita}, A.~A. and {Padilla}, C. and {Paltani}, S. and {Pasian}, F. and {Percival}, W.~J. and {Pires}, S. and {Polenta}, G. and {Poncet}, M. and {Pozzetti}, L. and {Racca}, G.~D. and {Raison}, F. and {Renzi}, A. and {Rhodes}, J. and {Romelli}, E. and {Roncarelli}, M. and {Rossetti}, E. and {Saglia}, R. and {Schneider}, P. and {Scottez}, V. and {Secroun}, A. and {Sirri}, G. and {Stanco}, L. and {Starck}, J.-L. and {Sureau}, F. and {Tallada-Cresp{\'\i}}, P. and {Tavagnacco}, D. and {Taylor}, A.~N. and {Tenti}, M. and {Tereno}, I. and {Toledo-Moreo}, R. and {Torradeflot}, F. and {Valenziano}, L. and {Vassallo}, T. and {Verdoes Kleijn}, G.~A. and {Viel}, M. and {Wang}, Y. and {Zacchei}, A. and {Zoubian}, J. and {Zucca}, E.},
        title = "{Euclid preparation. VII. Forecast validation for Euclid cosmological probes}",
      journal = {\aap},
         year = 2020,
        month = oct,
       volume = {642},
          eid = {A191},
        pages = {A191},
          doi = {10.1051/0004-6361/202038071},
archivePrefix = {arXiv},
       eprint = {1910.09273},
 primaryClass = {astro-ph.CO},
       adsurl = {https://ui.adsabs.harvard.edu/abs/2020A&A...642A.191E}
}

@ARTICLE{Corasaniti25,
       author = {{Corasaniti}, P.~S. and {Richardson}, T.~R.~G. and {Ettori}, S. and {De Petris}, M. and {Rasia}, E. and {Cui}, W. and {Yepes}, G. and {Gianfagna}, G. and {Le Brun}, A.~M.~C. and {Rasera}, Y.},
        title = "{The impact of baryons on the sparsity of simulated galaxy clusters from THE THREE HUNDRED Project: Astrophysical and cosmological implications}",
      journal = {\aap},
         year = 2025,
        month = may,
       volume = {697},
          eid = {A33},
        pages = {A33},
          doi = {10.1051/0004-6361/202553914},
archivePrefix = {arXiv},
       eprint = {2503.16379},
 primaryClass = {astro-ph.CO},
       adsurl = {https://ui.adsabs.harvard.edu/abs/2025A&A...697A..33C}
}

@ARTICLE{Raghunathan22b,
       author = {{Raghunathan}, Srinivasan},
        title = "{Assessing the Importance of Noise from Thermal Sunyaev-Zel'dovich Signals for CMB Cluster Surveys and Cluster Cosmology}",
      journal = {\apj},
         year = 2022,
        month = mar,
       volume = {928},
       number = {1},
          eid = {16},
        pages = {16},
          doi = {10.3847/1538-4357/ac510f},
archivePrefix = {arXiv},
       eprint = {2112.07656},
 primaryClass = {astro-ph.CO},
       adsurl = {https://ui.adsabs.harvard.edu/abs/2022ApJ...928...16R}
}

@ARTICLE{Abazajian_GW,
       author = {{Abazajian}, Kevork and {Addison}, Graeme E. and {Adshead}, Peter and {Ahmed}, Zeeshan and {Akerib}, Daniel and {Ali}, Aamir and {Allen}, Steven W. and {Alonso}, David and {Alvarez}, Marcelo and {Amin}, Mustafa A. and {Anderson}, Adam and {Arnold}, Kam S. and {Ashton}, Peter and {Baccigalupi}, Carlo and {Bard}, Debbie and {Barkats}, Denis and {Barron}, Darcy and {Barry}, Peter S. and {Bartlett}, James G. and {Basu Thakur}, Ritoban and {Battaglia}, Nicholas and {Bean}, Rachel and {Bebek}, Chris and {Bender}, Amy N. and {Benson}, Bradford A. and {Bianchini}, Federico and {Bischoff}, Colin A. and {Bleem}, Lindsey and {Bock}, James J. and {Bocquet}, Sebastian and {Boddy}, Kimberly K. and {Richard Bond}, J. and {Borrill}, Julian and {Bouchet}, Fran{\c{c}}ois R. and {Brinckmann}, Thejs and {Brown}, Michael L. and {Bryan}, Sean and {Buza}, Victor and {Byrum}, Karen and {Hervias Caimapo}, Carlos and {Calabrese}, Erminia and {Calafut}, Victoria and {Caldwell}, Robert and {Carlstrom}, John E. and {Carron}, Julien and {Cecil}, Thomas and {Challinor}, Anthony and {Chang}, Clarence L. and {Chinone}, Yuji and {Sherry Cho}, Hsiao-Mei and {Cooray}, Asantha and {Coulton}, Will and {Crawford}, Thomas M. and {Crites}, Abigail and {Cukierman}, Ari and {Cyr-Racine}, Francis-Yan and {de Haan}, Tijmen and {Delabrouille}, Jacques and {Devlin}, Mark and {Di Valentino}, Eleonora and {Dierickx}, Marion and {Dobbs}, Matt and {Duff}, Shannon and {Dvorkin}, Cora and {Eimer}, Joseph and {Elleflot}, Tucker and {Errard}, Josquin and {Essinger-Hileman}, Thomas and {Fabbian}, Giulio and {Feng}, Chang and {Ferraro}, Simone and {Filippini}, Jeffrey P. and {Flauger}, Raphael and {Flaugher}, Brenna and {Fraisse}, Aurelien A. and {Frolov}, Andrei and {Galitzki}, Nicholas and {Gallardo}, Patricio A. and {Galli}, Silvia and {Ganga}, Ken and {Gerbino}, Martina and {Gluscevic}, Vera and {Goeckner-Wald}, Neil and {Green}, Daniel and {Grin}, Daniel and {Grohs}, Evan and {Gualtieri}, Riccardo and {Gudmundsson}, Jon E. and {Gullett}, Ian and {Gupta}, Nikhel and {Habib}, Salman and {Halpern}, Mark and {Halverson}, Nils W. and {Hanany}, Shaul and {Harrington}, Kathleen and {Hasegawa}, Masaya and {Hasselfield}, Matthew and {Hazumi}, Masashi and {Heitmann}, Katrin and {Henderson}, Shawn and {Hensley}, Brandon and {Hill}, Charles and {Colin Hill}, J. and {Hlo{\v{z}}ek}, Ren{\'e}e and {Patty Ho}, Shuay-Pwu and {Hoang}, Thuong and {Holder}, Gil and {Holzapfel}, William and {Hood}, John and {Hubmayr}, Johannes and {Huffenberger}, Kevin M. and {Hui}, Howard and {Irwin}, Kent and {Jeong}, Oliver and {Johnson}, Bradley R. and {Jones}, William C. and {Hwan Kang}, Jae and {Karkare}, Kirit S. and {Katayama}, Nobuhiko and {Keskitalo}, Reijo and {Kisner}, Theodore and {Knox}, Lloyd and {Koopman}, Brian J. and {Kosowsky}, Arthur and {Kovac}, John and {Kovetz}, Ely D. and {Kuhlmann}, Steve and {Kuo}, Chao-lin and {Kusaka}, Akito and {L{\"a}hteenm{\"a}ki}, Anne and {Lawrence}, Charles R. and {Lee}, Adrian T. and {Lewis}, Antony and {Li}, Dale and {Linder}, Eric and {Loverde}, Marilena and {Lowitz}, Amy and {Lubin}, Phil and {Madhavacheril}, Mathew S. and {Mantz}, Adam and {Marques}, Gabriela and {Matsuda}, Frederick and {Mauskopf}, Philip and {McCarrick}, Heather and {McMahon}, Jeffrey and {Daniel Meerburg}, P. and {Melin}, Jean-Baptiste and {Menanteau}, Felipe and {Meyers}, Joel and {Millea}, Marius and {Mohr}, Joseph and {Moncelsi}, Lorenzo and {Monzani}, Maria and {Mroczkowski}, Tony and {Mukherjee}, Suvodip and {Nagy}, Johanna and {Namikawa}, Toshiya and {Nati}, Federico and {Natoli}, Tyler and {Newburgh}, Laura and {Niemack}, Michael D. and {Nishino}, Haruki and {Nord}, Brian and {Novosad}, Valentine and {O'Brient}, Roger and {Padin}, Stephen and {Palladino}, Steven and {Partridge}, Bruce and {Petravick}, Don and {Pierpaoli}, Elena and {Pogosian}, Levon and {Prabhu}, Karthik and {Pryke}, Clement and {Puglisi}, Giuseppe and {Racine}, Benjamin and {Rahlin}, Alexandra and {Sathyanarayana Rao}, Mayuri and {Raveri}, Marco and {Reichardt}, Christian L. and {Remazeilles}, Mathieu and {Rocha}, Graca and {Roe}, Natalie A. and {Roy}, Anirban and {Ruhl}, John E. and {Salatino}, Maria and {Saliwanchik}, Benjamin and {Schaan}, Emmanuel and {Schillaci}, Alessandro and {Schmitt}, Benjamin and {Schmittfull}, Marcel M. and {Scott}, Douglas and {Sehgal}, Neelima and {Shandera}, Sarah and {Sherwin}, Blake D. and {Shirokoff}, Erik and {Simon}, Sara M. and {Slosar}, An{\v{z}}e and {Spergel}, David and {St. Germaine}, Tyler and {Staggs}, Suzanne T.},
        title = "{CMB-S4: Forecasting Constraints on Primordial Gravitational Waves}",
      journal = {\apj},
         year = 2022,
        month = feb,
       volume = {926},
       number = {1},
          eid = {54},
        pages = {54},
          doi = {10.3847/1538-4357/ac1596},
archivePrefix = {arXiv},
       eprint = {2008.12619},
 primaryClass = {astro-ph.CO},
       adsurl = {https://ui.adsabs.harvard.edu/abs/2022ApJ...926...54A}
}

@ARTICLE{SPT3G,
       author = {{Camphuis}, E. and {Quan}, W. and {Balkenhol}, L. and {Khalife}, A.~R. and {Ge}, F. and {Guidi}, F. and {Huang}, N. and {Lynch}, G.~P. and {Omori}, Y. and {Trendafilova}, C. and {Anderson}, A.~J. and {Ansarinejad}, B. and {Archipley}, M. and {Barry}, P.~S. and {Benabed}, K. and {Bender}, A.~N. and {Benson}, B.~A. and {Bianchini}, F. and {Bleem}, L.~E. and {Bouchet}, F.~R. and {Bryant}, L. and {Campitiello}, M.~G. and {Carlstrom}, J.~E. and {Chang}, C.~L. and {Chaubal}, P. and {Chichura}, P.~M. and {Chokshi}, A. and {Chou}, T. -L. and {Coerver}, A. and {Crawford}, T.~M. and {Daley}, C. and {de Haan}, T. and {Dibert}, K.~R. and {Dobbs}, M.~A. and {Doohan}, M. and {Doussot}, A. and {Dutcher}, D. and {Everett}, W. and {Feng}, C. and {Ferguson}, K.~R. and {Fichman}, K. and {Foster}, A. and {Galli}, S. and {Gambrel}, A.~E. and {Gardner}, R.~W. and {Goeckner-Wald}, N. and {Gualtieri}, R. and {Guns}, S. and {Halverson}, N.~W. and {Hivon}, E. and {Holder}, G.~P. and {Holzapfel}, W.~L. and {Hood}, J.~C. and {Hryciuk}, A. and {K{\'e}ruzor{\'e}}, F. and {Knox}, L. and {Korman}, M. and {Kornoelje}, K. and {Kuo}, C. -L. and {Levy}, K. and {Lowitz}, A.~E. and {Lu}, C. and {Maniyar}, A. and {Martsen}, E.~S. and {Menanteau}, F. and {Millea}, M. and {Montgomery}, J. and {Nakato}, Y. and {Natoli}, T. and {Noble}, G.~I. and {Ouellette}, A. and {Pan}, Z. and {Paschos}, P. and {Phadke}, K.~A. and {Pollak}, A.~W. and {Prabhu}, K. and {Raghunathan}, S. and {Rahimi}, M. and {Rahlin}, A. and {Reichardt}, C.~L. and {Rouble}, M. and {Ruhl}, J.~E. and {Schiappucci}, E. and {Simpson}, A. and {Sobrin}, J.~A. and {Stark}, A.~A. and {Stephen}, J. and {Tandoi}, C. and {Thorne}, B. and {Umilta}, C. and {Vieira}, J.~D. and {Vitrier}, A. and {Wan}, Y. and {Whitehorn}, N. and {Wu}, W.~L.~K. and {Young}, M.~R. and {Zebrowski}, J.~A.},
        title = "{SPT-3G D1: CMB temperature and polarization power spectra and cosmology from 2019 and 2020 observations of the SPT-3G Main field}",
      journal = {arXiv e-prints},
         year = 2025,
        month = jun,
          eid = {arXiv:2506.20707},
        pages = {arXiv:2506.20707},
          doi = {10.48550/arXiv.2506.20707},
archivePrefix = {arXiv},
       eprint = {2506.20707},
 primaryClass = {astro-ph.CO},
       adsurl = {https://ui.adsabs.harvard.edu/abs/2025arXiv250620707C}
}

@ARTICLE{Raghunathan22a,
       author = {{Raghunathan}, Srinivasan and {Whitehorn}, Nathan and {Alvarez}, Marcelo A. and {Aung}, Han and {Battaglia}, Nicholas and {Holder}, Gilbert P. and {Nagai}, Daisuke and {Pierpaoli}, Elena and {Reichardt}, Christian L. and {Vieira}, Joaquin D.},
        title = "{Constraining Cluster Virialization Mechanism and Cosmology Using Thermal-SZ-selected Clusters from Future CMB Surveys}",
      journal = {\apj},
         year = 2022,
        month = feb,
       volume = {926},
       number = {2},
          eid = {172},
        pages = {172},
          doi = {10.3847/1538-4357/ac4712},
archivePrefix = {arXiv},
       eprint = {2107.10250},
 primaryClass = {astro-ph.CO},
       adsurl = {https://ui.adsabs.harvard.edu/abs/2022ApJ...926..172R}
}

@ARTICLE{Aymerich25,
       author = {{Aymerich}, G. and {Douspis}, M. and {Battaglia}, N. and {Aghanim}, N. and {Salvati}, L. and {Pratt}, G.~W. and {Fabbian}, G.},
        title = "{Implications of SPT and eROSITA cosmologies for Planck cluster number counts and t-SZ power spectrum}",
      journal = {arXiv e-prints},
         year = 2025,
        month = sep,
          eid = {arXiv:2509.08673},
        pages = {arXiv:2509.08673},
          doi = {10.48550/arXiv.2509.08673},
archivePrefix = {arXiv},
       eprint = {2509.08673},
 primaryClass = {astro-ph.CO},
       adsurl = {https://ui.adsabs.harvard.edu/abs/2025arXiv250908673A}
}

@INPROCEEDINGS{Bolliet20,
       author = {{Bolliet}, Boris},
        title = "{Cosmology with Thermal Sunyaev Zeldovich Power Spectrum and Cluster Counts: Consistency, Tensions and Prospects}",
    booktitle = {mm Universe @ NIKA2 - Observing the mm Universe with the NIKA2 Camera},
         year = 2020,
       series = {European Physical Journal Web of Conferences},
       volume = {228},
        month = jun,
    publisher = {EDP},
          eid = {00005},
        pages = {00005},
          doi = {10.1051/epjconf/202022800005},
       adsurl = {https://ui.adsabs.harvard.edu/abs/2020EPJWC.22800005B}
}

@ARTICLE{Tinker2008,
       author = {{Tinker}, Jeremy and {Kravtsov}, Andrey V. and {Klypin}, Anatoly and
         {Abazajian}, Kevork and {Warren}, Michael and {Yepes}, Gustavo and
         {Gottl{\"o}ber}, Stefan and {Holz}, Daniel E.},
        title = "{Toward a Halo Mass Function for Precision Cosmology: The Limits of Universality}",
      journal = {\apj},
         year = 2008,
        month = dec,
       volume = {688},
       number = {2},
        pages = {709-728},
          doi = {10.1086/591439},
archivePrefix = {arXiv},
       eprint = {0803.2706},
 primaryClass = {astro-ph},
       adsurl = {https://ui.adsabs.harvard.edu/abs/2008ApJ...688..709T}
}

@ARTICLE{Tinker2010,
       author = {{Tinker}, Jeremy L. and {Robertson}, Brant E. and {Kravtsov}, Andrey V. and
         {Klypin}, Anatoly and {Warren}, Michael S. and {Yepes}, Gustavo and
         {Gottl{\"o}ber}, Stefan},
        title = "{The Large-scale Bias of Dark Matter Halos: Numerical Calibration and Model Tests}",
      journal = {\apj},
         year = 2010,
        month = dec,
       volume = {724},
       number = {2},
        pages = {878-886},
          doi = {10.1088/0004-637X/724/2/878},
archivePrefix = {arXiv},
       eprint = {1001.3162},
 primaryClass = {astro-ph.CO},
       adsurl = {https://ui.adsabs.harvard.edu/abs/2010ApJ...724..878T}
}

@ARTICLE{Allen11,
       author = {{Allen}, Steven W. and {Evrard}, August E. and {Mantz}, Adam B.},
        title = "{Cosmological Parameters from Observations of Galaxy Clusters}",
      journal = {\araa},
         year = 2011,
        month = sep,
       volume = {49},
       number = {1},
        pages = {409-470},
          doi = {10.1146/annurev-astro-081710-102514},
archivePrefix = {arXiv},
       eprint = {1103.4829},
 primaryClass = {astro-ph.CO},
       adsurl = {https://ui.adsabs.harvard.edu/abs/2011ARA&A..49..409A}
}

@ARTICLE{navarro96,
       author = {{Navarro}, Julio F. and {Frenk}, Carlos S. and {White}, Simon D.~M.},
        title = "{The Structure of Cold Dark Matter Halos}",
      journal = {\apj},
         year = 1996,
        month = may,
       volume = {462},
        pages = {563},
          doi = {10.1086/177173},
archivePrefix = {arXiv},
       eprint = {astro-ph/9508025},
 primaryClass = {astro-ph},
       adsurl = {https://ui.adsabs.harvard.edu/abs/1996ApJ...462..563N}
}

@ARTICLE{Limber53,
       author = {{Limber}, D. Nelson},
        title = "{The Analysis of Counts of the Extragalactic Nebulae in Terms of a Fluctuating Density Field.}",
      journal = {\apj},
         year = 1953,
        month = jan,
       volume = {117},
        pages = {134},
          doi = {10.1086/145672},
       adsurl = {https://ui.adsabs.harvard.edu/abs/1953ApJ...117..134L}
}

@ARTICLE{Lewis2000,
       author = {{Lewis}, Antony and {Challinor}, Anthony and {Lasenby}, Anthony},
        title = "{Efficient Computation of Cosmic Microwave Background Anisotropies in Closed Friedmann-Robertson-Walker Models}",
      journal = {\apj},
         year = 2000,
        month = aug,
       volume = {538},
       number = {2},
        pages = {473-476},
          doi = {10.1086/309179},
archivePrefix = {arXiv},
       eprint = {astro-ph/9911177},
 primaryClass = {astro-ph},
       adsurl = {https://ui.adsabs.harvard.edu/abs/2000ApJ...538..473L}
}

@ARTICLE{Duffy2008,
       author = {{Duffy}, Alan R. and {Schaye}, Joop and {Kay}, Scott T. and
         {Dalla Vecchia}, Claudio},
        title = "{Dark matter halo concentrations in the Wilkinson Microwave Anisotropy Probe year 5 cosmology}",
      journal = {\mnras},
         year = 2008,
        month = oct,
       volume = {390},
       number = {1},
        pages = {L64-L68},
          doi = {10.1111/j.1745-3933.2008.00537.x},
archivePrefix = {arXiv},
       eprint = {0804.2486},
 primaryClass = {astro-ph},
       adsurl = {https://ui.adsabs.harvard.edu/abs/2008MNRAS.390L..64D}
}

@ARTICLE{SO_22,
       author = {{Hensley}, Brandon S. and {Clark}, Susan E. and {Fanfani}, Valentina and {Krachmalnicoff}, Nicoletta and {Fabbian}, Giulio and {Poletti}, Davide and {Puglisi}, Giuseppe and {Coppi}, Gabriele and {Nibauer}, Jacob and {Gerasimov}, Roman and {Galitzki}, Nicholas and {Choi}, Steve K. and {Ashton}, Peter C. and {Baccigalupi}, Carlo and {Baxter}, Eric and {Burkhart}, Blakesley and {Calabrese}, Erminia and {Chluba}, Jens and {Errard}, Josquin and {Frolov}, Andrei V. and {Herv{\'\i}as-Caimapo}, Carlos and {Huffenberger}, Kevin M. and {Johnson}, Bradley R. and {Jost}, Baptiste and {Keating}, Brian and {McCarrick}, Heather and {Nati}, Federico and {Sathyanarayana Rao}, Mayuri and {van Engelen}, Alexander and {Walker}, Samantha and {Wolz}, Kevin and {Xu}, Zhilei and {Zhu}, Ningfeng and {Zonca}, Andrea},
        title = "{The Simons Observatory: Galactic Science Goals and Forecasts}",
      journal = {\apj},
         year = 2022,
        month = apr,
       volume = {929},
       number = {2},
          eid = {166},
        pages = {166},
          doi = {10.3847/1538-4357/ac5e36},
archivePrefix = {arXiv},
       eprint = {2111.02425},
 primaryClass = {astro-ph.GA},
       adsurl = {https://ui.adsabs.harvard.edu/abs/2022ApJ...929..166H}
}

@ARTICLE{Ibitoye22,
       author = {{Ibitoye}, Ayodeji and {Tramonte}, Denis and {Ma}, Yin-Zhe and {Dai}, Wei-Ming},
        title = "{Cross Correlation between the Thermal Sunyaev-Zel'dovich Effect and Projected Galaxy Density Field}",
      journal = {\apj},
         year = 2022,
        month = aug,
       volume = {935},
       number = {1},
          eid = {18},
        pages = {18},
          doi = {10.3847/1538-4357/ac7b8c},
archivePrefix = {arXiv},
       eprint = {2206.05689},
 primaryClass = {astro-ph.CO},
       adsurl = {https://ui.adsabs.harvard.edu/abs/2022ApJ...935...18I}
}

@ARTICLE{Barranco26,
       author = {{Barranco-Llorca}, Isac and {Vall{\'e}s-P{\'e}rez}, David and {Planelles}, Susana and {Quilis}, Vicent},
        title = "{The signature of major mergers on the hydrostatic mass bias of galaxy clusters}",
      journal = {\aap},
         year = 2026,
        month = jan,
       volume = {705},
          eid = {A101},
        pages = {A101},
          doi = {10.1051/0004-6361/202556598},
archivePrefix = {arXiv},
       eprint = {2511.10730},
 primaryClass = {astro-ph.CO},
       adsurl = {https://ui.adsabs.harvard.edu/abs/2026A&A...705A.101B}
}

@ARTICLE{Arnaud2010,
       author = {{Arnaud}, M. and {Pratt}, G.~W. and {Piffaretti}, R. and
         {B{\"o}hringer}, H. and {Croston}, J.~H. and {Pointecouteau}, E.},
        title = "{The universal galaxy cluster pressure profile from a representative sample of nearby systems (REXCESS) and the Y$_{SZ}$ - M$_{500}$ relation}",
      journal = {\aap},
         year = 2010,
        month = jul,
       volume = {517},
          eid = {A92},
        pages = {A92},
          doi = {10.1051/0004-6361/200913416},
archivePrefix = {arXiv},
       eprint = {0910.1234},
 primaryClass = {astro-ph.CO},
       adsurl = {https://ui.adsabs.harvard.edu/abs/2010A&A...517A..92A}
}

@article{Planck:2018vyg,
    author = "Aghanim, Nabila and others",
    collaboration = "Planck",
    title = "{Planck 2018 results. VI. Cosmological parameters}",
    eprint = "1807.06209",
    archivePrefix = "arXiv",
    primaryClass = "astro-ph.CO",
    doi = "10.1051/0004-6361/201833910",
    journal = "Astron. Astrophys.",
    volume = "641",
    pages = "A6",
    year = "2020"
}

@ARTICLE{Mead21,
       author = {{Mead}, A.~J. and {Brieden}, S. and {Tr{\"o}ster}, T. and {Heymans}, C.},
        title = "{HMCODE-2020: improved modelling of non-linear cosmological power spectra with baryonic feedback}",
      journal = {\mnras},
         year = 2021,
        month = mar,
       volume = {502},
       number = {1},
        pages = {1401-1422},
          doi = {10.1093/mnras/stab082},
archivePrefix = {arXiv},
       eprint = {2009.01858},
 primaryClass = {astro-ph.CO},
       adsurl = {https://ui.adsabs.harvard.edu/abs/2021MNRAS.502.1401M}
}
\bibliographystyle{aasjournal}

\label{lastpage}
\end{document}